\documentclass[notitlepage,floatfix,aps,prl,twocolumn,superscriptaddress,10pt,floatfix]{revtex4-1}
\usepackage{graphicx}% Include figure files
\usepackage{dcolumn}% Align table columns on decimal point
\usepackage{bm}
\usepackage{braket}
\usepackage{amssymb}
\usepackage{hyperref}
\usepackage{xcolor}
\begin{document}

\title{Fundamental Limits to Coherent Photon Generation with Solid-State Atomlike Transitions}
\author{Z. X. Koong}
\email[Correspondence: ]{zk49@hw.ac.uk}
\affiliation{
 SUPA, Institute of Photonics and Quantum Sciences, Heriot-Watt University, Edinburgh EH14 4AS, Scotland, United Kingdom
}
\author{D. Scerri}
\affiliation{
 SUPA, Institute of Photonics and Quantum Sciences, Heriot-Watt University, Edinburgh EH14 4AS, Scotland, United Kingdom
}
\author{M. Rambach}
\affiliation{
 SUPA, Institute of Photonics and Quantum Sciences, Heriot-Watt University, Edinburgh EH14 4AS, Scotland, United Kingdom
}
\author{T. S. Santana}
\affiliation{%
Departamento de F\'{i}sica, Universidade Federal de Sergipe, Sergipe, 49100-000, Brazil
}%
\author{S. I. Park}
\affiliation{
 Center for Opto-Electronic Materials and Devices Research, Korea Institute of Science and Technology, Seoul 02792, Republic of Korea
}
\author{J. D. Song}
\affiliation{
 Center for Opto-Electronic Materials and Devices Research, Korea Institute of Science and Technology, Seoul 02792, Republic of Korea
}
\author{E. M. Gauger}
\affiliation{
 SUPA, Institute of Photonics and Quantum Sciences, Heriot-Watt University, Edinburgh EH14 4AS, Scotland, United Kingdom
}
\author{B. D. Gerardot}
\email[Correspondence: ]{b.d.gerardot@hw.ac.uk}
\affiliation{
 SUPA, Institute of Photonics and Quantum Sciences, Heriot-Watt University, Edinburgh EH14 4AS, Scotland, United Kingdom
}

\date{October 16, 2019}

\begin{abstract}
Coherent generation of indistinguishable single photons is crucial for many quantum communication and processing protocols. 
Solid-state realizations of two-level atomic transitions or three-level spin-$\Lambda$ systems offer significant advantages over their atomic counterparts for this purpose, albeit decoherence can arise due to environmental couplings. 
One popular approach to mitigate dephasing is to operate in the weak-excitation limit, where excited-state population is minimal and coherently scattered photons dominate over incoherent emission. 
Here we probe the coherence of photons produced using two-level and spin-$\Lambda$ solid-state systems. 
We observe that the coupling of the atomiclike transitions to the vibronic transitions of the crystal lattice is independent of the driving strength, even for detuned excitation using the spin-$\Lambda$ configuration.
We apply a polaron master equation to capture the non-Markovian dynamics of the vibrational manifolds. 
These results provide insight into the fundamental limitations to photon coherence from solid-state quantum emitters. 

\end{abstract}

\maketitle

Solid-state quantum emitters can mimic the behavior of few-level atomic systems. 
Two-level optical transitions can be driven resonantly for coherent manipulation~\cite{wrigge_efficient_2008,nick_vamivakas_spin-resolved_2009,flagg_resonantly_2009} which can be used to generate transform-limited single photons~\cite{kuhlmann_transform-limited_2015} with a high degree of indistinguishability from single~\cite{ding_-demand_2016,somaschi_near-optimal_2016} or multiple emitters~\cite{lettow_quantum_2010,bernien_two-photon_2012}. 
Coherent excitation and control can be extended to solid-state three-level spin-$\Lambda$ systems, which enables spin initialization, manipulation, and readout~\cite{press_complete_2008,yale_all-optical_2013,bechtold_three-stage_2015} as well as spin-photon entanglement~\cite{togan_quantum_2010,gao_observation_2012,de_greve_quantum-dot_2012} and indistinguishable single photon generation with tunable temporal and spectral properties~\cite{santori_indistinguishability_2009,fernandez_optically_2009,he_indistinguishable_2013,sun_measurement_2016,beguin_-demand_2018,pursley_picosecond_2018}. 
These advances can be applied to quantum information applications, for example, distribution of entanglement among independent quantum nodes~\cite{ZollerPRA1999,Kok2005,bernien_heralded_2013,delteil_generation_2016,stockill_phase-tuned_2017} or multiphoton boson sampling~\cite{wang_high-efficiency_2017,loredo_boson_2017}. 

At the heart of such quantum applications are photon-photon interactions, achieved when two single-photon wave packets interfere at a beam splitter~\cite{hong_measurement_1987}. 
Maximum interference visibility demands both coherent and indistinguishable photon wave packets. 
However, interactions between the atomlike eigenstates and the solid-state environment can degrade the coherence and indistinguishability of even coherently generated photons. 
The most prominent mechanisms involve charge~\cite{houel_probing_2012,kuhlmann_transform-limited_2015} and spin~\cite{kuhlmann_transform-limited_2015,bechtold_three-stage_2015} fluctuations and interactions with acoustic phonons~\cite{hameau_strong_1999,seebeck_polarons_2005,vagov_nonmonotonous_2004,borri_exciton_2005,kaer_microscopic_2013,roy-choudhury_theory_2015,thoma_exploring_2016, reigue_probing_2017, tighineanu_phonon_2018,  grange_reducing_2017, iles-smith_phonon_2017}. 
One popular approach to mitigate these effects is to use weak coherent excitation. 
In the weak-excitation regime of a transform-limited optical transition, coherent scattering dominates over incoherent (spontaneous) emission  due to minimal excited-state population~\cite{nguyen_ultra-coherent_2011, konthasinghe_coherent_2012,matthiesen_subnatural_2012,matthiesen_phase-locked_2013,proux_measuring_2015, schulte_quadrature_2015,malein_screening_2016,baudin_correlation_2019}. 
Likewise, spin-flip Raman scattered photons from solid-state spin-$\Lambda$ systems produce highly coherent photons; in the atomic picture such photons have coherence determined solely by the excitation source and ground (spin) state dephasing~\cite{santori_indistinguishability_2009,fernandez_optically_2009,he_indistinguishable_2013,sun_measurement_2016,beguin_-demand_2018,pursley_picosecond_2018}.

In this Letter, we experimentally test the assumption that detrimental interactions with longitudinal acoustic phonons can be minimized by suppressing the excited-state population in the weak-driving limit. 
We probe both two-level atomiclike transitions and spin-$\Lambda$ systems in a prototypical solid-state quantum emitter: a charge tunable semiconductor quantum dot (QD). 
We demonstrate experimentally that the vibrational environment--intrinsic to all solid-state emitters--imposes a fundamental limit on the coherence and indistinguishability of resonantly generated photons from a two-level quantum emitter, independent of driving strength or excited-state population, as recently predicted~\cite{iles-smith_limits_2017}. 
We proceed to show that this limit equally affects spin-flip Raman scattered photons, contrary to expectations~\cite{santori_indistinguishability_2009,fernandez_optically_2009,he_indistinguishable_2013,sun_measurement_2016,beguin_-demand_2018,pursley_picosecond_2018}. 

\begin{figure}
\includegraphics[width=0.5\textwidth]{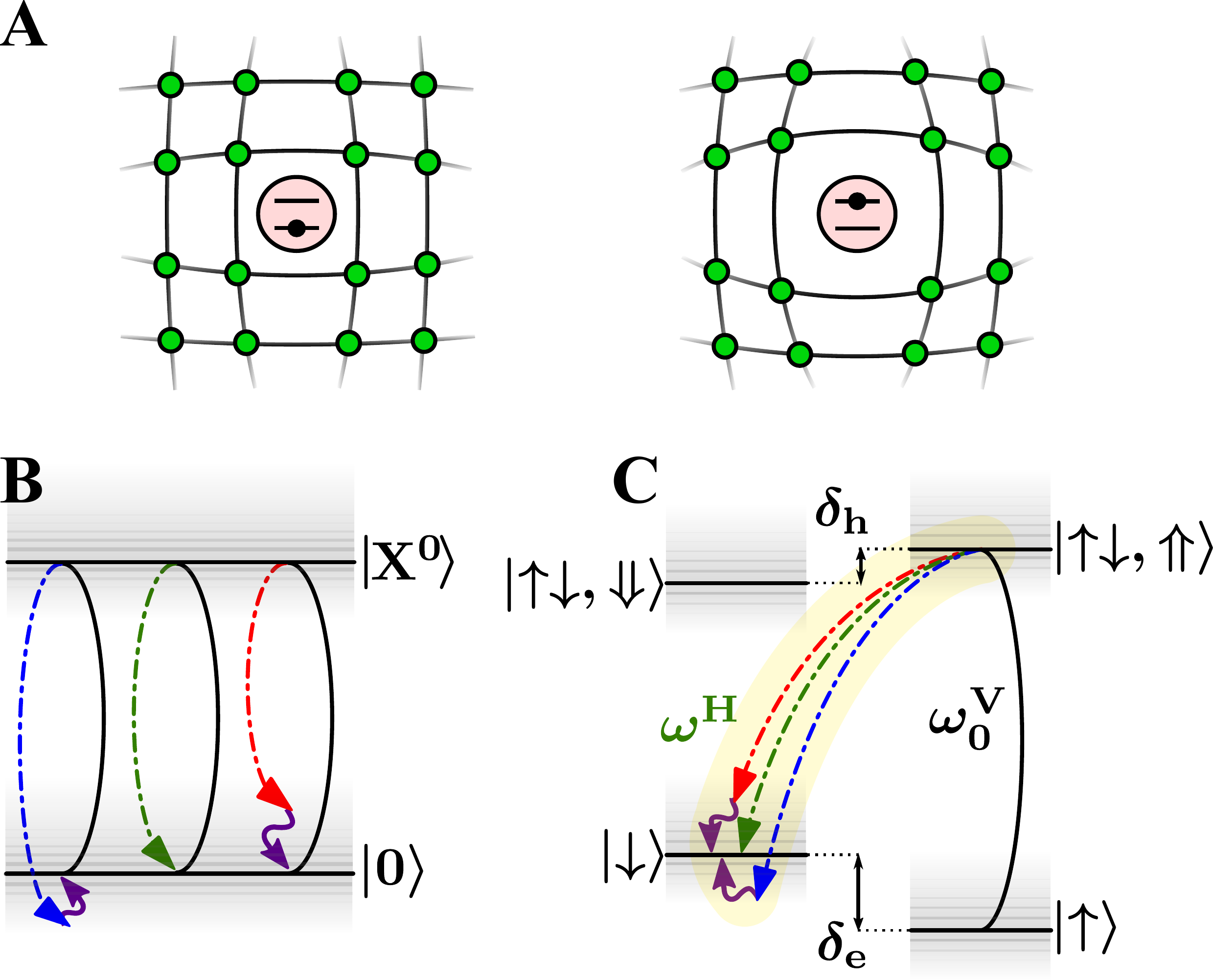}
\caption{\textbf{A quantum dot coupled to vibrational modes.}
\textbf{(A)} The equilibrium position of crystal lattice ions depends on the charge state of the quantum dot (schematic illustration) due to the deformation potential electron-phonon interaction. 
\textbf{(B)} Energy level schematic for the neutral exciton ($\mathrm{ X^0}$) in the weak-driving regime, showing elastic scattering from the excited-state (green) and the inelastic Stokes (red) or anti-Stokes (blue) scattering due to the electronic relaxation from the excited-state to the ground-state vibrational manifold. 
\textbf{(C)} Spin-$\Lambda$ energy level structure for the negatively charged exciton ($\mathrm{ X^{1-}}$) in the weak-driving regime, showing the inelastic zero-phonon (green) and stokes (red) and anti-stokes (blue) scatterings for the Raman spin-flip transition (yellow). 
The scattering from the spin-preserving transition is not shown. Single (double) arrows represent electron (heavy-hole) spin states. The energy separation between the Zeeman split ground (excited) states is given by $\delta_e$ ($\delta_h$).
}
\label{fig:1}
\end{figure}

Using a non-Markovian master equation model, we quantitatively explain the coherence as a function of driving strength for both the two-level and spin-$\Lambda$ systems, and we interpret the nonvanishing fraction of incoherently scattered photons as attributable to  
the vibrational manifolds dressing both the excited and the ground-state, and thus also affecting the optical dipole operator (see Fig.~\ref{fig:1}), even in the absence of excited electronic population.
Our results imply that spectral filtering is necessary for perfectly indistinguishable photons, even in the weak-driving and Raman-detuned regimes. This introduces a probabilistic element, hindering the use of solid-state emitters in deterministic single-shot protocols~\cite{lindner_proposal_2009,buterakos_deterministic_2017,gimeno-segovia_deterministic_2018}, even in optimized solid-state phononic or photonic structures~\cite{iles-smith_phonon_2017,grange_reducing_2017,tighineanu_phonon_2018}. 
Protocols embracing probabilistic operation~\cite{rudolph_why_2017,scerri_frequency-encoded_2018}, or where the detection of a photon heralds success~\cite{ZollerPRA1999,Kok2005,bernien_heralded_2013,delteil_generation_2016,stockill_phase-tuned_2017}, provide mitigation but this limitation remains detrimental.

Fig.~\ref{fig:1}A graphically illustrates the electron-phonon coupling, where the charge state of the QD is dressed by lattice displacements. 
In the strong-driving regime, this interaction is sometimes modeled using a Markovian weak-coupling approach which correctly captures excitation-induced dephasing~\cite{ramsay_damping_2010} and phonon-induced Rabi frequency renormalization~\cite{ramsay_phonon-induced_2010}. 
However, the standard weak-coupling approach generally fails to adequately resolve the electron-phonon interaction, and this is particularly evident in the weak-driving regime: here it becomes necessary to treat the electron lattice interaction in terms of phonon-dressed electronic states, so-called polaron quasiparticles. 
This leads to intrinsically non-Markovian dynamics for the evolution of the QD charge state, which is also reflected in its optical properties.
In the Supplemental Material \cite{[{See Supplemental Material at \url{http://link.aps.org/supplemental/10.1103/PhysRevLett.123.167402} for description of the experimental setup, details on the postselect two-photon interference measurement, coalescence time window, the nuclear spin coupling to Raman photons, and {the derivation of the associated polaron master equation for a spin-$\Lambda$ system}, which includes Refs.~\cite{Ficke2005,lebreton2013,urbaszek_nuclear_2013,Scerri2017,Hughes2015b,Breuer2007}}] SupMat} (SI), we extend the non-Markovian polaron model for the neutral exciton~\cite{nazir_modelling_2016} to account for the vibrational effects on the charged exciton (which in our case reduces to a spin-$\Lambda$ system).
To briefly summarize here: we begin with the canonical Lang-Firsov transformation~\cite{lang_kinetic_1962} applied to our Hamiltonian prior to solving its dynamics. 
In this frame the original electron-phonon coupling term disappears from the effective Hamiltonian (see Eq.~(20-22) in the SI~\cite{SupMat}). 
Instead, we are left with a weaker remnant vibrational coupling term, which we proceed to treat perturbatively. 
Importantly, both ground and excited-states now possess vibrational manifolds, in analogy to the Franck-Condon model~\cite{Mahan}. 

\begin{figure*}
\includegraphics[width=1\textwidth]{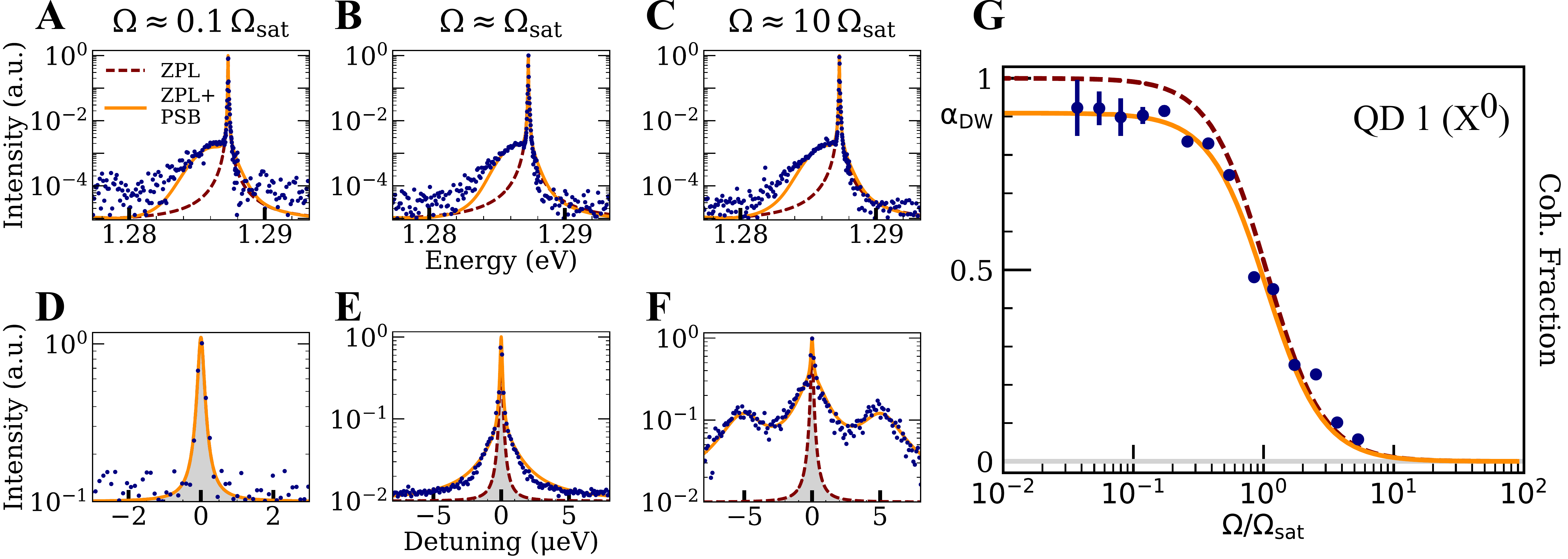}
\caption{\textbf{Resonance fluorescence spectra from a solid-state two-level transition.}
\textbf{(A, B, C)} Scattered photons at three different Rabi frequencies ($\Omega\approx 0.1\,{\Omega_\mathrm{ sat}}$, $\Omega_\mathrm{sat}$, $10\,{\Omega_\mathrm{ sat}}$) show a narrow ZPL accompanied by a broad PSB. 
The fit is produced by non-Markovian model, using a super-Ohmic spectral density~\cite{leggett_dynamics_1987} with system-phonon coupling strength $\alpha = 0.03\,\mathrm{ps}^2$ and frequency cutoff $\omega_c = 2.2\,\mathrm{ps}^{-1}$.
\textbf{(D, E, F)} Higher resolution ($\approx 0.1\,\mu \mathrm{ eV}$)  spectra, as a function of the detuning from the ZPL (1.287~eV) show the evolution from a coherent elastic peak at $\Omega \ll \Omega_\mathrm{sat}$ to a Mollow triplet at $\Omega \gg \Omega_\mathrm{ sat}$, matching the behavior of a coherently driven two-level system.
\textbf{(G)} Our theoretical curve (orange) matches the data and confirms that in the weak-driving regime ($\Omega\ll \Omega_\mathrm{ sat}$), the coherence of the transform-limited photons is limited by the branching ratio, given by the Debye-Waller factor, $\alpha_\mathrm{ DW}\approx 0.91$.
Blue data points are experimental data, obtained by computing the ratio of elastic peak (shaded region in (D-F)) to total spectrum.
The red line shows the theoretical behavior for an atomic two-level system.
}
\label{fig:2ls}
\end{figure*}

We experimentally probe effective two-level optical transitions and spin-$\Lambda$ systems using the neutral ($\mathrm{X^0}$) and negatively charged ($\mathrm{X^{1-}}$) excitons respectively, from a charge tunable QD device~\cite{malein_screening_2016} at a temperature of 4~K. 
For $\mathrm{X^0}$, we resonantly excite and collect from just one of the fine-structure split peaks using orthogonal linear polarizers to suppress the scattered laser background. 
For $\mathrm{ X^{1-}}$, we apply an in plane magnetic field (Voigt geometry) of 4~T to mix the spin states and create ``diagonal'' spin-flipping transitions (e.g. $\ket{\downarrow} \leftrightarrow \ket{\uparrow \downarrow, \Uparrow}$) with horizontal polarization ($\omega^H$) and equal oscillator strength to the spin-conserving transitions (e.g. $\ket{\uparrow} \leftrightarrow \ket{\uparrow \downarrow, \Uparrow}$) with vertical polarization ($\omega_0^V$).
Here, the single (double) arrows refer to electron (heavy-hole) spin states. 
We resonantly excite the spin-conserving transition with $\omega_0^V$ and collect $\omega^H$ from the spin-flipping transition, as shown in Fig.~\ref{fig:1}C. 
Photon spectra are characterized by a $\sim 30\,\mu \mathrm{ eV}$ resolution spectrometer and a $\sim 0.1\,\mu \mathrm{ eV}$ resolution scanning Fabry-P\'erot interferometer. 
A Hanbury Brown-Twiss interferometer and an unbalanced Mach-Zehnder interferometer (with an interferometric delay of $49.7\,\mathrm{ ns}$) are used to characterize the intensity correlation $g^{(2)}(\tau)$ and postselected, two-photon interference, respectively.

Fig.~\ref{fig:2ls}(A-F) show the resonantly scattered photon spectra from the $\ket{0} \leftrightarrow \ket{\mathrm{ X^0}}$ transition at three different driving regimes: $\Omega\approx\,0.1\,\Omega_\mathrm{ sat}$, $\Omega_\mathrm{ sat}$, and $10\,\Omega_\mathrm{ sat}$, where $\Omega$ ($\Omega_\mathrm{ sat}$) is the Rabi (saturation) frequency. 
The spectra exhibit identical features as measured by the spectrometer: a narrow zero-phonon line (ZPL, red dashed line) and a broad shoulder near the ZPL---the phonon sideband (PSB).
High-resolution spectra of the ZPL reveal a single resolution-limited elastic peak at $\Omega \approx 0.1 \,\Omega_\mathrm{ sat}$  ~\cite{proux_measuring_2015,matthiesen_phase-locked_2013,konthasinghe_coherent_2012,nguyen_ultra-coherent_2011,matthiesen_subnatural_2012}, the emergence of a broad incoherent spectrum at $\Omega \approx \Omega_\mathrm{ sat}$, and a Mollow triplet at $\Omega > \Omega_\mathrm{ sat}$~\cite{wrigge_efficient_2008,nick_vamivakas_spin-resolved_2009,flagg_resonantly_2009}. 
The orange lines in Fig.~\ref{fig:2ls}D to \ref{fig:2ls}F are fits to the experimental data using the theoretical functions as described in Ref.~\cite{scully_quantum_1997} by fixing the lifetime $T_1=0.625\,\mathrm{ ns}$ and coherence time $T_2=2\,T_1=1.250\,\mathrm{ns}$. 
The lifetime is independently measured using time-resolved resonance fluorescence~\cite{SupMat}. 

Fig.~\ref{fig:2ls}(D-F) exhibit textbook atomiclike behaviour of resonance fluorescence of a two-level system for the ZPL. 
On the other hand, the consistent PSB regardless of $\Omega$ in Fig.~\ref{fig:2ls}(A-C) reveals a fundamental departure from atomiclike behavior, which can be modeled using the polaron master equation. 
To tune our microscopic theoretical model to the specific properties of this QD, we have { extracted the system-phonon coupling strength and frequency cutoff from the fits to the data in Fig.~\ref{fig:2ls} to be $\alpha = 0.03\,\mathrm{ps}^2$ and $\omega_c = 2.2\,\mathrm{ps}^{-1}$, respectively.}  
The frequency cutoff provides an indication on the size and confinement of the QD~\cite{iles-smith_limits_2017}. 
By fitting the experimental data, we extract the ratio of coherent to total (coherent and incoherent) light in the spectrum in the range $0.04\,\Omega_\mathrm{sat} \leq \Omega \leq 6\,\Omega_\mathrm{sat}$, as shown in Fig.~\ref{fig:2ls}G, which, at low power, gives the fraction of photons coherently scattered in the ZPL.
For an atomic two-level system, this coherent fraction (CF) is determined by~\cite{cohentannoudji_atomphoton_1998}
\begin{equation}
    CF = \frac{T_2}{2\,T_1}\frac{1}{1+\Omega/\Omega_\mathrm{ sat}}~,
\label{eqn:CF}
\end{equation}
represented by the dashed red curve in Fig.~\ref{fig:2ls}G. 
The experimental data depart from Eq.~(\ref{eqn:CF}) in the weak-excitation regime; to fit the data, we modify the coherent fraction according to 
\begin{equation}
    CF^\prime = \alpha_\mathrm{DW}\,CF,
\end{equation}
where $\alpha_\mathrm{DW}$ is the Debye-Waller coefficient, which quantifies the influence of the vibrational manifold on the nature of scattering process~\cite{lipkin_physics_2004}
and is equivalent to the square of the Franck-Condon factor $\left\langle B \right\rangle$~\cite{Mahan}~\footnote{The equivalence between {$ \langle B \rangle$} and the Franck-Condon factor stems from the fact that both quantities emerge as vibrational prefactors when taking the transition probability of the system coupled to the vibrational environment.}, i.e., $\alpha_\mathrm{DW}=\left\langle B \right\rangle^2$ (cf.~Refs.~\cite{iles-smith_limits_2017,iles-smith_phonon_2017}).
Based on the fits of the PSB in the spectra, the theoretical model gives an upper bound of $\alpha_\mathrm{DW} = 0.91$ as the fraction of maximum coherence in the weak-driving regime. 

The fact that the coherent fraction of a two-level system in this QD is capped at $\alpha_\mathrm{DW}\approx 0.91$ in the weak-driving regime shows that a substantial number of emitted photons still interact with the phonon bath, despite the QD population remaining in the ground state throughout. 
This demonstrates that the phonon sideband is independent of excited-state occupation.
Hence, the phonon sideband remains a detrimental effect in exploiting the properties of the photons, namely, the long coherence time of the resonantly scattered photons in the weak-driving regime, contrary to previous claims~\cite{nguyen_ultra-coherent_2011, konthasinghe_coherent_2012,matthiesen_subnatural_2012,matthiesen_phase-locked_2013}. 

\begin{figure}
\includegraphics[width=0.5\textwidth]{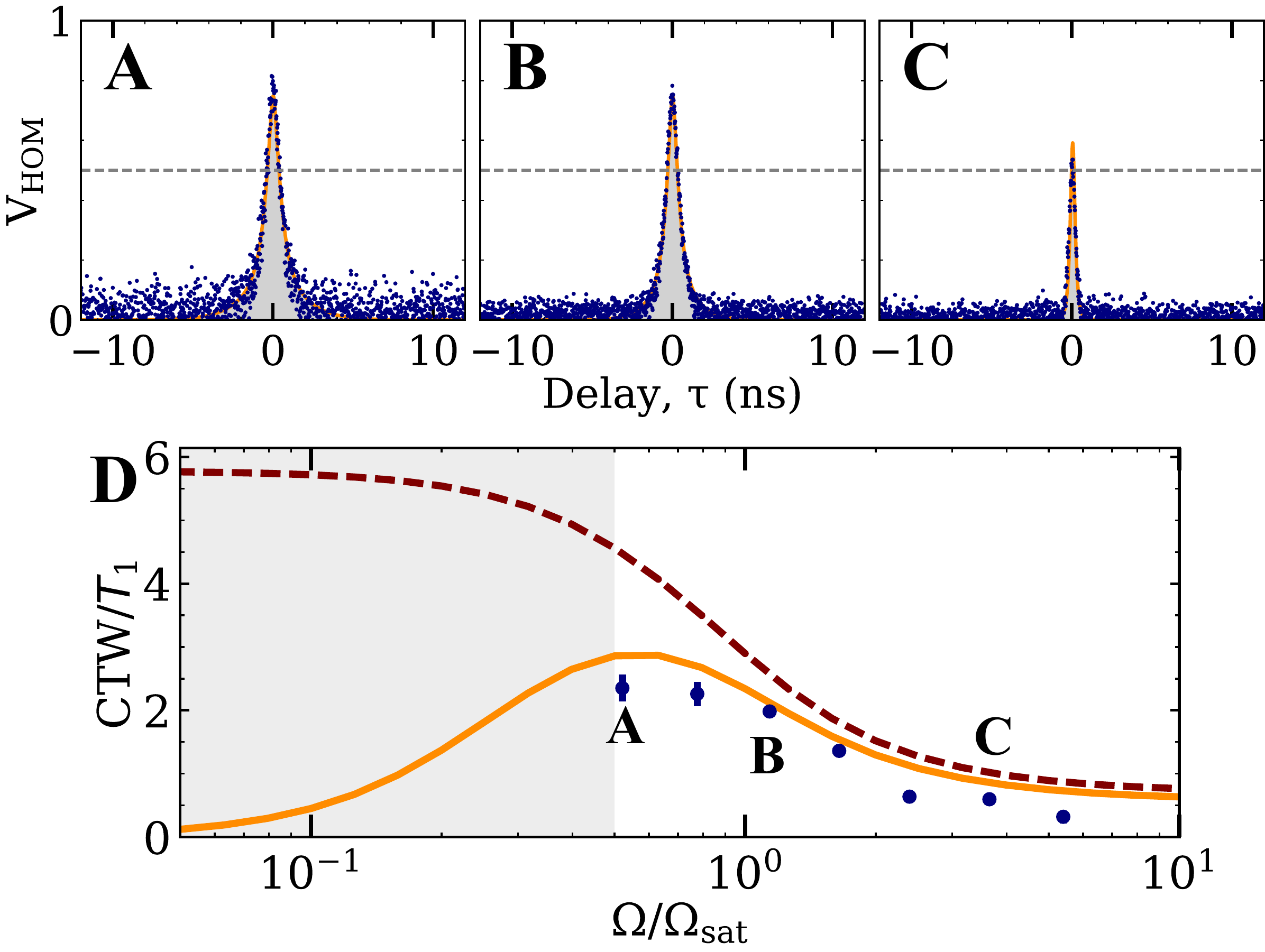}
\caption{\textbf{Two-photon interference between scattered photons { from a neutral exciton $\mathrm{X^0}$} as a function of driving strength.}
\textbf{(A, B, C)} Two-photon interference visibilities, $V_\mathrm{HOM}(\tau)$ at $\Omega\approx\,0.5\,\Omega_\mathrm{sat}$, $\Omega\approx\,\Omega_\mathrm{sat}$ and $\Omega\approx\,3\,\Omega_\mathrm{sat}$ respectively. 
\textbf{(D)} The coalescence time window normalized by the lifetime of the emitter ($T_1 = 0.625\,\mathrm{ns}$) deduced from the experimental (dots) and theoretical (solid line) CTW as a function of driving Rabi frequency $\Omega$. The dashed line represents the theoretical CTW curve without the contribution from phonons.
The shaded region represents the region inaccessible in experiment.
}
\label{fig:3}
\end{figure}

To verify the spectroscopic observations regarding exciton-phonon coupling and scattered photon coherence, we investigate the two-photon interference. 
First, we measure the second-order correlation $g^{(2)}(\tau)$ of the scattered photons, which exhibits a suppressed multiphoton emission probability of $g^{(2)}(0)=0.046\,(13)$ at $\Omega\approx 0.1 \,\Omega_\mathrm{sat}$ (see  Fig.~S2 in the SI~\cite{SupMat}). 
Next, we perform Hong-Ou-Mandel (HOM)-type two-photon interference with the unbalanced Mach-Zehnder interferometer. 
As the two-photon interference visibility $V_\mathrm{HOM}(\tau)$ at zero-time delay is solely determined by the detector response time under continuous wave excitation, its maximum value is not indicative of indistinguishability of the photon wave packets~\cite{legero_quantum_2004}.
Therefore, we instead consider the coalescence time window $\mathrm{CTW} =\int V_\mathrm{ HOM}(\tau) \,d\tau$, given by the shaded area in Fig.~\ref{fig:3}(A-C), which depends on the full duration of $V_\mathrm{HOM}(\tau)$ and is thus independent of detector jitter~\cite{proux_measuring_2015, baudin_correlation_2019}.
A detailed analysis of CTW for atomic two-level systems and solid-state nanostructures interacting with a vibrational environment is described in the SI~\cite{SupMat}.
We now investigate the effect of PSB on the CTW as shown in Fig.~\ref{fig:3}D. 
The experimental data points around $\Omega \gtrsim \Omega_\mathrm{sat}$ show qualitative agreement with our theoretical model (solid line): the quantitative discrepancy originates from experimental imperfections (e.g. spatial mode overlap or unbalanced splitting ratio of the beam splitter \cite{giesz_cavity-enhanced_2015}). 
In the absence of phonons (dashed curve), we find that it differs from our data around $\Omega_\mathrm{sat}$ and leads to very different predicted limiting behavior for $\Omega \ll \Omega_\mathrm{sat}$. 
The saturation at low-excitation power is due to the lifetime limit of the QD: the photon coherence is not solely determined  by the laser coherence in this power regime. Thus, the width of the visibility dip should converge as $\Omega \rightarrow 0$, resulting in the convergence of the CTW to a finite value, as opposed to increasing indefinitely, even for an ideal continuous wave source \cite{proux_measuring_2015,baudin_correlation_2019}.

Unfortunately, the shaded region $\Omega \lesssim 0.5\,\Omega_\mathrm{sat}$ is inaccessible in our experimental setup: here the predominantly Rayleigh scattered photons inherit the laser coherence ($\sim 100\,\mathrm{kHz}$). 
This coherence exceeds the interferometric delay in our unbalanced Mach-Zehnder setup, and undesired one-photon interference dominates the measurement~\cite{proux_measuring_2015,baudin_correlation_2019}.
Nonetheless, the agreement between measured data and the theoretical model for $\Omega \gtrapprox \Omega_\mathrm{sat}$ supports the validity of the polaron model and justifies the extrapolation into the low-excitation regime. 
The decreasing CTW is consistent with the reduced $V_\mathrm{HOM}(0)$ in Ref.~\cite{iles-smith_limits_2017}, owing to the large separation of timescales between phonon dynamics and average time between scattering events. 
Our results thus show that the vibrational environment of solid-state emitters degrades the achievable indistinguishability of resonantly scattered photons. 

\begin{figure}
\includegraphics[width=0.5\textwidth]{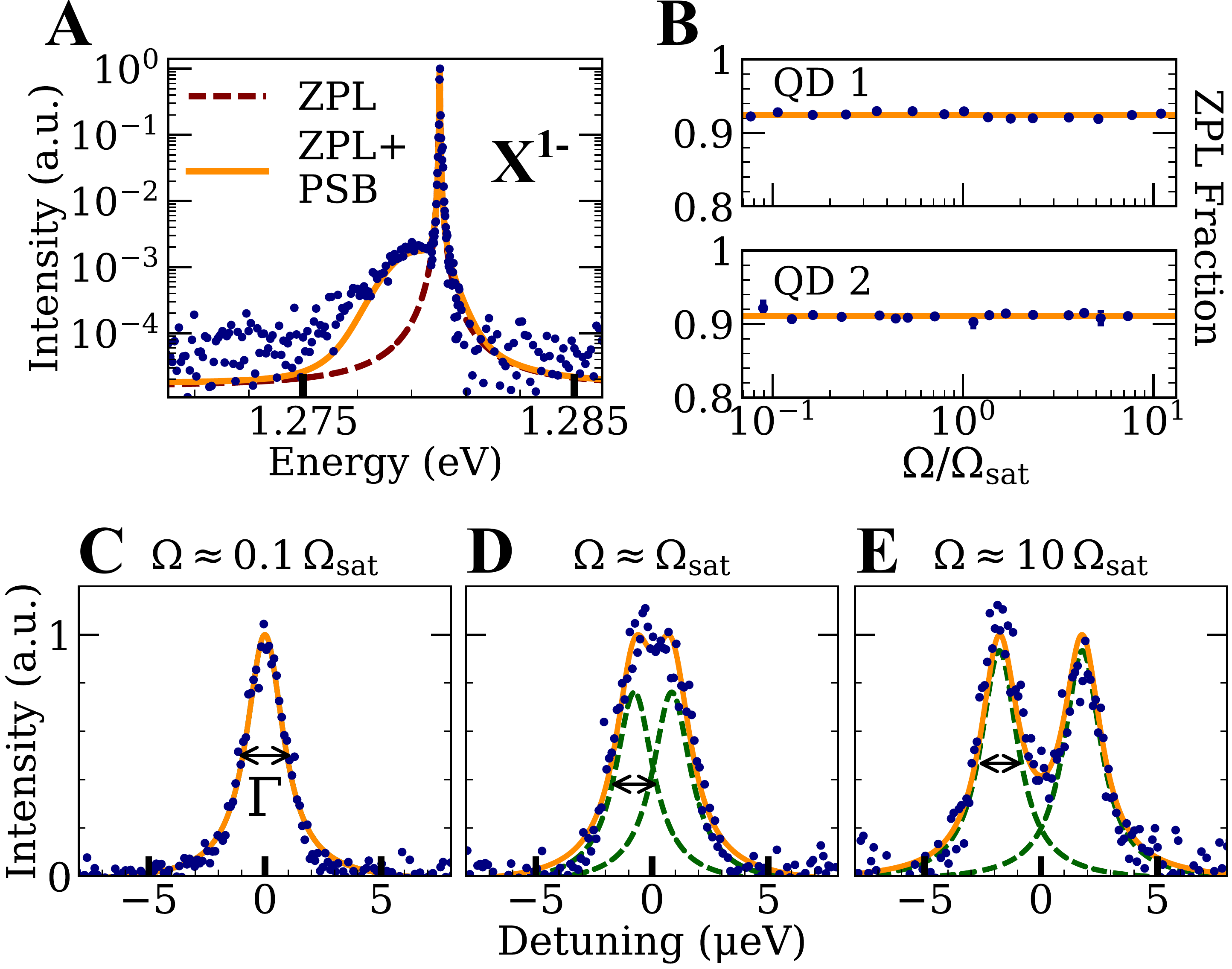}
\caption{\textbf{Spin-flip Raman photon spectra from a solid-state spin-$\Lambda$ system.}
\textbf{(A)} Scattered Raman photons from QD1 shows a narrow ZPL and a broad PSB at $\Omega \approx 0.1\,\Omega_\mathrm{ sat}$.
The fit is produced by the non-Markovian model using the same set of parameters as in Fig.~\ref{fig:2ls}.
\textbf{(B)} The ZPL fraction is constant over 2 orders of magnitude in Rabi frequencies. 
It gives $\alpha_\mathrm{DW} = 0.924\,(4)$ (QD1) and $\alpha_\mathrm{DW} = 0.911\,(4)$ (QD2). 
\textbf{(C, D, E)} Emission spectra of the scattered Raman photons from QD2 at three different Rabi frequencies ($\Omega\approx 0.1\,{\Omega_\mathrm{ sat}}$, $\Omega_\mathrm{ sat}$, $10\,{\Omega_\mathrm{ sat}}$) show an Autler-Townes splitting at $ \Omega \gtrapprox \Omega_\mathrm{ sat}$.
Zero detuning corresponds to energy of ZPL at $1.277\,\mathrm{eV}$.
The broad linewidth ($\Gamma\approx 2\, \mu{\mathrm{eV}}$) originates from the spin ground-state dephasing due to nuclear spin fluctuations.}
\label{fig:4}
\end{figure}

Motivated by the observation that the vibrational environment impacts the ground state of the two-level transition, we proceed to investigate how the phonon interaction affects the spin-flip Raman photon spectrum. 
The three-level spin-$\Lambda$ system for $\mathrm{ X^{1-}}$ with an in plane magnetic field is shown in Fig.~\ref{fig:1}C. 

The vertical transition $\Ket{\uparrow}\leftrightarrow \Ket{\uparrow \downarrow,\Uparrow}$ is resonantly excited and we collect the scattered Raman photons from the diagonal transition $\Ket{\downarrow} \leftrightarrow \Ket{\uparrow \downarrow,\Uparrow} $.
Fig.~\ref{fig:4}A shows the spin-flip Raman photon spectrum in the weak-driving regime, $\Omega\approx\,0.1\,\Omega_\mathrm{sat}$. We observe a narrow ZPL and a broad PSB, independent of $\Omega$, in the low-resolution spectra--similar to the two-level case. The ZPL fraction, given by the ratio of the integrated intensity of the ZPL to the total emission spectrum (ZPL + PSB) is shown in Fig.~\ref{fig:4}B for two QDs, with $\alpha_\mathrm{ DW}=0.924\,(4)$ and $\alpha_\mathrm{ DW}=0.911\,(4)$ for QD1 and QD2, respectively.
The observation that the ratio of ZPL to PSB of the Raman photons remains constant over 2 orders of magnitude in Rabi frequency leads to the conclusion that the ground spin states are dressed by the vibrational couplings. This contradicts the common consensus that Raman photons are highly coherent and limited only by ground-state spin dephasing~\cite{santori_indistinguishability_2009,fernandez_optically_2009,he_indistinguishable_2013,sun_measurement_2016,beguin_-demand_2018,pursley_picosecond_2018}. 
We note that the detrimental effect of the electron-phonon interaction persists even when there is only negligible excited QD population, so that even Raman red-detuned excitation of the spin-$\Lambda$ system will not eliminate the phonon sideband. Furthermore, we expect the same effect on the wave packet indistinguishability (CTW) as we have discussed for the two-level transition.

To further characterize the coherence of the ZPL of the Raman photons, we measure the spectra from QD2 with the high-resolution  Fabry-P\'erot interferometer. The emergence of a single Gaussian peak with width $\approx 2\, \mu \mathrm{eV}$  (FWHM) at $\Omega \approx 0.1\,\Omega_\mathrm{ sat}$ (Fig.~\ref{fig:4}C) and an Autler-Townes doublet at $\Omega \gtrapprox \Omega_\mathrm{ sat}$ (Fig.~\ref{fig:4}D and \ref{fig:4}E) confirm the nature of coherent driving in the spin-$\Lambda$ system. As explained in the SI~\cite{SupMat}, the relatively broad linewidth ($\Gamma\approx 2\,\mu\mathrm{ eV}$) of the spin-flip Raman photon is due to the spin ground-state dephasing which is dominated by the coupling of electron spin to the nuclear spin bath ~\cite{malein_screening_2016,sun_measurement_2016}.

In summary, our experimental and theoretical results contradict the expectation that perfectly coherent photons can be obtained from a solid-state emitter, either by weak resonant driving of a two-level transition or as Raman scattered photons. 
Instead, we have shown that the solid-state environment, and the associated exciton-phonon interaction, invariably limit the coherence of resonantly scattered photons: a minimum fraction $\alpha_\mathrm{DW}$ of photons are scattered incoherently, for any excitation power and scheme.    
We argue that these phonon-induced effects are due to the relaxation of the phonon bath in the excitonic ground-state, explaining why neither the weak-driving nor Raman red-detuned excitation regimes mitigate the interaction. 
While it is possible to filter the phonon sideband, this introduces a probabilistic element to the success rate of obtaining indistinguishable photons. 
Furthermore, despite being an intuitive solution, embedding the QD inside a cavity or waveguide for strong Purcell enhancement can only partly reduce the adverse effects of the vibrational environment, and, in the case of a cavity-embedded QD, at the cost of efficiency \cite{iles-smith_phonon_2017}.
Our non-Markovian polaron frame model agrees well with the experimental data for both the two- and three-level solid-state systems, showing that the presence of the vibrational environment impacts the emitter's dynamics even in the low-power, Raman red-detuned regime.

\begin{acknowledgments}
This work was supported by the EPSRC (Grants No. EP/L015110/1, No. EP/M013472/1, and No. EP/P029892/1), the ERC (Grant No. 725920), and the EU Horizon 2020 research and innovation program under Grant Agreement no. 820423. 
B.D.G. thanks the Royal Society for a Wolfson Merit Award, and the Royal Academy of Engineering for a Chair in Emerging Technology, and E.M.G. acknowledges financial support from the Royal Society of Edinburgh and the Scottish Government. 
T.S.S. acknowledges PNPD/CAPES for financial support.
The authors in KIST acknowledge the support from the KIST institutional program and the program of quantum sensor core technology through IITP.

\textit{Note added.}--Recently, we become aware of related results~\cite{brash_light_2019}.
We thank A. J. Brash for bringing this to our attention.
\end{acknowledgments}

\bibliographystyle{apsrev4-1}
\bibliography{reference2}

%merlin.mbs apsrev4-1.bst 2010-07-25 4.21a (PWD, AO, DPC) hacked
%Control: key (0)
%Control: author (72) initials jnrlst
%Control: editor formatted (1) identically to author
%Control: production of article title (-1) disabled
%Control: page (0) single
%Control: year (1) truncated
%Control: production of eprint (0) enabled
\begin{thebibliography}{74}%
\makeatletter
\providecommand \@ifxundefined [1]{%
 \@ifx{#1\undefined}
}%
\providecommand \@ifnum [1]{%
 \ifnum #1\expandafter \@firstoftwo
 \else \expandafter \@secondoftwo
 \fi
}%
\providecommand \@ifx [1]{%
 \ifx #1\expandafter \@firstoftwo
 \else \expandafter \@secondoftwo
 \fi
}%
\providecommand \natexlab [1]{#1}%
\providecommand \enquote  [1]{``#1''}%
\providecommand \bibnamefont  [1]{#1}%
\providecommand \bibfnamefont [1]{#1}%
\providecommand \citenamefont [1]{#1}%
\providecommand \href@noop [0]{\@secondoftwo}%
\providecommand \href [0]{\begingroup \@sanitize@url \@href}%
\providecommand \@href[1]{\@@startlink{#1}\@@href}%
\providecommand \@@href[1]{\endgroup#1\@@endlink}%
\providecommand \@sanitize@url [0]{\catcode `\\12\catcode `\$12\catcode
  `\&12\catcode `\#12\catcode `\^12\catcode `\_12\catcode `\%12\relax}%
\providecommand \@@startlink[1]{}%
\providecommand \@@endlink[0]{}%
\providecommand \url  [0]{\begingroup\@sanitize@url \@url }%
\providecommand \@url [1]{\endgroup\@href {#1}{\urlprefix }}%
\providecommand \urlprefix  [0]{URL }%
\providecommand \Eprint [0]{\href }%
\providecommand \doibase [0]{http://dx.doi.org/}%
\providecommand \selectlanguage [0]{\@gobble}%
\providecommand \bibinfo  [0]{\@secondoftwo}%
\providecommand \bibfield  [0]{\@secondoftwo}%
\providecommand \translation [1]{[#1]}%
\providecommand \BibitemOpen [0]{}%
\providecommand \bibitemStop [0]{}%
\providecommand \bibitemNoStop [0]{.\EOS\space}%
\providecommand \EOS [0]{\spacefactor3000\relax}%
\providecommand \BibitemShut  [1]{\csname bibitem#1\endcsname}%
\let\auto@bib@innerbib\@empty
%</preamble>
\bibitem [{\citenamefont {Wrigge}\ \emph {et~al.}(2008)\citenamefont {Wrigge},
  \citenamefont {Gerhardt}, \citenamefont {Hwang}, \citenamefont {Zumofen},\
  and\ \citenamefont {Sandoghdar}}]{wrigge_efficient_2008}%
  \BibitemOpen
  \bibfield  {author} {\bibinfo {author} {\bibfnamefont {G.}~\bibnamefont
  {Wrigge}}, \bibinfo {author} {\bibfnamefont {I.}~\bibnamefont {Gerhardt}},
  \bibinfo {author} {\bibfnamefont {J.}~\bibnamefont {Hwang}}, \bibinfo
  {author} {\bibfnamefont {G.}~\bibnamefont {Zumofen}}, \ and\ \bibinfo
  {author} {\bibfnamefont {V.}~\bibnamefont {Sandoghdar}},\ }\href {\doibase
  10.1038/nphys812} {\bibfield  {journal} {\bibinfo  {journal} {Nat. Phys.}\
  }\textbf {\bibinfo {volume} {4}},\ \bibinfo {pages} {60} (\bibinfo {year}
  {2008})}\BibitemShut {NoStop}%
\bibitem [{\citenamefont {Nick~Vamivakas}\ \emph {et~al.}(2009)\citenamefont
  {Nick~Vamivakas}, \citenamefont {Zhao}, \citenamefont {Lu},\ and\
  \citenamefont {Atat\"{u}re}}]{nick_vamivakas_spin-resolved_2009}%
  \BibitemOpen
  \bibfield  {author} {\bibinfo {author} {\bibfnamefont {A.}~\bibnamefont
  {Nick~Vamivakas}}, \bibinfo {author} {\bibfnamefont {Y.}~\bibnamefont
  {Zhao}}, \bibinfo {author} {\bibfnamefont {C.-Y.}\ \bibnamefont {Lu}}, \ and\
  \bibinfo {author} {\bibfnamefont {M.}~\bibnamefont {Atat\"{u}re}},\ }\href
  {\doibase 10.1038/nphys1182} {\bibfield  {journal} {\bibinfo  {journal} {Nat.
  Phys.}\ }\textbf {\bibinfo {volume} {5}},\ \bibinfo {pages} {198} (\bibinfo
  {year} {2009})}\BibitemShut {NoStop}%
\bibitem [{\citenamefont {Flagg}\ \emph {et~al.}(2009)\citenamefont {Flagg},
  \citenamefont {Muller}, \citenamefont {Robertson}, \citenamefont {Founta},
  \citenamefont {Deppe}, \citenamefont {Xiao}, \citenamefont {Ma},
  \citenamefont {Salamo},\ and\ \citenamefont {Shih}}]{flagg_resonantly_2009}%
  \BibitemOpen
  \bibfield  {author} {\bibinfo {author} {\bibfnamefont {E.~B.}\ \bibnamefont
  {Flagg}}, \bibinfo {author} {\bibfnamefont {A.}~\bibnamefont {Muller}},
  \bibinfo {author} {\bibfnamefont {J.~W.}\ \bibnamefont {Robertson}}, \bibinfo
  {author} {\bibfnamefont {S.}~\bibnamefont {Founta}}, \bibinfo {author}
  {\bibfnamefont {D.~G.}\ \bibnamefont {Deppe}}, \bibinfo {author}
  {\bibfnamefont {M.}~\bibnamefont {Xiao}}, \bibinfo {author} {\bibfnamefont
  {W.}~\bibnamefont {Ma}}, \bibinfo {author} {\bibfnamefont {G.~J.}\
  \bibnamefont {Salamo}}, \ and\ \bibinfo {author} {\bibfnamefont {C.~K.}\
  \bibnamefont {Shih}},\ }\href {\doibase 10.1038/nphys1184} {\bibfield
  {journal} {\bibinfo  {journal} {Nat. Phys.}\ }\textbf {\bibinfo {volume}
  {5}},\ \bibinfo {pages} {203} (\bibinfo {year} {2009})}\BibitemShut {NoStop}%
\bibitem [{\citenamefont {Kuhlmann}\ \emph {et~al.}(2015)\citenamefont
  {Kuhlmann}, \citenamefont {Prechtel}, \citenamefont {Houel}, \citenamefont
  {Ludwig}, \citenamefont {Reuter}, \citenamefont {Wieck},\ and\ \citenamefont
  {Warburton}}]{kuhlmann_transform-limited_2015}%
  \BibitemOpen
  \bibfield  {author} {\bibinfo {author} {\bibfnamefont {A.~V.}\ \bibnamefont
  {Kuhlmann}}, \bibinfo {author} {\bibfnamefont {J.~H.}\ \bibnamefont
  {Prechtel}}, \bibinfo {author} {\bibfnamefont {J.}~\bibnamefont {Houel}},
  \bibinfo {author} {\bibfnamefont {A.}~\bibnamefont {Ludwig}}, \bibinfo
  {author} {\bibfnamefont {D.}~\bibnamefont {Reuter}}, \bibinfo {author}
  {\bibfnamefont {A.~D.}\ \bibnamefont {Wieck}}, \ and\ \bibinfo {author}
  {\bibfnamefont {R.~J.}\ \bibnamefont {Warburton}},\ }\href {\doibase
  10.1038/ncomms9204} {\bibfield  {journal} {\bibinfo  {journal} {Nat.
  Commun.}\ }\textbf {\bibinfo {volume} {6}},\ \bibinfo {pages} {8204}
  (\bibinfo {year} {2015})}\BibitemShut {NoStop}%
\bibitem [{\citenamefont {Ding}\ \emph {et~al.}(2016)\citenamefont {Ding},
  \citenamefont {He}, \citenamefont {Duan}, \citenamefont {Gregersen},
  \citenamefont {Chen}, \citenamefont {Unsleber}, \citenamefont {Maier},
  \citenamefont {Schneider}, \citenamefont {Kamp}, \citenamefont {H\"{o}fling},
  \citenamefont {Lu},\ and\ \citenamefont {Pan}}]{ding_-demand_2016}%
  \BibitemOpen
  \bibfield  {author} {\bibinfo {author} {\bibfnamefont {X.}~\bibnamefont
  {Ding}}, \bibinfo {author} {\bibfnamefont {Y.}~\bibnamefont {He}}, \bibinfo
  {author} {\bibfnamefont {Z.-C.}\ \bibnamefont {Duan}}, \bibinfo {author}
  {\bibfnamefont {N.}~\bibnamefont {Gregersen}}, \bibinfo {author}
  {\bibfnamefont {M.-C.}\ \bibnamefont {Chen}}, \bibinfo {author}
  {\bibfnamefont {S.}~\bibnamefont {Unsleber}}, \bibinfo {author}
  {\bibfnamefont {S.}~\bibnamefont {Maier}}, \bibinfo {author} {\bibfnamefont
  {C.}~\bibnamefont {Schneider}}, \bibinfo {author} {\bibfnamefont
  {M.}~\bibnamefont {Kamp}}, \bibinfo {author} {\bibfnamefont {S.}~\bibnamefont
  {H\"{o}fling}}, \bibinfo {author} {\bibfnamefont {C.-Y.}\ \bibnamefont {Lu}},
  \ and\ \bibinfo {author} {\bibfnamefont {J.-W.}\ \bibnamefont {Pan}},\ }\href
  {\doibase 10.1103/PhysRevLett.116.020401} {\bibfield  {journal} {\bibinfo
  {journal} {Phys. Rev. Lett.}\ }\textbf {\bibinfo {volume} {116}},\ \bibinfo
  {pages} {020401} (\bibinfo {year} {2016})}\BibitemShut {NoStop}%
\bibitem [{\citenamefont {Somaschi}\ \emph {et~al.}(2016)\citenamefont
  {Somaschi}, \citenamefont {Giesz}, \citenamefont {De~Santis}, \citenamefont
  {Loredo}, \citenamefont {Almeida}, \citenamefont {Hornecker}, \citenamefont
  {Portalupi}, \citenamefont {Grange}, \citenamefont {Ant\'{o}n}, \citenamefont
  {Demory}, \citenamefont {G\'{o}mez}, \citenamefont {Sagnes}, \citenamefont
  {Lanzillotti-Kimura}, \citenamefont {Lemaítre}, \citenamefont {Auffeves},
  \citenamefont {White}, \citenamefont {Lanco},\ and\ \citenamefont
  {Senellart}}]{somaschi_near-optimal_2016}%
  \BibitemOpen
  \bibfield  {author} {\bibinfo {author} {\bibfnamefont {N.}~\bibnamefont
  {Somaschi}}, \bibinfo {author} {\bibfnamefont {V.}~\bibnamefont {Giesz}},
  \bibinfo {author} {\bibfnamefont {L.}~\bibnamefont {De~Santis}}, \bibinfo
  {author} {\bibfnamefont {J.~C.}\ \bibnamefont {Loredo}}, \bibinfo {author}
  {\bibfnamefont {M.~P.}\ \bibnamefont {Almeida}}, \bibinfo {author}
  {\bibfnamefont {G.}~\bibnamefont {Hornecker}}, \bibinfo {author}
  {\bibfnamefont {S.~L.}\ \bibnamefont {Portalupi}}, \bibinfo {author}
  {\bibfnamefont {T.}~\bibnamefont {Grange}}, \bibinfo {author} {\bibfnamefont
  {C.}~\bibnamefont {Ant\'{o}n}}, \bibinfo {author} {\bibfnamefont
  {J.}~\bibnamefont {Demory}}, \bibinfo {author} {\bibfnamefont
  {C.}~\bibnamefont {G\'{o}mez}}, \bibinfo {author} {\bibfnamefont
  {I.}~\bibnamefont {Sagnes}}, \bibinfo {author} {\bibfnamefont {N.~D.}\
  \bibnamefont {Lanzillotti-Kimura}}, \bibinfo {author} {\bibfnamefont
  {A.}~\bibnamefont {Lemaítre}}, \bibinfo {author} {\bibfnamefont
  {A.}~\bibnamefont {Auffeves}}, \bibinfo {author} {\bibfnamefont {A.~G.}\
  \bibnamefont {White}}, \bibinfo {author} {\bibfnamefont {L.}~\bibnamefont
  {Lanco}}, \ and\ \bibinfo {author} {\bibfnamefont {P.}~\bibnamefont
  {Senellart}},\ }\href {\doibase 10.1038/nphoton.2016.23} {\bibfield
  {journal} {\bibinfo  {journal} {Nat. Photonics}\ }\textbf {\bibinfo {volume}
  {10}},\ \bibinfo {pages} {340} (\bibinfo {year} {2016})}\BibitemShut
  {NoStop}%
\bibitem [{\citenamefont {Lettow}\ \emph {et~al.}(2010)\citenamefont {Lettow},
  \citenamefont {Rezus}, \citenamefont {Renn}, \citenamefont {Zumofen},
  \citenamefont {Ikonen}, \citenamefont {G\"{o}tzinger},\ and\ \citenamefont
  {Sandoghdar}}]{lettow_quantum_2010}%
  \BibitemOpen
  \bibfield  {author} {\bibinfo {author} {\bibfnamefont {R.}~\bibnamefont
  {Lettow}}, \bibinfo {author} {\bibfnamefont {Y.~L.~A.}\ \bibnamefont
  {Rezus}}, \bibinfo {author} {\bibfnamefont {A.}~\bibnamefont {Renn}},
  \bibinfo {author} {\bibfnamefont {G.}~\bibnamefont {Zumofen}}, \bibinfo
  {author} {\bibfnamefont {E.}~\bibnamefont {Ikonen}}, \bibinfo {author}
  {\bibfnamefont {S.}~\bibnamefont {G\"{o}tzinger}}, \ and\ \bibinfo {author}
  {\bibfnamefont {V.}~\bibnamefont {Sandoghdar}},\ }\href {\doibase
  10.1103/PhysRevLett.104.123605} {\bibfield  {journal} {\bibinfo  {journal}
  {Phys. Rev. Lett.}\ }\textbf {\bibinfo {volume} {104}},\ \bibinfo {pages}
  {123605} (\bibinfo {year} {2010})}\BibitemShut {NoStop}%
\bibitem [{\citenamefont {Bernien}\ \emph {et~al.}(2012)\citenamefont
  {Bernien}, \citenamefont {Childress}, \citenamefont {Robledo}, \citenamefont
  {Markham}, \citenamefont {Twitchen},\ and\ \citenamefont
  {Hanson}}]{bernien_two-photon_2012}%
  \BibitemOpen
  \bibfield  {author} {\bibinfo {author} {\bibfnamefont {H.}~\bibnamefont
  {Bernien}}, \bibinfo {author} {\bibfnamefont {L.}~\bibnamefont {Childress}},
  \bibinfo {author} {\bibfnamefont {L.}~\bibnamefont {Robledo}}, \bibinfo
  {author} {\bibfnamefont {M.}~\bibnamefont {Markham}}, \bibinfo {author}
  {\bibfnamefont {D.}~\bibnamefont {Twitchen}}, \ and\ \bibinfo {author}
  {\bibfnamefont {R.}~\bibnamefont {Hanson}},\ }\href {\doibase
  10.1103/PhysRevLett.108.043604} {\bibfield  {journal} {\bibinfo  {journal}
  {Phys. Rev. Lett.}\ }\textbf {\bibinfo {volume} {108}},\ \bibinfo {pages}
  {043604} (\bibinfo {year} {2012})}\BibitemShut {NoStop}%
\bibitem [{\citenamefont {Press}\ \emph {et~al.}(2008)\citenamefont {Press},
  \citenamefont {Ladd}, \citenamefont {Zhang},\ and\ \citenamefont
  {Yamamoto}}]{press_complete_2008}%
  \BibitemOpen
  \bibfield  {author} {\bibinfo {author} {\bibfnamefont {D.}~\bibnamefont
  {Press}}, \bibinfo {author} {\bibfnamefont {T.~D.}\ \bibnamefont {Ladd}},
  \bibinfo {author} {\bibfnamefont {B.}~\bibnamefont {Zhang}}, \ and\ \bibinfo
  {author} {\bibfnamefont {Y.}~\bibnamefont {Yamamoto}},\ }\href {\doibase
  10.1038/nature07530} {\bibfield  {journal} {\bibinfo  {journal} {Nature
  (London)}\ }\textbf {\bibinfo {volume} {456}},\ \bibinfo {pages} {218}
  (\bibinfo {year} {2008})}\BibitemShut {NoStop}%
\bibitem [{\citenamefont {Yale}\ \emph {et~al.}(2013)\citenamefont {Yale},
  \citenamefont {Buckley}, \citenamefont {Christle}, \citenamefont {Burkard},
  \citenamefont {Heremans}, \citenamefont {Bassett},\ and\ \citenamefont
  {Awschalom}}]{yale_all-optical_2013}%
  \BibitemOpen
  \bibfield  {author} {\bibinfo {author} {\bibfnamefont {C.~G.}\ \bibnamefont
  {Yale}}, \bibinfo {author} {\bibfnamefont {B.~B.}\ \bibnamefont {Buckley}},
  \bibinfo {author} {\bibfnamefont {D.~J.}\ \bibnamefont {Christle}}, \bibinfo
  {author} {\bibfnamefont {G.}~\bibnamefont {Burkard}}, \bibinfo {author}
  {\bibfnamefont {F.~J.}\ \bibnamefont {Heremans}}, \bibinfo {author}
  {\bibfnamefont {L.~C.}\ \bibnamefont {Bassett}}, \ and\ \bibinfo {author}
  {\bibfnamefont {D.~D.}\ \bibnamefont {Awschalom}},\ }\href {\doibase
  10.1073/pnas.1305920110} {\bibfield  {journal} {\bibinfo  {journal} {Proc.
  Natl. Acad. Sci. U.S.A.}\ }\textbf {\bibinfo {volume} {110}},\ \bibinfo
  {pages} {7595} (\bibinfo {year} {2013})}\BibitemShut {NoStop}%
\bibitem [{\citenamefont {Bechtold}\ \emph {et~al.}(2015)\citenamefont
  {Bechtold}, \citenamefont {Rauch}, \citenamefont {Li}, \citenamefont
  {Simmet}, \citenamefont {Ardelt}, \citenamefont {Regler}, \citenamefont
  {M\"{u}ller}, \citenamefont {Sinitsyn},\ and\ \citenamefont
  {Finley}}]{bechtold_three-stage_2015}%
  \BibitemOpen
  \bibfield  {author} {\bibinfo {author} {\bibfnamefont {A.}~\bibnamefont
  {Bechtold}}, \bibinfo {author} {\bibfnamefont {D.}~\bibnamefont {Rauch}},
  \bibinfo {author} {\bibfnamefont {F.}~\bibnamefont {Li}}, \bibinfo {author}
  {\bibfnamefont {T.}~\bibnamefont {Simmet}}, \bibinfo {author} {\bibfnamefont
  {P.-L.}\ \bibnamefont {Ardelt}}, \bibinfo {author} {\bibfnamefont
  {A.}~\bibnamefont {Regler}}, \bibinfo {author} {\bibfnamefont
  {K.}~\bibnamefont {M\"{u}ller}}, \bibinfo {author} {\bibfnamefont {N.~A.}\
  \bibnamefont {Sinitsyn}}, \ and\ \bibinfo {author} {\bibfnamefont {J.~J.}\
  \bibnamefont {Finley}},\ }\href {\doibase 10.1038/nphys3470} {\bibfield
  {journal} {\bibinfo  {journal} {Nat. Phys.}\ }\textbf {\bibinfo {volume}
  {11}},\ \bibinfo {pages} {1005} (\bibinfo {year} {2015})}\BibitemShut
  {NoStop}%
\bibitem [{\citenamefont {Togan}\ \emph {et~al.}(2010)\citenamefont {Togan},
  \citenamefont {Chu}, \citenamefont {Trifonov}, \citenamefont {Jiang},
  \citenamefont {Maze}, \citenamefont {Childress}, \citenamefont {Dutt},
  \citenamefont {S{\o}rensen}, \citenamefont {Hemmer}, \citenamefont {Zibrov},\
  and\ \citenamefont {Lukin}}]{togan_quantum_2010}%
  \BibitemOpen
  \bibfield  {author} {\bibinfo {author} {\bibfnamefont {E.}~\bibnamefont
  {Togan}}, \bibinfo {author} {\bibfnamefont {Y.}~\bibnamefont {Chu}}, \bibinfo
  {author} {\bibfnamefont {A.~S.}\ \bibnamefont {Trifonov}}, \bibinfo {author}
  {\bibfnamefont {L.}~\bibnamefont {Jiang}}, \bibinfo {author} {\bibfnamefont
  {J.}~\bibnamefont {Maze}}, \bibinfo {author} {\bibfnamefont {L.}~\bibnamefont
  {Childress}}, \bibinfo {author} {\bibfnamefont {M.~V.~G.}\ \bibnamefont
  {Dutt}}, \bibinfo {author} {\bibfnamefont {A.~S.}\ \bibnamefont
  {S{\o}rensen}}, \bibinfo {author} {\bibfnamefont {P.~R.}\ \bibnamefont
  {Hemmer}}, \bibinfo {author} {\bibfnamefont {A.~S.}\ \bibnamefont {Zibrov}},
  \ and\ \bibinfo {author} {\bibfnamefont {M.~D.}\ \bibnamefont {Lukin}},\
  }\href {\doibase 10.1038/nature09256} {\bibfield  {journal} {\bibinfo
  {journal} {Nature (London)}\ }\textbf {\bibinfo {volume} {466}},\ \bibinfo
  {pages} {730} (\bibinfo {year} {2010})}\BibitemShut {NoStop}%
\bibitem [{\citenamefont {Gao}\ \emph {et~al.}(2012)\citenamefont {Gao},
  \citenamefont {Fallahi}, \citenamefont {Togan}, \citenamefont
  {Miguel-Sanchez},\ and\ \citenamefont {Imamo{\u
  g}lu}}]{gao_observation_2012}%
  \BibitemOpen
  \bibfield  {author} {\bibinfo {author} {\bibfnamefont {W.~B.}\ \bibnamefont
  {Gao}}, \bibinfo {author} {\bibfnamefont {P.}~\bibnamefont {Fallahi}},
  \bibinfo {author} {\bibfnamefont {E.}~\bibnamefont {Togan}}, \bibinfo
  {author} {\bibfnamefont {J.}~\bibnamefont {Miguel-Sanchez}}, \ and\ \bibinfo
  {author} {\bibfnamefont {A.}~\bibnamefont {Imamo{\u g}lu}},\ }\href {\doibase
  10.1038/nature11573} {\bibfield  {journal} {\bibinfo  {journal} {Nature
  (London)}\ }\textbf {\bibinfo {volume} {491}},\ \bibinfo {pages} {426}
  (\bibinfo {year} {2012})}\BibitemShut {NoStop}%
\bibitem [{\citenamefont {De~Greve}\ \emph {et~al.}(2012)\citenamefont
  {De~Greve}, \citenamefont {Yu}, \citenamefont {McMahon}, \citenamefont
  {Pelc}, \citenamefont {Natarajan}, \citenamefont {Kim}, \citenamefont {Abe},
  \citenamefont {Maier}, \citenamefont {Schneider}, \citenamefont {Kamp},
  \citenamefont {H\"ofling}, \citenamefont {Hadfield}, \citenamefont {Forchel},
  \citenamefont {Fejer},\ and\ \citenamefont
  {Yamamoto}}]{de_greve_quantum-dot_2012}%
  \BibitemOpen
  \bibfield  {author} {\bibinfo {author} {\bibfnamefont {K.}~\bibnamefont
  {De~Greve}}, \bibinfo {author} {\bibfnamefont {L.}~\bibnamefont {Yu}},
  \bibinfo {author} {\bibfnamefont {P.~L.}\ \bibnamefont {McMahon}}, \bibinfo
  {author} {\bibfnamefont {J.~S.}\ \bibnamefont {Pelc}}, \bibinfo {author}
  {\bibfnamefont {C.~M.}\ \bibnamefont {Natarajan}}, \bibinfo {author}
  {\bibfnamefont {N.~Y.}\ \bibnamefont {Kim}}, \bibinfo {author} {\bibfnamefont
  {E.}~\bibnamefont {Abe}}, \bibinfo {author} {\bibfnamefont {S.}~\bibnamefont
  {Maier}}, \bibinfo {author} {\bibfnamefont {C.}~\bibnamefont {Schneider}},
  \bibinfo {author} {\bibfnamefont {M.}~\bibnamefont {Kamp}}, \bibinfo {author}
  {\bibfnamefont {S.}~\bibnamefont {H\"ofling}}, \bibinfo {author}
  {\bibfnamefont {R.~H.}\ \bibnamefont {Hadfield}}, \bibinfo {author}
  {\bibfnamefont {A.}~\bibnamefont {Forchel}}, \bibinfo {author} {\bibfnamefont
  {M.~M.}\ \bibnamefont {Fejer}}, \ and\ \bibinfo {author} {\bibfnamefont
  {Y.}~\bibnamefont {Yamamoto}},\ }\href {\doibase 10.1038/nature11577}
  {\bibfield  {journal} {\bibinfo  {journal} {Nature (London)}\ }\textbf
  {\bibinfo {volume} {491}},\ \bibinfo {pages} {421} (\bibinfo {year}
  {2012})}\BibitemShut {NoStop}%
\bibitem [{\citenamefont {Santori}\ \emph {et~al.}(2009)\citenamefont
  {Santori}, \citenamefont {Fattal}, \citenamefont {Fu}, \citenamefont
  {Barclay},\ and\ \citenamefont
  {Beausoleil}}]{santori_indistinguishability_2009}%
  \BibitemOpen
  \bibfield  {author} {\bibinfo {author} {\bibfnamefont {C.}~\bibnamefont
  {Santori}}, \bibinfo {author} {\bibfnamefont {D.}~\bibnamefont {Fattal}},
  \bibinfo {author} {\bibfnamefont {K.-M.~C.}\ \bibnamefont {Fu}}, \bibinfo
  {author} {\bibfnamefont {P.~E.}\ \bibnamefont {Barclay}}, \ and\ \bibinfo
  {author} {\bibfnamefont {R.~G.}\ \bibnamefont {Beausoleil}},\ }\href
  {\doibase 10.1088/1367-2630/11/12/123009} {\bibfield  {journal} {\bibinfo
  {journal} {New Journal of Physics}\ }\textbf {\bibinfo {volume} {11}},\
  \bibinfo {pages} {123009} (\bibinfo {year} {2009})}\BibitemShut {NoStop}%
\bibitem [{\citenamefont {Fernandez}\ \emph {et~al.}(2009)\citenamefont
  {Fernandez}, \citenamefont {Volz}, \citenamefont {Desbuquois}, \citenamefont
  {Badolato},\ and\ \citenamefont {Imamo{\u g}lu}}]{fernandez_optically_2009}%
  \BibitemOpen
  \bibfield  {author} {\bibinfo {author} {\bibfnamefont {G.}~\bibnamefont
  {Fernandez}}, \bibinfo {author} {\bibfnamefont {T.}~\bibnamefont {Volz}},
  \bibinfo {author} {\bibfnamefont {R.}~\bibnamefont {Desbuquois}}, \bibinfo
  {author} {\bibfnamefont {A.}~\bibnamefont {Badolato}}, \ and\ \bibinfo
  {author} {\bibfnamefont {A.}~\bibnamefont {Imamo{\u g}lu}},\ }\href {\doibase
  10.1103/PhysRevLett.103.087406} {\bibfield  {journal} {\bibinfo  {journal}
  {Phys. Rev. Lett.}\ }\textbf {\bibinfo {volume} {103}},\ \bibinfo {pages}
  {087406} (\bibinfo {year} {2009})}\BibitemShut {NoStop}%
\bibitem [{\citenamefont {He}\ \emph {et~al.}(2013)\citenamefont {He},
  \citenamefont {He}, \citenamefont {Wei}, \citenamefont {Jiang}, \citenamefont
  {Chen}, \citenamefont {Xiong}, \citenamefont {Zhao}, \citenamefont
  {Schneider}, \citenamefont {Kamp}, \citenamefont {H\"{o}fling}, \citenamefont
  {Lu},\ and\ \citenamefont {Pan}}]{he_indistinguishable_2013}%
  \BibitemOpen
  \bibfield  {author} {\bibinfo {author} {\bibfnamefont {Y.}~\bibnamefont
  {He}}, \bibinfo {author} {\bibfnamefont {Y.-M.}\ \bibnamefont {He}}, \bibinfo
  {author} {\bibfnamefont {Y.-J.}\ \bibnamefont {Wei}}, \bibinfo {author}
  {\bibfnamefont {X.}~\bibnamefont {Jiang}}, \bibinfo {author} {\bibfnamefont
  {M.-C.}\ \bibnamefont {Chen}}, \bibinfo {author} {\bibfnamefont {F.-L.}\
  \bibnamefont {Xiong}}, \bibinfo {author} {\bibfnamefont {Y.}~\bibnamefont
  {Zhao}}, \bibinfo {author} {\bibfnamefont {C.}~\bibnamefont {Schneider}},
  \bibinfo {author} {\bibfnamefont {M.}~\bibnamefont {Kamp}}, \bibinfo {author}
  {\bibfnamefont {S.}~\bibnamefont {H\"{o}fling}}, \bibinfo {author}
  {\bibfnamefont {C.-Y.}\ \bibnamefont {Lu}}, \ and\ \bibinfo {author}
  {\bibfnamefont {J.-W.}\ \bibnamefont {Pan}},\ }\href {\doibase
  10.1103/PhysRevLett.111.237403} {\bibfield  {journal} {\bibinfo  {journal}
  {Phys. Rev. Lett.}\ }\textbf {\bibinfo {volume} {111}},\ \bibinfo {pages}
  {237403} (\bibinfo {year} {2013})}\BibitemShut {NoStop}%
\bibitem [{\citenamefont {Sun}\ \emph {et~al.}(2016)\citenamefont {Sun},
  \citenamefont {Delteil}, \citenamefont {Faelt},\ and\ \citenamefont {Imamo{\u
  g}lu}}]{sun_measurement_2016}%
  \BibitemOpen
  \bibfield  {author} {\bibinfo {author} {\bibfnamefont {Z.}~\bibnamefont
  {Sun}}, \bibinfo {author} {\bibfnamefont {A.}~\bibnamefont {Delteil}},
  \bibinfo {author} {\bibfnamefont {S.}~\bibnamefont {Faelt}}, \ and\ \bibinfo
  {author} {\bibfnamefont {A.}~\bibnamefont {Imamo{\u g}lu}},\ }\href {\doibase
  10.1103/PhysRevB.93.241302} {\bibfield  {journal} {\bibinfo  {journal} {Phys.
  Rev. B}\ }\textbf {\bibinfo {volume} {93}},\ \bibinfo {pages} {241302(R)}
  (\bibinfo {year} {2016})}\BibitemShut {NoStop}%
\bibitem [{\citenamefont {B\'{e}guin}\ \emph {et~al.}(2018)\citenamefont
  {B\'{e}guin}, \citenamefont {Jahn}, \citenamefont {Wolters}, \citenamefont
  {Reindl}, \citenamefont {Huo}, \citenamefont {Trotta}, \citenamefont
  {Rastelli}, \citenamefont {Ding}, \citenamefont {Schmidt}, \citenamefont
  {Treutlein},\ and\ \citenamefont {Warburton}}]{beguin_-demand_2018}%
  \BibitemOpen
  \bibfield  {author} {\bibinfo {author} {\bibfnamefont {L.}~\bibnamefont
  {B\'{e}guin}}, \bibinfo {author} {\bibfnamefont {J.-P.}\ \bibnamefont
  {Jahn}}, \bibinfo {author} {\bibfnamefont {J.}~\bibnamefont {Wolters}},
  \bibinfo {author} {\bibfnamefont {M.}~\bibnamefont {Reindl}}, \bibinfo
  {author} {\bibfnamefont {Y.}~\bibnamefont {Huo}}, \bibinfo {author}
  {\bibfnamefont {R.}~\bibnamefont {Trotta}}, \bibinfo {author} {\bibfnamefont
  {A.}~\bibnamefont {Rastelli}}, \bibinfo {author} {\bibfnamefont
  {F.}~\bibnamefont {Ding}}, \bibinfo {author} {\bibfnamefont {O.~G.}\
  \bibnamefont {Schmidt}}, \bibinfo {author} {\bibfnamefont {P.}~\bibnamefont
  {Treutlein}}, \ and\ \bibinfo {author} {\bibfnamefont {R.~J.}\ \bibnamefont
  {Warburton}},\ }\href {\doibase 10.1103/PhysRevB.97.205304} {\bibfield
  {journal} {\bibinfo  {journal} {Phys. Rev. B}\ }\textbf {\bibinfo {volume}
  {97}},\ \bibinfo {pages} {205304} (\bibinfo {year} {2018})}\BibitemShut
  {NoStop}%
\bibitem [{\citenamefont {Pursley}\ \emph {et~al.}(2018)\citenamefont
  {Pursley}, \citenamefont {Carter}, \citenamefont {Yakes}, \citenamefont
  {Bracker},\ and\ \citenamefont {Gammon}}]{pursley_picosecond_2018}%
  \BibitemOpen
  \bibfield  {author} {\bibinfo {author} {\bibfnamefont {B.~C.}\ \bibnamefont
  {Pursley}}, \bibinfo {author} {\bibfnamefont {S.~G.}\ \bibnamefont {Carter}},
  \bibinfo {author} {\bibfnamefont {M.~K.}\ \bibnamefont {Yakes}}, \bibinfo
  {author} {\bibfnamefont {A.~S.}\ \bibnamefont {Bracker}}, \ and\ \bibinfo
  {author} {\bibfnamefont {D.}~\bibnamefont {Gammon}},\ }\href {\doibase
  10.1038/s41467-017-02552-7} {\bibfield  {journal} {\bibinfo  {journal} {Nat.
  Commun.}\ }\textbf {\bibinfo {volume} {9}},\ \bibinfo {pages} {115} (\bibinfo
  {year} {2018})}\BibitemShut {NoStop}%
\bibitem [{\citenamefont {Cabrillo}\ \emph {et~al.}(1999)\citenamefont
  {Cabrillo}, \citenamefont {Cirac}, \citenamefont {Garc\'{\i}a-Fern\'andez},\
  and\ \citenamefont {Zoller}}]{ZollerPRA1999}%
  \BibitemOpen
  \bibfield  {author} {\bibinfo {author} {\bibfnamefont {C.}~\bibnamefont
  {Cabrillo}}, \bibinfo {author} {\bibfnamefont {J.~I.}\ \bibnamefont {Cirac}},
  \bibinfo {author} {\bibfnamefont {P.}~\bibnamefont
  {Garc\'{\i}a-Fern\'andez}}, \ and\ \bibinfo {author} {\bibfnamefont
  {P.}~\bibnamefont {Zoller}},\ }\href {\doibase 10.1103/PhysRevA.59.1025}
  {\bibfield  {journal} {\bibinfo  {journal} {Phys. Rev. A}\ }\textbf {\bibinfo
  {volume} {59}},\ \bibinfo {pages} {1025} (\bibinfo {year}
  {1999})}\BibitemShut {NoStop}%
\bibitem [{\citenamefont {Barrett}\ and\ \citenamefont {Kok}(2005)}]{Kok2005}%
  \BibitemOpen
  \bibfield  {author} {\bibinfo {author} {\bibfnamefont {S.~D.}\ \bibnamefont
  {Barrett}}\ and\ \bibinfo {author} {\bibfnamefont {P.}~\bibnamefont {Kok}},\
  }\href {\doibase 10.1103/PhysRevA.71.060310} {\bibfield  {journal} {\bibinfo
  {journal} {Phys. Rev. A}\ }\textbf {\bibinfo {volume} {71}},\ \bibinfo
  {pages} {060310(R)} (\bibinfo {year} {2005})}\BibitemShut {NoStop}%
\bibitem [{\citenamefont {Bernien}\ \emph {et~al.}(2013)\citenamefont
  {Bernien}, \citenamefont {Hensen}, \citenamefont {Pfaff}, \citenamefont
  {Koolstra}, \citenamefont {Blok}, \citenamefont {Robledo}, \citenamefont
  {Taminiau}, \citenamefont {Markham}, \citenamefont {Twitchen}, \citenamefont
  {Childress},\ and\ \citenamefont {Hanson}}]{bernien_heralded_2013}%
  \BibitemOpen
  \bibfield  {author} {\bibinfo {author} {\bibfnamefont {H.}~\bibnamefont
  {Bernien}}, \bibinfo {author} {\bibfnamefont {B.}~\bibnamefont {Hensen}},
  \bibinfo {author} {\bibfnamefont {W.}~\bibnamefont {Pfaff}}, \bibinfo
  {author} {\bibfnamefont {G.}~\bibnamefont {Koolstra}}, \bibinfo {author}
  {\bibfnamefont {M.~S.}\ \bibnamefont {Blok}}, \bibinfo {author}
  {\bibfnamefont {L.}~\bibnamefont {Robledo}}, \bibinfo {author} {\bibfnamefont
  {T.~H.}\ \bibnamefont {Taminiau}}, \bibinfo {author} {\bibfnamefont
  {M.}~\bibnamefont {Markham}}, \bibinfo {author} {\bibfnamefont {D.~J.}\
  \bibnamefont {Twitchen}}, \bibinfo {author} {\bibfnamefont {L.}~\bibnamefont
  {Childress}}, \ and\ \bibinfo {author} {\bibfnamefont {R.}~\bibnamefont
  {Hanson}},\ }\href {\doibase 10.1038/nature12016} {\bibfield  {journal}
  {\bibinfo  {journal} {Nature (London)}\ }\textbf {\bibinfo {volume} {497}},\
  \bibinfo {pages} {86} (\bibinfo {year} {2013})}\BibitemShut {NoStop}%
\bibitem [{\citenamefont {Delteil}\ \emph {et~al.}(2016)\citenamefont
  {Delteil}, \citenamefont {Sun}, \citenamefont {Gao}, \citenamefont {Togan},
  \citenamefont {Faelt},\ and\ \citenamefont {Imamo{\u
  g}lu}}]{delteil_generation_2016}%
  \BibitemOpen
  \bibfield  {author} {\bibinfo {author} {\bibfnamefont {A.}~\bibnamefont
  {Delteil}}, \bibinfo {author} {\bibfnamefont {Z.}~\bibnamefont {Sun}},
  \bibinfo {author} {\bibfnamefont {W.-b.}\ \bibnamefont {Gao}}, \bibinfo
  {author} {\bibfnamefont {E.}~\bibnamefont {Togan}}, \bibinfo {author}
  {\bibfnamefont {S.}~\bibnamefont {Faelt}}, \ and\ \bibinfo {author}
  {\bibfnamefont {A.}~\bibnamefont {Imamo{\u g}lu}},\ }\href {\doibase
  10.1038/nphys3605} {\bibfield  {journal} {\bibinfo  {journal} {Nat. Phys.}\
  }\textbf {\bibinfo {volume} {12}},\ \bibinfo {pages} {218} (\bibinfo {year}
  {2016})}\BibitemShut {NoStop}%
\bibitem [{\citenamefont {Stockill}\ \emph {et~al.}(2017)\citenamefont
  {Stockill}, \citenamefont {Stanley}, \citenamefont {Huthmacher},
  \citenamefont {Clarke}, \citenamefont {Hugues}, \citenamefont {Miller},
  \citenamefont {Matthiesen}, \citenamefont {Le~Gall},\ and\ \citenamefont
  {Atat\"{u}re}}]{stockill_phase-tuned_2017}%
  \BibitemOpen
  \bibfield  {author} {\bibinfo {author} {\bibfnamefont {R.}~\bibnamefont
  {Stockill}}, \bibinfo {author} {\bibfnamefont {M.~J.}\ \bibnamefont
  {Stanley}}, \bibinfo {author} {\bibfnamefont {L.}~\bibnamefont {Huthmacher}},
  \bibinfo {author} {\bibfnamefont {E.}~\bibnamefont {Clarke}}, \bibinfo
  {author} {\bibfnamefont {M.}~\bibnamefont {Hugues}}, \bibinfo {author}
  {\bibfnamefont {A.~J.}\ \bibnamefont {Miller}}, \bibinfo {author}
  {\bibfnamefont {C.}~\bibnamefont {Matthiesen}}, \bibinfo {author}
  {\bibfnamefont {C.}~\bibnamefont {Le~Gall}}, \ and\ \bibinfo {author}
  {\bibfnamefont {M.}~\bibnamefont {Atat\"{u}re}},\ }\href {\doibase
  10.1103/PhysRevLett.119.010503} {\bibfield  {journal} {\bibinfo  {journal}
  {Phys. Rev. Lett.}\ }\textbf {\bibinfo {volume} {119}},\ \bibinfo {pages}
  {010503} (\bibinfo {year} {2017})}\BibitemShut {NoStop}%
\bibitem [{\citenamefont {Wang}\ \emph {et~al.}(2017)\citenamefont {Wang},
  \citenamefont {He}, \citenamefont {Li}, \citenamefont {Su}, \citenamefont
  {Li}, \citenamefont {Huang}, \citenamefont {Ding}, \citenamefont {Chen},
  \citenamefont {Liu}, \citenamefont {Qin}, \citenamefont {Li}, \citenamefont
  {He}, \citenamefont {Schneider}, \citenamefont {Kamp}, \citenamefont {Peng},
  \citenamefont {H\"{o}fling}, \citenamefont {Lu},\ and\ \citenamefont
  {Pan}}]{wang_high-efficiency_2017}%
  \BibitemOpen
  \bibfield  {author} {\bibinfo {author} {\bibfnamefont {H.}~\bibnamefont
  {Wang}}, \bibinfo {author} {\bibfnamefont {Y.}~\bibnamefont {He}}, \bibinfo
  {author} {\bibfnamefont {Y.-H.}\ \bibnamefont {Li}}, \bibinfo {author}
  {\bibfnamefont {Z.-E.}\ \bibnamefont {Su}}, \bibinfo {author} {\bibfnamefont
  {B.}~\bibnamefont {Li}}, \bibinfo {author} {\bibfnamefont {H.-L.}\
  \bibnamefont {Huang}}, \bibinfo {author} {\bibfnamefont {X.}~\bibnamefont
  {Ding}}, \bibinfo {author} {\bibfnamefont {M.-C.}\ \bibnamefont {Chen}},
  \bibinfo {author} {\bibfnamefont {C.}~\bibnamefont {Liu}}, \bibinfo {author}
  {\bibfnamefont {J.}~\bibnamefont {Qin}}, \bibinfo {author} {\bibfnamefont
  {J.-P.}\ \bibnamefont {Li}}, \bibinfo {author} {\bibfnamefont {Y.-M.}\
  \bibnamefont {He}}, \bibinfo {author} {\bibfnamefont {C.}~\bibnamefont
  {Schneider}}, \bibinfo {author} {\bibfnamefont {M.}~\bibnamefont {Kamp}},
  \bibinfo {author} {\bibfnamefont {C.-Z.}\ \bibnamefont {Peng}}, \bibinfo
  {author} {\bibfnamefont {S.}~\bibnamefont {H\"{o}fling}}, \bibinfo {author}
  {\bibfnamefont {C.-Y.}\ \bibnamefont {Lu}}, \ and\ \bibinfo {author}
  {\bibfnamefont {J.-W.}\ \bibnamefont {Pan}},\ }\href {\doibase
  10.1038/nphoton.2017.63} {\bibfield  {journal} {\bibinfo  {journal} {Nat.
  Photonics}\ }\textbf {\bibinfo {volume} {11}},\ \bibinfo {pages} {361}
  (\bibinfo {year} {2017})}\BibitemShut {NoStop}%
\bibitem [{\citenamefont {Loredo}\ \emph {et~al.}(2017)\citenamefont {Loredo},
  \citenamefont {Broome}, \citenamefont {Hilaire}, \citenamefont {Gazzano},
  \citenamefont {Sagnes}, \citenamefont {Lemaitre}, \citenamefont {Almeida},
  \citenamefont {Senellart},\ and\ \citenamefont {White}}]{loredo_boson_2017}%
  \BibitemOpen
  \bibfield  {author} {\bibinfo {author} {\bibfnamefont {J.~C.}\ \bibnamefont
  {Loredo}}, \bibinfo {author} {\bibfnamefont {M.~A.}\ \bibnamefont {Broome}},
  \bibinfo {author} {\bibfnamefont {P.}~\bibnamefont {Hilaire}}, \bibinfo
  {author} {\bibfnamefont {O.}~\bibnamefont {Gazzano}}, \bibinfo {author}
  {\bibfnamefont {I.}~\bibnamefont {Sagnes}}, \bibinfo {author} {\bibfnamefont
  {A.}~\bibnamefont {Lemaitre}}, \bibinfo {author} {\bibfnamefont {M.~P.}\
  \bibnamefont {Almeida}}, \bibinfo {author} {\bibfnamefont {P.}~\bibnamefont
  {Senellart}}, \ and\ \bibinfo {author} {\bibfnamefont {A.~G.}\ \bibnamefont
  {White}},\ }\href {\doibase 10.1103/PhysRevLett.118.130503} {\bibfield
  {journal} {\bibinfo  {journal} {Phys. Rev. Lett.}\ }\textbf {\bibinfo
  {volume} {118}},\ \bibinfo {pages} {130503} (\bibinfo {year}
  {2017})}\BibitemShut {NoStop}%
\bibitem [{\citenamefont {Hong}\ \emph {et~al.}(1987)\citenamefont {Hong},
  \citenamefont {Ou},\ and\ \citenamefont {Mandel}}]{hong_measurement_1987}%
  \BibitemOpen
  \bibfield  {author} {\bibinfo {author} {\bibfnamefont {C.~K.}\ \bibnamefont
  {Hong}}, \bibinfo {author} {\bibfnamefont {Z.~Y.}\ \bibnamefont {Ou}}, \ and\
  \bibinfo {author} {\bibfnamefont {L.}~\bibnamefont {Mandel}},\ }\href
  {\doibase 10.1103/PhysRevLett.59.2044} {\bibfield  {journal} {\bibinfo
  {journal} {Phys. Rev. Lett.}\ }\textbf {\bibinfo {volume} {59}},\ \bibinfo
  {pages} {2044} (\bibinfo {year} {1987})}\BibitemShut {NoStop}%
\bibitem [{\citenamefont {Houel}\ \emph {et~al.}(2012)\citenamefont {Houel},
  \citenamefont {Kuhlmann}, \citenamefont {Greuter}, \citenamefont {Xue},
  \citenamefont {Poggio}, \citenamefont {Gerardot}, \citenamefont {Dalgarno},
  \citenamefont {Badolato}, \citenamefont {Petroff}, \citenamefont {Ludwig},
  \citenamefont {Reuter}, \citenamefont {Wieck},\ and\ \citenamefont
  {Warburton}}]{houel_probing_2012}%
  \BibitemOpen
  \bibfield  {author} {\bibinfo {author} {\bibfnamefont {J.}~\bibnamefont
  {Houel}}, \bibinfo {author} {\bibfnamefont {A.~V.}\ \bibnamefont {Kuhlmann}},
  \bibinfo {author} {\bibfnamefont {L.}~\bibnamefont {Greuter}}, \bibinfo
  {author} {\bibfnamefont {F.}~\bibnamefont {Xue}}, \bibinfo {author}
  {\bibfnamefont {M.}~\bibnamefont {Poggio}}, \bibinfo {author} {\bibfnamefont
  {B.~D.}\ \bibnamefont {Gerardot}}, \bibinfo {author} {\bibfnamefont {P.~A.}\
  \bibnamefont {Dalgarno}}, \bibinfo {author} {\bibfnamefont {A.}~\bibnamefont
  {Badolato}}, \bibinfo {author} {\bibfnamefont {P.~M.}\ \bibnamefont
  {Petroff}}, \bibinfo {author} {\bibfnamefont {A.}~\bibnamefont {Ludwig}},
  \bibinfo {author} {\bibfnamefont {D.}~\bibnamefont {Reuter}}, \bibinfo
  {author} {\bibfnamefont {A.~D.}\ \bibnamefont {Wieck}}, \ and\ \bibinfo
  {author} {\bibfnamefont {R.~J.}\ \bibnamefont {Warburton}},\ }\href {\doibase
  10.1103/PhysRevLett.108.107401} {\bibfield  {journal} {\bibinfo  {journal}
  {Phys. Rev. Lett.}\ }\textbf {\bibinfo {volume} {108}},\ \bibinfo {pages}
  {107401} (\bibinfo {year} {2012})}\BibitemShut {NoStop}%
\bibitem [{\citenamefont {Hameau}\ \emph {et~al.}(1999)\citenamefont {Hameau},
  \citenamefont {Guldner}, \citenamefont {Verzelen}, \citenamefont {Ferreira},
  \citenamefont {Bastard}, \citenamefont {Zeman}, \citenamefont
  {Lema{\^i}tre},\ and\ \citenamefont {G{\'e}rard}}]{hameau_strong_1999}%
  \BibitemOpen
  \bibfield  {author} {\bibinfo {author} {\bibfnamefont {S.}~\bibnamefont
  {Hameau}}, \bibinfo {author} {\bibfnamefont {Y.}~\bibnamefont {Guldner}},
  \bibinfo {author} {\bibfnamefont {O.}~\bibnamefont {Verzelen}}, \bibinfo
  {author} {\bibfnamefont {R.}~\bibnamefont {Ferreira}}, \bibinfo {author}
  {\bibfnamefont {G.}~\bibnamefont {Bastard}}, \bibinfo {author} {\bibfnamefont
  {J.}~\bibnamefont {Zeman}}, \bibinfo {author} {\bibfnamefont
  {A.}~\bibnamefont {Lema{\^i}tre}}, \ and\ \bibinfo {author} {\bibfnamefont
  {J.~M.}\ \bibnamefont {G{\'e}rard}},\ }\href {\doibase
  10.1103/PhysRevLett.83.4152} {\bibfield  {journal} {\bibinfo  {journal}
  {Phys. Rev. Lett.}\ }\textbf {\bibinfo {volume} {83}},\ \bibinfo {pages}
  {4152} (\bibinfo {year} {1999})}\BibitemShut {NoStop}%
\bibitem [{\citenamefont {Seebeck}\ \emph {et~al.}(2005)\citenamefont
  {Seebeck}, \citenamefont {Nielsen}, \citenamefont {Gartner},\ and\
  \citenamefont {Jahnke}}]{seebeck_polarons_2005}%
  \BibitemOpen
  \bibfield  {author} {\bibinfo {author} {\bibfnamefont {J.}~\bibnamefont
  {Seebeck}}, \bibinfo {author} {\bibfnamefont {T.~R.}\ \bibnamefont
  {Nielsen}}, \bibinfo {author} {\bibfnamefont {P.}~\bibnamefont {Gartner}}, \
  and\ \bibinfo {author} {\bibfnamefont {F.}~\bibnamefont {Jahnke}},\ }\href
  {\doibase 10.1103/PhysRevB.71.125327} {\bibfield  {journal} {\bibinfo
  {journal} {Phys. Rev. B}\ }\textbf {\bibinfo {volume} {71}},\ \bibinfo
  {pages} {125327} (\bibinfo {year} {2005})}\BibitemShut {NoStop}%
\bibitem [{\citenamefont {Vagov}\ \emph {et~al.}(2004)\citenamefont {Vagov},
  \citenamefont {Axt}, \citenamefont {Kuhn}, \citenamefont {Langbein},
  \citenamefont {Borri},\ and\ \citenamefont
  {Woggon}}]{vagov_nonmonotonous_2004}%
  \BibitemOpen
  \bibfield  {author} {\bibinfo {author} {\bibfnamefont {A.}~\bibnamefont
  {Vagov}}, \bibinfo {author} {\bibfnamefont {V.~M.}\ \bibnamefont {Axt}},
  \bibinfo {author} {\bibfnamefont {T.}~\bibnamefont {Kuhn}}, \bibinfo {author}
  {\bibfnamefont {W.}~\bibnamefont {Langbein}}, \bibinfo {author}
  {\bibfnamefont {P.}~\bibnamefont {Borri}}, \ and\ \bibinfo {author}
  {\bibfnamefont {U.}~\bibnamefont {Woggon}},\ }\href {\doibase
  10.1103/PhysRevB.70.201305} {\bibfield  {journal} {\bibinfo  {journal} {Phys.
  Rev. B}\ }\textbf {\bibinfo {volume} {70}},\ \bibinfo {pages} {201305(R)}
  (\bibinfo {year} {2004})}\BibitemShut {NoStop}%
\bibitem [{\citenamefont {Borri}\ \emph {et~al.}(2005)\citenamefont {Borri},
  \citenamefont {Langbein}, \citenamefont {Woggon}, \citenamefont {Stavarache},
  \citenamefont {Reuter},\ and\ \citenamefont {Wieck}}]{borri_exciton_2005}%
  \BibitemOpen
  \bibfield  {author} {\bibinfo {author} {\bibfnamefont {P.}~\bibnamefont
  {Borri}}, \bibinfo {author} {\bibfnamefont {W.}~\bibnamefont {Langbein}},
  \bibinfo {author} {\bibfnamefont {U.}~\bibnamefont {Woggon}}, \bibinfo
  {author} {\bibfnamefont {V.}~\bibnamefont {Stavarache}}, \bibinfo {author}
  {\bibfnamefont {D.}~\bibnamefont {Reuter}}, \ and\ \bibinfo {author}
  {\bibfnamefont {A.~D.}\ \bibnamefont {Wieck}},\ }\href {\doibase
  10.1103/PhysRevB.71.115328} {\bibfield  {journal} {\bibinfo  {journal} {Phys.
  Rev. B}\ }\textbf {\bibinfo {volume} {71}},\ \bibinfo {pages} {115328}
  (\bibinfo {year} {2005})}\BibitemShut {NoStop}%
\bibitem [{\citenamefont {Kaer}\ \emph {et~al.}(2013)\citenamefont {Kaer},
  \citenamefont {Lodahl}, \citenamefont {Jauho},\ and\ \citenamefont
  {Mork}}]{kaer_microscopic_2013}%
  \BibitemOpen
  \bibfield  {author} {\bibinfo {author} {\bibfnamefont {P.}~\bibnamefont
  {Kaer}}, \bibinfo {author} {\bibfnamefont {P.}~\bibnamefont {Lodahl}},
  \bibinfo {author} {\bibfnamefont {A.-P.}\ \bibnamefont {Jauho}}, \ and\
  \bibinfo {author} {\bibfnamefont {J.}~\bibnamefont {Mork}},\ }\href {\doibase
  10.1103/PhysRevB.87.081308} {\bibfield  {journal} {\bibinfo  {journal} {Phys.
  Rev. B}\ }\textbf {\bibinfo {volume} {87}},\ \bibinfo {pages} {081308(R)}
  (\bibinfo {year} {2013})}\BibitemShut {NoStop}%
\bibitem [{\citenamefont {{Roy-Choudhury}}\ and\ \citenamefont
  {Hughes}(2015)}]{roy-choudhury_theory_2015}%
  \BibitemOpen
  \bibfield  {author} {\bibinfo {author} {\bibfnamefont {K.}~\bibnamefont
  {{Roy-Choudhury}}}\ and\ \bibinfo {author} {\bibfnamefont {S.}~\bibnamefont
  {Hughes}},\ }\href {\doibase 10.1364/OL.40.001838} {\bibfield  {journal}
  {\bibinfo  {journal} {Opt. Lett.}\ }\textbf {\bibinfo {volume} {40}},\
  \bibinfo {pages} {1838} (\bibinfo {year} {2015})}\BibitemShut {NoStop}%
\bibitem [{\citenamefont {Thoma}\ \emph {et~al.}(2016)\citenamefont {Thoma},
  \citenamefont {Schnauber}, \citenamefont {Gschrey}, \citenamefont {Seifried},
  \citenamefont {Wolters}, \citenamefont {Schulze}, \citenamefont
  {Strittmatter}, \citenamefont {Rodt}, \citenamefont {Carmele}, \citenamefont
  {Knorr}, \citenamefont {Heindel},\ and\ \citenamefont
  {Reitzenstein}}]{thoma_exploring_2016}%
  \BibitemOpen
  \bibfield  {author} {\bibinfo {author} {\bibfnamefont {A.}~\bibnamefont
  {Thoma}}, \bibinfo {author} {\bibfnamefont {P.}~\bibnamefont {Schnauber}},
  \bibinfo {author} {\bibfnamefont {M.}~\bibnamefont {Gschrey}}, \bibinfo
  {author} {\bibfnamefont {M.}~\bibnamefont {Seifried}}, \bibinfo {author}
  {\bibfnamefont {J.}~\bibnamefont {Wolters}}, \bibinfo {author} {\bibfnamefont
  {J.-H.}\ \bibnamefont {Schulze}}, \bibinfo {author} {\bibfnamefont
  {A.}~\bibnamefont {Strittmatter}}, \bibinfo {author} {\bibfnamefont
  {S.}~\bibnamefont {Rodt}}, \bibinfo {author} {\bibfnamefont {A.}~\bibnamefont
  {Carmele}}, \bibinfo {author} {\bibfnamefont {A.}~\bibnamefont {Knorr}},
  \bibinfo {author} {\bibfnamefont {T.}~\bibnamefont {Heindel}}, \ and\
  \bibinfo {author} {\bibfnamefont {S.}~\bibnamefont {Reitzenstein}},\ }\href
  {\doibase 10.1103/PhysRevLett.116.033601} {\bibfield  {journal} {\bibinfo
  {journal} {Phys. Rev. Lett.}\ }\textbf {\bibinfo {volume} {116}},\ \bibinfo
  {pages} {033601} (\bibinfo {year} {2016})}\BibitemShut {NoStop}%
\bibitem [{\citenamefont {Reigue}\ \emph {et~al.}(2017)\citenamefont {Reigue},
  \citenamefont {{Iles-Smith}}, \citenamefont {Lux}, \citenamefont {Monniello},
  \citenamefont {Bernard}, \citenamefont {Margaillan}, \citenamefont
  {Lemaitre}, \citenamefont {Martinez}, \citenamefont {McCutcheon},
  \citenamefont {M\o{}rk}, \citenamefont {Hostein},\ and\ \citenamefont
  {Voliotis}}]{reigue_probing_2017}%
  \BibitemOpen
  \bibfield  {author} {\bibinfo {author} {\bibfnamefont {A.}~\bibnamefont
  {Reigue}}, \bibinfo {author} {\bibfnamefont {J.}~\bibnamefont
  {{Iles-Smith}}}, \bibinfo {author} {\bibfnamefont {F.}~\bibnamefont {Lux}},
  \bibinfo {author} {\bibfnamefont {L.}~\bibnamefont {Monniello}}, \bibinfo
  {author} {\bibfnamefont {M.}~\bibnamefont {Bernard}}, \bibinfo {author}
  {\bibfnamefont {F.}~\bibnamefont {Margaillan}}, \bibinfo {author}
  {\bibfnamefont {A.}~\bibnamefont {Lemaitre}}, \bibinfo {author}
  {\bibfnamefont {A.}~\bibnamefont {Martinez}}, \bibinfo {author}
  {\bibfnamefont {D.~P.~S.}\ \bibnamefont {McCutcheon}}, \bibinfo {author}
  {\bibfnamefont {J.}~\bibnamefont {M\o{}rk}}, \bibinfo {author} {\bibfnamefont
  {R.}~\bibnamefont {Hostein}}, \ and\ \bibinfo {author} {\bibfnamefont
  {V.}~\bibnamefont {Voliotis}},\ }\href {\doibase
  10.1103/PhysRevLett.118.233602} {\bibfield  {journal} {\bibinfo  {journal}
  {Phys. Rev. Lett.}\ }\textbf {\bibinfo {volume} {118}},\ \bibinfo {pages}
  {233602} (\bibinfo {year} {2017})}\BibitemShut {NoStop}%
\bibitem [{\citenamefont {Tighineanu}\ \emph {et~al.}(2018)\citenamefont
  {Tighineanu}, \citenamefont {Dree\ss{}en}, \citenamefont {Flindt},
  \citenamefont {Lodahl},\ and\ \citenamefont
  {S\o{}rensen}}]{tighineanu_phonon_2018}%
  \BibitemOpen
  \bibfield  {author} {\bibinfo {author} {\bibfnamefont {P.}~\bibnamefont
  {Tighineanu}}, \bibinfo {author} {\bibfnamefont {C.~L.}\ \bibnamefont
  {Dree\ss{}en}}, \bibinfo {author} {\bibfnamefont {C.}~\bibnamefont {Flindt}},
  \bibinfo {author} {\bibfnamefont {P.}~\bibnamefont {Lodahl}}, \ and\ \bibinfo
  {author} {\bibfnamefont {A.~S.}\ \bibnamefont {S\o{}rensen}},\ }\href
  {\doibase 10.1103/PhysRevLett.120.257401} {\bibfield  {journal} {\bibinfo
  {journal} {Phys. Rev. Lett.}\ }\textbf {\bibinfo {volume} {120}},\ \bibinfo
  {pages} {257401} (\bibinfo {year} {2018})}\BibitemShut {NoStop}%
\bibitem [{\citenamefont {Grange}\ \emph {et~al.}(2017)\citenamefont {Grange},
  \citenamefont {Somaschi}, \citenamefont {Ant\'on}, \citenamefont {De~Santis},
  \citenamefont {Coppola}, \citenamefont {Giesz}, \citenamefont {Lema\^itre},
  \citenamefont {Sagnes}, \citenamefont {Auff\`eves},\ and\ \citenamefont
  {Senellart}}]{grange_reducing_2017}%
  \BibitemOpen
  \bibfield  {author} {\bibinfo {author} {\bibfnamefont {T.}~\bibnamefont
  {Grange}}, \bibinfo {author} {\bibfnamefont {N.}~\bibnamefont {Somaschi}},
  \bibinfo {author} {\bibfnamefont {C.}~\bibnamefont {Ant\'on}}, \bibinfo
  {author} {\bibfnamefont {L.}~\bibnamefont {De~Santis}}, \bibinfo {author}
  {\bibfnamefont {G.}~\bibnamefont {Coppola}}, \bibinfo {author} {\bibfnamefont
  {V.}~\bibnamefont {Giesz}}, \bibinfo {author} {\bibfnamefont
  {A.}~\bibnamefont {Lema\^itre}}, \bibinfo {author} {\bibfnamefont
  {I.}~\bibnamefont {Sagnes}}, \bibinfo {author} {\bibfnamefont
  {A.}~\bibnamefont {Auff\`eves}}, \ and\ \bibinfo {author} {\bibfnamefont
  {P.}~\bibnamefont {Senellart}},\ }\href {\doibase
  10.1103/PhysRevLett.118.253602} {\bibfield  {journal} {\bibinfo  {journal}
  {Phys. Rev. Lett.}\ }\textbf {\bibinfo {volume} {118}},\ \bibinfo {pages}
  {253602} (\bibinfo {year} {2017})}\BibitemShut {NoStop}%
\bibitem [{\citenamefont {Iles-Smith}\ \emph
  {et~al.}(2017{\natexlab{a}})\citenamefont {Iles-Smith}, \citenamefont
  {McCutcheon}, \citenamefont {Nazir},\ and\ \citenamefont
  {M{\o}rk}}]{iles-smith_phonon_2017}%
  \BibitemOpen
  \bibfield  {author} {\bibinfo {author} {\bibfnamefont {J.}~\bibnamefont
  {Iles-Smith}}, \bibinfo {author} {\bibfnamefont {D.~P.~S.}\ \bibnamefont
  {McCutcheon}}, \bibinfo {author} {\bibfnamefont {A.}~\bibnamefont {Nazir}}, \
  and\ \bibinfo {author} {\bibfnamefont {J.}~\bibnamefont {M{\o}rk}},\ }\href
  {\doibase 10.1038/nphoton.2017.101} {\bibfield  {journal} {\bibinfo
  {journal} {Nat. Photonics}\ }\textbf {\bibinfo {volume} {11}},\ \bibinfo
  {pages} {521} (\bibinfo {year} {2017}{\natexlab{a}})}\BibitemShut {NoStop}%
\bibitem [{\citenamefont {Nguyen}\ \emph {et~al.}(2011)\citenamefont {Nguyen},
  \citenamefont {Sallen}, \citenamefont {Voisin}, \citenamefont {Roussignol},
  \citenamefont {Diederichs},\ and\ \citenamefont
  {Cassabois}}]{nguyen_ultra-coherent_2011}%
  \BibitemOpen
  \bibfield  {author} {\bibinfo {author} {\bibfnamefont {H.~S.}\ \bibnamefont
  {Nguyen}}, \bibinfo {author} {\bibfnamefont {G.}~\bibnamefont {Sallen}},
  \bibinfo {author} {\bibfnamefont {C.}~\bibnamefont {Voisin}}, \bibinfo
  {author} {\bibfnamefont {P.}~\bibnamefont {Roussignol}}, \bibinfo {author}
  {\bibfnamefont {C.}~\bibnamefont {Diederichs}}, \ and\ \bibinfo {author}
  {\bibfnamefont {G.}~\bibnamefont {Cassabois}},\ }\href {\doibase
  10.1063/1.3672034} {\bibfield  {journal} {\bibinfo  {journal} {Appl. Phys.
  Lett.}\ }\textbf {\bibinfo {volume} {99}},\ \bibinfo {pages} {261904}
  (\bibinfo {year} {2011})}\BibitemShut {NoStop}%
\bibitem [{\citenamefont {Konthasinghe}\ \emph {et~al.}(2012)\citenamefont
  {Konthasinghe}, \citenamefont {Walker}, \citenamefont {Peiris}, \citenamefont
  {Shih}, \citenamefont {Yu}, \citenamefont {Li}, \citenamefont {He},
  \citenamefont {Wang}, \citenamefont {Ni}, \citenamefont {Niu},\ and\
  \citenamefont {Muller}}]{konthasinghe_coherent_2012}%
  \BibitemOpen
  \bibfield  {author} {\bibinfo {author} {\bibfnamefont {K.}~\bibnamefont
  {Konthasinghe}}, \bibinfo {author} {\bibfnamefont {J.}~\bibnamefont
  {Walker}}, \bibinfo {author} {\bibfnamefont {M.}~\bibnamefont {Peiris}},
  \bibinfo {author} {\bibfnamefont {C.~K.}\ \bibnamefont {Shih}}, \bibinfo
  {author} {\bibfnamefont {Y.}~\bibnamefont {Yu}}, \bibinfo {author}
  {\bibfnamefont {M.~F.}\ \bibnamefont {Li}}, \bibinfo {author} {\bibfnamefont
  {J.~F.}\ \bibnamefont {He}}, \bibinfo {author} {\bibfnamefont {L.~J.}\
  \bibnamefont {Wang}}, \bibinfo {author} {\bibfnamefont {H.~Q.}\ \bibnamefont
  {Ni}}, \bibinfo {author} {\bibfnamefont {Z.~C.}\ \bibnamefont {Niu}}, \ and\
  \bibinfo {author} {\bibfnamefont {A.}~\bibnamefont {Muller}},\ }\href
  {\doibase 10.1103/PhysRevB.85.235315} {\bibfield  {journal} {\bibinfo
  {journal} {Phys. Rev. B}\ }\textbf {\bibinfo {volume} {85}},\ \bibinfo
  {pages} {235315} (\bibinfo {year} {2012})}\BibitemShut {NoStop}%
\bibitem [{\citenamefont {Matthiesen}\ \emph {et~al.}(2012)\citenamefont
  {Matthiesen}, \citenamefont {Vamivakas},\ and\ \citenamefont
  {Atat\"{u}re}}]{matthiesen_subnatural_2012}%
  \BibitemOpen
  \bibfield  {author} {\bibinfo {author} {\bibfnamefont {C.}~\bibnamefont
  {Matthiesen}}, \bibinfo {author} {\bibfnamefont {A.~N.}\ \bibnamefont
  {Vamivakas}}, \ and\ \bibinfo {author} {\bibfnamefont {M.}~\bibnamefont
  {Atat\"{u}re}},\ }\href {\doibase 10.1103/PhysRevLett.108.093602} {\bibfield
  {journal} {\bibinfo  {journal} {Phys. Rev. Lett.}\ }\textbf {\bibinfo
  {volume} {108}},\ \bibinfo {pages} {093602} (\bibinfo {year}
  {2012})}\BibitemShut {NoStop}%
\bibitem [{\citenamefont {Matthiesen}\ \emph {et~al.}(2013)\citenamefont
  {Matthiesen}, \citenamefont {Geller}, \citenamefont {Schulte}, \citenamefont
  {Le~Gall}, \citenamefont {Hansom}, \citenamefont {Li}, \citenamefont
  {Hugues}, \citenamefont {Clarke},\ and\ \citenamefont
  {Atat\"{u}re}}]{matthiesen_phase-locked_2013}%
  \BibitemOpen
  \bibfield  {author} {\bibinfo {author} {\bibfnamefont {C.}~\bibnamefont
  {Matthiesen}}, \bibinfo {author} {\bibfnamefont {M.}~\bibnamefont {Geller}},
  \bibinfo {author} {\bibfnamefont {C.~H.~H.}\ \bibnamefont {Schulte}},
  \bibinfo {author} {\bibfnamefont {C.}~\bibnamefont {Le~Gall}}, \bibinfo
  {author} {\bibfnamefont {J.}~\bibnamefont {Hansom}}, \bibinfo {author}
  {\bibfnamefont {Z.}~\bibnamefont {Li}}, \bibinfo {author} {\bibfnamefont
  {M.}~\bibnamefont {Hugues}}, \bibinfo {author} {\bibfnamefont
  {E.}~\bibnamefont {Clarke}}, \ and\ \bibinfo {author} {\bibfnamefont
  {M.}~\bibnamefont {Atat\"{u}re}},\ }\href {\doibase 10.1038/ncomms2601}
  {\bibfield  {journal} {\bibinfo  {journal} {Nat. Commun.}\ }\textbf {\bibinfo
  {volume} {4}},\ \bibinfo {pages} {1600} (\bibinfo {year} {2013})}\BibitemShut
  {NoStop}%
\bibitem [{\citenamefont {Proux}\ \emph {et~al.}(2015)\citenamefont {Proux},
  \citenamefont {Maragkou}, \citenamefont {Baudin}, \citenamefont {Voisin},
  \citenamefont {Roussignol},\ and\ \citenamefont
  {Diederichs}}]{proux_measuring_2015}%
  \BibitemOpen
  \bibfield  {author} {\bibinfo {author} {\bibfnamefont {R.}~\bibnamefont
  {Proux}}, \bibinfo {author} {\bibfnamefont {M.}~\bibnamefont {Maragkou}},
  \bibinfo {author} {\bibfnamefont {E.}~\bibnamefont {Baudin}}, \bibinfo
  {author} {\bibfnamefont {C.}~\bibnamefont {Voisin}}, \bibinfo {author}
  {\bibfnamefont {P.}~\bibnamefont {Roussignol}}, \ and\ \bibinfo {author}
  {\bibfnamefont {C.}~\bibnamefont {Diederichs}},\ }\href {\doibase
  10.1103/PhysRevLett.114.067401} {\bibfield  {journal} {\bibinfo  {journal}
  {Phys. Rev. Lett.}\ }\textbf {\bibinfo {volume} {114}},\ \bibinfo {pages}
  {067401} (\bibinfo {year} {2015})}\BibitemShut {NoStop}%
\bibitem [{\citenamefont {Schulte}\ \emph {et~al.}(2015)\citenamefont
  {Schulte}, \citenamefont {Hansom}, \citenamefont {Jones}, \citenamefont
  {Matthiesen}, \citenamefont {Le~Gall},\ and\ \citenamefont
  {Atat\"{u}re}}]{schulte_quadrature_2015}%
  \BibitemOpen
  \bibfield  {author} {\bibinfo {author} {\bibfnamefont {C.~H.~H.}\
  \bibnamefont {Schulte}}, \bibinfo {author} {\bibfnamefont {J.}~\bibnamefont
  {Hansom}}, \bibinfo {author} {\bibfnamefont {A.~E.}\ \bibnamefont {Jones}},
  \bibinfo {author} {\bibfnamefont {C.}~\bibnamefont {Matthiesen}}, \bibinfo
  {author} {\bibfnamefont {C.}~\bibnamefont {Le~Gall}}, \ and\ \bibinfo
  {author} {\bibfnamefont {M.}~\bibnamefont {Atat\"{u}re}},\ }\href {\doibase
  10.1038/nature14868} {\bibfield  {journal} {\bibinfo  {journal} {Nature
  (London)}\ }\textbf {\bibinfo {volume} {525}},\ \bibinfo {pages} {222}
  (\bibinfo {year} {2015})}\BibitemShut {NoStop}%
\bibitem [{\citenamefont {Malein}\ \emph {et~al.}(2016)\citenamefont {Malein},
  \citenamefont {Santana}, \citenamefont {Zajac}, \citenamefont {Dada},
  \citenamefont {Gauger}, \citenamefont {Petroff}, \citenamefont {Lim},
  \citenamefont {Song},\ and\ \citenamefont
  {Gerardot}}]{malein_screening_2016}%
  \BibitemOpen
  \bibfield  {author} {\bibinfo {author} {\bibfnamefont {R.~N.~E.}\
  \bibnamefont {Malein}}, \bibinfo {author} {\bibfnamefont {T.~S.}\
  \bibnamefont {Santana}}, \bibinfo {author} {\bibfnamefont {J.~M.}\
  \bibnamefont {Zajac}}, \bibinfo {author} {\bibfnamefont {A.~C.}\ \bibnamefont
  {Dada}}, \bibinfo {author} {\bibfnamefont {E.~M.}\ \bibnamefont {Gauger}},
  \bibinfo {author} {\bibfnamefont {P.~M.}\ \bibnamefont {Petroff}}, \bibinfo
  {author} {\bibfnamefont {J.~Y.}\ \bibnamefont {Lim}}, \bibinfo {author}
  {\bibfnamefont {J.~D.}\ \bibnamefont {Song}}, \ and\ \bibinfo {author}
  {\bibfnamefont {B.~D.}\ \bibnamefont {Gerardot}},\ }\href {\doibase
  10.1103/PhysRevLett.116.257401} {\bibfield  {journal} {\bibinfo  {journal}
  {Phys. Rev. Lett.}\ }\textbf {\bibinfo {volume} {116}},\ \bibinfo {pages}
  {257401} (\bibinfo {year} {2016})}\BibitemShut {NoStop}%
\bibitem [{\citenamefont {Baudin}\ \emph {et~al.}(2019)\citenamefont {Baudin},
  \citenamefont {Proux}, \citenamefont {Maragkou}, \citenamefont {Roussignol},\
  and\ \citenamefont {Diederichs}}]{baudin_correlation_2019}%
  \BibitemOpen
  \bibfield  {author} {\bibinfo {author} {\bibfnamefont {E.}~\bibnamefont
  {Baudin}}, \bibinfo {author} {\bibfnamefont {R.}~\bibnamefont {Proux}},
  \bibinfo {author} {\bibfnamefont {M.}~\bibnamefont {Maragkou}}, \bibinfo
  {author} {\bibfnamefont {P.}~\bibnamefont {Roussignol}}, \ and\ \bibinfo
  {author} {\bibfnamefont {C.}~\bibnamefont {Diederichs}},\ }\href {\doibase
  10.1103/PhysRevA.99.013842} {\bibfield  {journal} {\bibinfo  {journal} {Phys.
  Rev. A}\ }\textbf {\bibinfo {volume} {99}},\ \bibinfo {pages} {013842}
  (\bibinfo {year} {2019})}\BibitemShut {NoStop}%
\bibitem [{\citenamefont {Iles-Smith}\ \emph
  {et~al.}(2017{\natexlab{b}})\citenamefont {Iles-Smith}, \citenamefont
  {McCutcheon}, \citenamefont {M{\o}rk},\ and\ \citenamefont
  {Nazir}}]{iles-smith_limits_2017}%
  \BibitemOpen
  \bibfield  {author} {\bibinfo {author} {\bibfnamefont {J.}~\bibnamefont
  {Iles-Smith}}, \bibinfo {author} {\bibfnamefont {D.~P.~S.}\ \bibnamefont
  {McCutcheon}}, \bibinfo {author} {\bibfnamefont {J.}~\bibnamefont {M{\o}rk}},
  \ and\ \bibinfo {author} {\bibfnamefont {A.}~\bibnamefont {Nazir}},\ }\href
  {\doibase 10.1103/PhysRevB.95.201305} {\bibfield  {journal} {\bibinfo
  {journal} {Phys. Rev. B}\ }\textbf {\bibinfo {volume} {95}},\ \bibinfo
  {pages} {201305(R)} (\bibinfo {year} {2017}{\natexlab{b}})}\BibitemShut
  {NoStop}%
\bibitem [{\citenamefont {Lindner}\ and\ \citenamefont
  {Rudolph}(2009)}]{lindner_proposal_2009}%
  \BibitemOpen
  \bibfield  {author} {\bibinfo {author} {\bibfnamefont {N.~H.}\ \bibnamefont
  {Lindner}}\ and\ \bibinfo {author} {\bibfnamefont {T.}~\bibnamefont
  {Rudolph}},\ }\href {\doibase 10.1103/PhysRevLett.103.113602} {\bibfield
  {journal} {\bibinfo  {journal} {Phys. Rev. Lett.}\ }\textbf {\bibinfo
  {volume} {103}},\ \bibinfo {pages} {113602} (\bibinfo {year}
  {2009})}\BibitemShut {NoStop}%
\bibitem [{\citenamefont {Buterakos}\ \emph {et~al.}(2017)\citenamefont
  {Buterakos}, \citenamefont {Barnes},\ and\ \citenamefont
  {Economou}}]{buterakos_deterministic_2017}%
  \BibitemOpen
  \bibfield  {author} {\bibinfo {author} {\bibfnamefont {D.}~\bibnamefont
  {Buterakos}}, \bibinfo {author} {\bibfnamefont {E.}~\bibnamefont {Barnes}}, \
  and\ \bibinfo {author} {\bibfnamefont {S.~E.}\ \bibnamefont {Economou}},\
  }\href {\doibase 10.1103/PhysRevX.7.041023} {\bibfield  {journal} {\bibinfo
  {journal} {Phys. Rev. X}\ }\textbf {\bibinfo {volume} {7}},\ \bibinfo {pages}
  {041023} (\bibinfo {year} {2017})}\BibitemShut {NoStop}%
\bibitem [{\citenamefont {Gimeno-Segovia}\ \emph {et~al.}(2019)\citenamefont
  {Gimeno-Segovia}, \citenamefont {Rudolph},\ and\ \citenamefont
  {Economou}}]{gimeno-segovia_deterministic_2018}%
  \BibitemOpen
  \bibfield  {author} {\bibinfo {author} {\bibfnamefont {M.}~\bibnamefont
  {Gimeno-Segovia}}, \bibinfo {author} {\bibfnamefont {T.}~\bibnamefont
  {Rudolph}}, \ and\ \bibinfo {author} {\bibfnamefont {S.~E.}\ \bibnamefont
  {Economou}},\ }\href {http://arxiv.org/abs/1801.02599} {\bibfield  {journal}
  {\bibinfo  {journal} {Phys. Rev. Lett.}\ ,\ \bibinfo {pages} {070501}}
  (\bibinfo {year} {2019})}\BibitemShut {NoStop}%
\bibitem [{\citenamefont {Rudolph}(2017)}]{rudolph_why_2017}%
  \BibitemOpen
  \bibfield  {author} {\bibinfo {author} {\bibfnamefont {T.}~\bibnamefont
  {Rudolph}},\ }\href {\doibase 10.1063/1.4976737} {\bibfield  {journal}
  {\bibinfo  {journal} {APL Photonics}\ }\textbf {\bibinfo {volume} {2}},\
  \bibinfo {pages} {030901} (\bibinfo {year} {2017})}\BibitemShut {NoStop}%
\bibitem [{\citenamefont {Scerri}\ \emph {et~al.}(2018)\citenamefont {Scerri},
  \citenamefont {Malein}, \citenamefont {Gerardot},\ and\ \citenamefont
  {Gauger}}]{scerri_frequency-encoded_2018}%
  \BibitemOpen
  \bibfield  {author} {\bibinfo {author} {\bibfnamefont {D.}~\bibnamefont
  {Scerri}}, \bibinfo {author} {\bibfnamefont {R.~N.~E.}\ \bibnamefont
  {Malein}}, \bibinfo {author} {\bibfnamefont {B.~D.}\ \bibnamefont
  {Gerardot}}, \ and\ \bibinfo {author} {\bibfnamefont {E.~M.}\ \bibnamefont
  {Gauger}},\ }\href {\doibase 10.1103/PhysRevA.98.022318} {\bibfield
  {journal} {\bibinfo  {journal} {Phys. Rev. A}\ }\textbf {\bibinfo {volume}
  {98}},\ \bibinfo {pages} {022318} (\bibinfo {year} {2018})}\BibitemShut
  {NoStop}%
\bibitem [{\citenamefont {Ramsay}\ \emph
  {et~al.}(2010{\natexlab{a}})\citenamefont {Ramsay}, \citenamefont {Gopal},
  \citenamefont {Gauger}, \citenamefont {Nazir}, \citenamefont {Lovett},
  \citenamefont {Fox},\ and\ \citenamefont {Skolnick}}]{ramsay_damping_2010}%
  \BibitemOpen
  \bibfield  {author} {\bibinfo {author} {\bibfnamefont {A.~J.}\ \bibnamefont
  {Ramsay}}, \bibinfo {author} {\bibfnamefont {A.~V.}\ \bibnamefont {Gopal}},
  \bibinfo {author} {\bibfnamefont {E.~M.}\ \bibnamefont {Gauger}}, \bibinfo
  {author} {\bibfnamefont {A.}~\bibnamefont {Nazir}}, \bibinfo {author}
  {\bibfnamefont {B.~W.}\ \bibnamefont {Lovett}}, \bibinfo {author}
  {\bibfnamefont {A.~M.}\ \bibnamefont {Fox}}, \ and\ \bibinfo {author}
  {\bibfnamefont {M.~S.}\ \bibnamefont {Skolnick}},\ }\href {\doibase
  10.1103/PhysRevLett.104.017402} {\bibfield  {journal} {\bibinfo  {journal}
  {Phys. Rev. Lett.}\ }\textbf {\bibinfo {volume} {104}},\ \bibinfo {pages}
  {017402} (\bibinfo {year} {2010}{\natexlab{a}})}\BibitemShut {NoStop}%
\bibitem [{\citenamefont {Ramsay}\ \emph
  {et~al.}(2010{\natexlab{b}})\citenamefont {Ramsay}, \citenamefont {Godden},
  \citenamefont {Boyle}, \citenamefont {Gauger}, \citenamefont {Nazir},
  \citenamefont {Lovett}, \citenamefont {Fox},\ and\ \citenamefont
  {Skolnick}}]{ramsay_phonon-induced_2010}%
  \BibitemOpen
  \bibfield  {author} {\bibinfo {author} {\bibfnamefont {A.~J.}\ \bibnamefont
  {Ramsay}}, \bibinfo {author} {\bibfnamefont {T.~M.}\ \bibnamefont {Godden}},
  \bibinfo {author} {\bibfnamefont {S.~J.}\ \bibnamefont {Boyle}}, \bibinfo
  {author} {\bibfnamefont {E.~M.}\ \bibnamefont {Gauger}}, \bibinfo {author}
  {\bibfnamefont {A.}~\bibnamefont {Nazir}}, \bibinfo {author} {\bibfnamefont
  {B.~W.}\ \bibnamefont {Lovett}}, \bibinfo {author} {\bibfnamefont {A.~M.}\
  \bibnamefont {Fox}}, \ and\ \bibinfo {author} {\bibfnamefont {M.~S.}\
  \bibnamefont {Skolnick}},\ }\href {\doibase 10.1103/PhysRevLett.105.177402}
  {\bibfield  {journal} {\bibinfo  {journal} {Phys. Rev. Lett.}\ }\textbf
  {\bibinfo {volume} {105}},\ \bibinfo {pages} {177402} (\bibinfo {year}
  {2010}{\natexlab{b}})}\BibitemShut {NoStop}%
\bibitem [{Sup()}]{SupMat}%
  \BibitemOpen
  \href@noop {} {}\BibitemShut {NoStop}%
\bibitem [{\citenamefont {Nazir}\ and\ \citenamefont
  {McCutcheon}(2016)}]{nazir_modelling_2016}%
  \BibitemOpen
  \bibfield  {author} {\bibinfo {author} {\bibfnamefont {A.}~\bibnamefont
  {Nazir}}\ and\ \bibinfo {author} {\bibfnamefont {D.~P.~S.}\ \bibnamefont
  {McCutcheon}},\ }\href {\doibase 10.1088/0953-8984/28/10/103002} {\bibfield
  {journal} {\bibinfo  {journal} {Journal of Physics: Condensed Matter}\
  }\textbf {\bibinfo {volume} {28}},\ \bibinfo {pages} {103002} (\bibinfo
  {year} {2016})}\BibitemShut {NoStop}%
\bibitem [{\citenamefont {Lang}\ and\ \citenamefont
  {Firsov}(1962)}]{lang_kinetic_1962}%
  \BibitemOpen
  \bibfield  {author} {\bibinfo {author} {\bibfnamefont {I.}~\bibnamefont
  {Lang}}\ and\ \bibinfo {author} {\bibfnamefont {Y.~A.}\ \bibnamefont
  {Firsov}},\ }\href@noop {} {\bibfield  {journal} {\bibinfo  {journal} {J.
  Exp. Theor. Phys.}\ }\textbf {\bibinfo {volume} {43}},\ \bibinfo {pages}
  {1301} (\bibinfo {year} {1962})}\BibitemShut {NoStop}%
\bibitem [{\citenamefont {Mahan}(2000)}]{Mahan}%
  \BibitemOpen
  \bibfield  {author} {\bibinfo {author} {\bibfnamefont {G.}~\bibnamefont
  {Mahan}},\ }\href {https://books.google.co.uk/books?id=xzSgZ4-yyMEC} {\emph
  {\bibinfo {title} {Many-Particle Physics}}},\ Physics of Solids and Liquids\
  (\bibinfo  {publisher} {Springer, New York},\ \bibinfo {year}
  {2000})\BibitemShut {NoStop}%
\bibitem [{\citenamefont {Leggett}\ \emph {et~al.}(1987)\citenamefont
  {Leggett}, \citenamefont {Chakravarty}, \citenamefont {Dorsey}, \citenamefont
  {Fisher}, \citenamefont {Garg},\ and\ \citenamefont
  {Zwerger}}]{leggett_dynamics_1987}%
  \BibitemOpen
  \bibfield  {author} {\bibinfo {author} {\bibfnamefont {A.~J.}\ \bibnamefont
  {Leggett}}, \bibinfo {author} {\bibfnamefont {S.}~\bibnamefont
  {Chakravarty}}, \bibinfo {author} {\bibfnamefont {A.~T.}\ \bibnamefont
  {Dorsey}}, \bibinfo {author} {\bibfnamefont {M.~P.~A.}\ \bibnamefont
  {Fisher}}, \bibinfo {author} {\bibfnamefont {A.}~\bibnamefont {Garg}}, \ and\
  \bibinfo {author} {\bibfnamefont {W.}~\bibnamefont {Zwerger}},\ }\href
  {\doibase 10.1103/RevModPhys.59.1} {\bibfield  {journal} {\bibinfo  {journal}
  {Rev. Mod. Phys.}\ }\textbf {\bibinfo {volume} {59}},\ \bibinfo {pages} {1}
  (\bibinfo {year} {1987})}\BibitemShut {NoStop}%
\bibitem [{\citenamefont {Scully}\ and\ \citenamefont
  {Zubairy}(1997)}]{scully_quantum_1997}%
  \BibitemOpen
  \bibfield  {author} {\bibinfo {author} {\bibfnamefont {M.~O.}\ \bibnamefont
  {Scully}}\ and\ \bibinfo {author} {\bibfnamefont {M.~S.}\ \bibnamefont
  {Zubairy}},\ }\href {\doibase 10.1017/CBO9780511813993} {\emph {\bibinfo
  {title} {Quantum Optics}}}\ (\bibinfo  {publisher} {Cambridge University
  Press, England},\ \bibinfo {address} {Cambridge},\ \bibinfo {year}
  {1997})\BibitemShut {NoStop}%
\bibitem [{\citenamefont {Cohen-Tannoudji}\ \emph {et~al.}(1998)\citenamefont
  {Cohen-Tannoudji}, \citenamefont {Dupont-Roc},\ and\ \citenamefont
  {Grynberg}}]{cohentannoudji_atomphoton_1998}%
  \BibitemOpen
  \bibfield  {author} {\bibinfo {author} {\bibfnamefont {C.}~\bibnamefont
  {Cohen-Tannoudji}}, \bibinfo {author} {\bibfnamefont {J.}~\bibnamefont
  {Dupont-Roc}}, \ and\ \bibinfo {author} {\bibfnamefont {G.}~\bibnamefont
  {Grynberg}},\ }\href {\doibase 10.1002/9783527617197} {\emph {\bibinfo
  {title} {Atom\textemdash{{Photon Interactions}}: {{Basic Process}} and
  {{Applications}}}}},\ \bibinfo {edition} {1st}\ ed.\ (\bibinfo  {publisher}
  {{Wiley, New York}},\ \bibinfo {year} {1998})\ pp.\ \bibinfo {pages}
  {369--383}\BibitemShut {NoStop}%
\bibitem [{\citenamefont {Lipkin}()}]{lipkin_physics_2004}%
  \BibitemOpen
  \bibfield  {author} {\bibinfo {author} {\bibfnamefont {H.~J.}\ \bibnamefont
  {Lipkin}},\ }\href {http://arxiv.org/abs/cond-mat/0405023} {\bibinfo
  {journal} {arXiv:cond-mat/0405023}\ }\BibitemShut {NoStop}%
\bibitem [{Note1()}]{Note1}%
  \BibitemOpen
\bibfield  {journal} {  }\bibinfo {note} {The equivalence between {$ \delimiter
  "426830A B \delimiter "526930B $} and the Franck-Condon factor stems from the
  fact that both quantities emerge as vibrational prefactors when taking the
  transition probability of the system coupled to the vibrational
  environment.}\BibitemShut {Stop}%
\bibitem [{\citenamefont {Legero}\ \emph {et~al.}(2004)\citenamefont {Legero},
  \citenamefont {Wilk}, \citenamefont {Hennrich}, \citenamefont {Rempe},\ and\
  \citenamefont {Kuhn}}]{legero_quantum_2004}%
  \BibitemOpen
  \bibfield  {author} {\bibinfo {author} {\bibfnamefont {T.}~\bibnamefont
  {Legero}}, \bibinfo {author} {\bibfnamefont {T.}~\bibnamefont {Wilk}},
  \bibinfo {author} {\bibfnamefont {M.}~\bibnamefont {Hennrich}}, \bibinfo
  {author} {\bibfnamefont {G.}~\bibnamefont {Rempe}}, \ and\ \bibinfo {author}
  {\bibfnamefont {A.}~\bibnamefont {Kuhn}},\ }\href {\doibase
  10.1103/PhysRevLett.93.070503} {\bibfield  {journal} {\bibinfo  {journal}
  {Phys. Rev. Lett.}\ }\textbf {\bibinfo {volume} {93}},\ \bibinfo {pages}
  {070503} (\bibinfo {year} {2004})}\BibitemShut {NoStop}%
\bibitem [{\citenamefont {Giesz}\ \emph {et~al.}(2015)\citenamefont {Giesz},
  \citenamefont {Portalupi}, \citenamefont {Grange}, \citenamefont {Ant{\'o}n},
  \citenamefont {De~Santis}, \citenamefont {Demory}, \citenamefont {Somaschi},
  \citenamefont {Sagnes}, \citenamefont {Lema{\^i}tre}, \citenamefont {Lanco},
  \citenamefont {Auff{\`e}ves},\ and\ \citenamefont
  {Senellart}}]{giesz_cavity-enhanced_2015}%
  \BibitemOpen
  \bibfield  {author} {\bibinfo {author} {\bibfnamefont {V.}~\bibnamefont
  {Giesz}}, \bibinfo {author} {\bibfnamefont {S.~L.}\ \bibnamefont
  {Portalupi}}, \bibinfo {author} {\bibfnamefont {T.}~\bibnamefont {Grange}},
  \bibinfo {author} {\bibfnamefont {C.}~\bibnamefont {Ant{\'o}n}}, \bibinfo
  {author} {\bibfnamefont {L.}~\bibnamefont {De~Santis}}, \bibinfo {author}
  {\bibfnamefont {J.}~\bibnamefont {Demory}}, \bibinfo {author} {\bibfnamefont
  {N.}~\bibnamefont {Somaschi}}, \bibinfo {author} {\bibfnamefont
  {I.}~\bibnamefont {Sagnes}}, \bibinfo {author} {\bibfnamefont
  {A.}~\bibnamefont {Lema{\^i}tre}}, \bibinfo {author} {\bibfnamefont
  {L.}~\bibnamefont {Lanco}}, \bibinfo {author} {\bibfnamefont
  {A.}~\bibnamefont {Auff{\`e}ves}}, \ and\ \bibinfo {author} {\bibfnamefont
  {P.}~\bibnamefont {Senellart}},\ }\href {\doibase 10.1103/PhysRevB.92.161302}
  {\bibfield  {journal} {\bibinfo  {journal} {Phys. Rev. B}\ }\textbf {\bibinfo
  {volume} {92}},\ \bibinfo {pages} {161302(R)} (\bibinfo {year}
  {2015})}\BibitemShut {NoStop}%
\bibitem [{\citenamefont {Brash}\ \emph {et~al.}(2019)\citenamefont {Brash},
  \citenamefont {Iles-Smith}, \citenamefont {Phillips}, \citenamefont
  {McCutcheon}, \citenamefont {O'Hara}, \citenamefont {Clarke}, \citenamefont
  {Royall}, \citenamefont {Wilson}, \citenamefont {M\o{}rk}, \citenamefont
  {Skolnick}, \citenamefont {Fox},\ and\ \citenamefont
  {Nazir}}]{brash_light_2019}%
  \BibitemOpen
  \bibfield  {author} {\bibinfo {author} {\bibfnamefont {A.~J.}\ \bibnamefont
  {Brash}}, \bibinfo {author} {\bibfnamefont {J.}~\bibnamefont {Iles-Smith}},
  \bibinfo {author} {\bibfnamefont {C.~L.}\ \bibnamefont {Phillips}}, \bibinfo
  {author} {\bibfnamefont {D.~P.~S.}\ \bibnamefont {McCutcheon}}, \bibinfo
  {author} {\bibfnamefont {J.}~\bibnamefont {O'Hara}}, \bibinfo {author}
  {\bibfnamefont {E.}~\bibnamefont {Clarke}}, \bibinfo {author} {\bibfnamefont
  {B.}~\bibnamefont {Royall}}, \bibinfo {author} {\bibfnamefont {L.~R.}\
  \bibnamefont {Wilson}}, \bibinfo {author} {\bibfnamefont {J.}~\bibnamefont
  {M\o{}rk}}, \bibinfo {author} {\bibfnamefont {M.~S.}\ \bibnamefont
  {Skolnick}}, \bibinfo {author} {\bibfnamefont {A.~M.}\ \bibnamefont {Fox}}, \
  and\ \bibinfo {author} {\bibfnamefont {A.}~\bibnamefont {Nazir}},\ }\href
  {\doibase 10.1103/PhysRevLett.123.167403} {\bibfield  {journal} {\bibinfo
  {journal} {Phys. Rev. Lett.}\ }\textbf {\bibinfo {volume} {123}},\ \bibinfo
  {pages} {167403} (\bibinfo {year} {2019})}\BibitemShut {NoStop}%
\bibitem [{\citenamefont {Ficek}\ \emph {et~al.}(2005)\citenamefont {Ficek},
  \citenamefont {Asakura}, \citenamefont {Swain}, \citenamefont {Brenner},
  \citenamefont {H{\"a}nsch}, \citenamefont {Kamiya}, \citenamefont {Krausz},
  \citenamefont {Monemar}, \citenamefont {Rhodes}, \citenamefont {Venghaus}
  \emph {et~al.}}]{Ficke2005}%
  \BibitemOpen
  \bibfield  {author} {\bibinfo {author} {\bibfnamefont {Z.}~\bibnamefont
  {Ficek}}, \bibinfo {author} {\bibfnamefont {T.}~\bibnamefont {Asakura}},
  \bibinfo {author} {\bibfnamefont {S.}~\bibnamefont {Swain}}, \bibinfo
  {author} {\bibfnamefont {K.}~\bibnamefont {Brenner}}, \bibinfo {author}
  {\bibfnamefont {T.}~\bibnamefont {H{\"a}nsch}}, \bibinfo {author}
  {\bibfnamefont {T.}~\bibnamefont {Kamiya}}, \bibinfo {author} {\bibfnamefont
  {F.}~\bibnamefont {Krausz}}, \bibinfo {author} {\bibfnamefont
  {B.}~\bibnamefont {Monemar}}, \bibinfo {author} {\bibfnamefont
  {W.}~\bibnamefont {Rhodes}}, \bibinfo {author} {\bibfnamefont
  {H.}~\bibnamefont {Venghaus}},  \emph {et~al.},\ }\href
  {https://books.google.co.uk/books?id=AYhF6-K2OKcC} {\emph {\bibinfo {title}
  {Quantum Interference and Coherence: Theory and Experiments}}},\ Springer
  Series in Optical Sciences\ (\bibinfo  {publisher} {Springer, New York},\
  \bibinfo {year} {2005})\BibitemShut {NoStop}%
\bibitem [{\citenamefont {Lebreton}\ \emph {et~al.}(2013)\citenamefont
  {Lebreton}, \citenamefont {Abram}, \citenamefont {Braive}, \citenamefont
  {Sagnes}, \citenamefont {Robert-Philip},\ and\ \citenamefont
  {Beveratos}}]{lebreton2013}%
  \BibitemOpen
  \bibfield  {author} {\bibinfo {author} {\bibfnamefont {A.}~\bibnamefont
  {Lebreton}}, \bibinfo {author} {\bibfnamefont {I.}~\bibnamefont {Abram}},
  \bibinfo {author} {\bibfnamefont {R.}~\bibnamefont {Braive}}, \bibinfo
  {author} {\bibfnamefont {I.}~\bibnamefont {Sagnes}}, \bibinfo {author}
  {\bibfnamefont {I.}~\bibnamefont {Robert-Philip}}, \ and\ \bibinfo {author}
  {\bibfnamefont {A.}~\bibnamefont {Beveratos}},\ }\href {\doibase
  10.1103/PhysRevA.88.013801} {\bibfield  {journal} {\bibinfo  {journal} {Phys.
  Rev. A}\ }\textbf {\bibinfo {volume} {88}},\ \bibinfo {pages} {013801}
  (\bibinfo {year} {2013})}\BibitemShut {NoStop}%
\bibitem [{\citenamefont {Urbaszek}\ \emph {et~al.}(2013)\citenamefont
  {Urbaszek}, \citenamefont {Marie}, \citenamefont {Amand}, \citenamefont
  {Krebs}, \citenamefont {Voisin}, \citenamefont {Malentinsky}, \citenamefont
  {H\"{o}gele},\ and\ \citenamefont {Imamo{\u g}lu}}]{urbaszek_nuclear_2013}%
  \BibitemOpen
  \bibfield  {author} {\bibinfo {author} {\bibfnamefont {B.}~\bibnamefont
  {Urbaszek}}, \bibinfo {author} {\bibfnamefont {X.}~\bibnamefont {Marie}},
  \bibinfo {author} {\bibfnamefont {T.}~\bibnamefont {Amand}}, \bibinfo
  {author} {\bibfnamefont {O.}~\bibnamefont {Krebs}}, \bibinfo {author}
  {\bibfnamefont {P.}~\bibnamefont {Voisin}}, \bibinfo {author} {\bibfnamefont
  {P.}~\bibnamefont {Malentinsky}}, \bibinfo {author} {\bibfnamefont
  {A.}~\bibnamefont {H\"{o}gele}}, \ and\ \bibinfo {author} {\bibfnamefont
  {A.}~\bibnamefont {Imamo{\u g}lu}},\ }\href {\doibase
  10.1103/RevModPhys.85.79} {\bibfield  {journal} {\bibinfo  {journal} {Rev.
  Mod. Phys.}\ }\textbf {\bibinfo {volume} {85}},\ \bibinfo {pages} {79}
  (\bibinfo {year} {2013})}\BibitemShut {NoStop}%
\bibitem [{\citenamefont {Scerri}\ \emph {et~al.}(2017)\citenamefont {Scerri},
  \citenamefont {Santana}, \citenamefont {Gerardot},\ and\ \citenamefont
  {Gauger}}]{Scerri2017}%
  \BibitemOpen
  \bibfield  {author} {\bibinfo {author} {\bibfnamefont {D.}~\bibnamefont
  {Scerri}}, \bibinfo {author} {\bibfnamefont {T.~S.}\ \bibnamefont {Santana}},
  \bibinfo {author} {\bibfnamefont {B.~D.}\ \bibnamefont {Gerardot}}, \ and\
  \bibinfo {author} {\bibfnamefont {E.~M.}\ \bibnamefont {Gauger}},\ }\href
  {\doibase 10.1103/PhysRevB.95.165403} {\bibfield  {journal} {\bibinfo
  {journal} {Phys. Rev. B}\ }\textbf {\bibinfo {volume} {95}},\ \bibinfo
  {pages} {165403} (\bibinfo {year} {2017})}\BibitemShut {NoStop}%
\bibitem [{\citenamefont {Roy-Choudhury}\ and\ \citenamefont
  {Hughes}(2015)}]{Hughes2015b}%
  \BibitemOpen
  \bibfield  {author} {\bibinfo {author} {\bibfnamefont {K.}~\bibnamefont
  {Roy-Choudhury}}\ and\ \bibinfo {author} {\bibfnamefont {S.}~\bibnamefont
  {Hughes}},\ }\href {\doibase 10.1103/PhysRevB.92.205406} {\bibfield
  {journal} {\bibinfo  {journal} {Phys. Rev. B}\ }\textbf {\bibinfo {volume}
  {92}},\ \bibinfo {pages} {205406} (\bibinfo {year} {2015})}\BibitemShut
  {NoStop}%
\bibitem [{\citenamefont {Breuer}\ and\ \citenamefont
  {Petruccione}(2007)}]{Breuer2007}%
  \BibitemOpen
  \bibfield  {author} {\bibinfo {author} {\bibfnamefont {H.}~\bibnamefont
  {Breuer}}\ and\ \bibinfo {author} {\bibfnamefont {F.}~\bibnamefont
  {Petruccione}},\ }\href {https://books.google.co.uk/books?id=DkcJPwAACAAJ}
  {\emph {\bibinfo {title} {The Theory of Open Quantum Systems}}}\ (\bibinfo
  {publisher} {Oxford University Press, Oxford},\ \bibinfo {year}
  {2007})\BibitemShut {NoStop}%
\end{thebibliography}%


%apsrev4-2.bst 2019-01-14 (MD) hand-edited version of apsrev4-1.bst
%Control: key (0)
%Control: author (72) initials jnrlst
%Control: editor formatted (1) identically to author
%Control: production of article title (-1) disabled
%Control: page (0) single
%Control: year (1) truncated
%Control: production of eprint (0) enabled
\begin{thebibliography}{23}%
\makeatletter
\providecommand \@ifxundefined [1]{%
 \@ifx{#1\undefined}
}%
\providecommand \@ifnum [1]{%
 \ifnum #1\expandafter \@firstoftwo
 \else \expandafter \@secondoftwo
 \fi
}%
\providecommand \@ifx [1]{%
 \ifx #1\expandafter \@firstoftwo
 \else \expandafter \@secondoftwo
 \fi
}%
\providecommand \natexlab [1]{#1}%
\providecommand \enquote  [1]{``#1''}%
\providecommand \bibnamefont  [1]{#1}%
\providecommand \bibfnamefont [1]{#1}%
\providecommand \citenamefont [1]{#1}%
\providecommand \href@noop [0]{\@secondoftwo}%
\providecommand \href [0]{\begingroup \@sanitize@url \@href}%
\providecommand \@href[1]{\@@startlink{#1}\@@href}%
\providecommand \@@href[1]{\endgroup#1\@@endlink}%
\providecommand \@sanitize@url [0]{\catcode `\\12\catcode `\$12\catcode
  `\&12\catcode `\#12\catcode `\^12\catcode `\_12\catcode `\%12\relax}%
\providecommand \@@startlink[1]{}%
\providecommand \@@endlink[0]{}%
\providecommand \url  [0]{\begingroup\@sanitize@url \@url }%
\providecommand \@url [1]{\endgroup\@href {#1}{\urlprefix }}%
\providecommand \urlprefix  [0]{URL }%
\providecommand \Eprint [0]{\href }%
\providecommand \doibase [0]{https://doi.org/}%
\providecommand \selectlanguage [0]{\@gobble}%
\providecommand \bibinfo  [0]{\@secondoftwo}%
\providecommand \bibfield  [0]{\@secondoftwo}%
\providecommand \translation [1]{[#1]}%
\providecommand \BibitemOpen [0]{}%
\providecommand \bibitemStop [0]{}%
\providecommand \bibitemNoStop [0]{.\EOS\space}%
\providecommand \EOS [0]{\spacefactor3000\relax}%
\providecommand \BibitemShut  [1]{\csname bibitem#1\endcsname}%
\let\auto@bib@innerbib\@empty
%</preamble>
\bibitem [{\citenamefont {Hong}\ \emph {et~al.}(1987)\citenamefont {Hong},
  \citenamefont {Ou},\ and\ \citenamefont {Mandel}}]{hong_measurement_1987}%
  \BibitemOpen
  \bibfield  {author} {\bibinfo {author} {\bibfnamefont {C.~K.}\ \bibnamefont
  {Hong}}, \bibinfo {author} {\bibfnamefont {Z.~Y.}\ \bibnamefont {Ou}},\ and\
  \bibinfo {author} {\bibfnamefont {L.}~\bibnamefont {Mandel}},\ }\href
  {https://doi.org/10.1103/PhysRevLett.59.2044} {\bibfield  {journal} {\bibinfo
   {journal} {Physical Review Letters}\ }\textbf {\bibinfo {volume} {59}},\
  \bibinfo {pages} {2044} (\bibinfo {year} {1987})}\BibitemShut {NoStop}%
\bibitem [{\citenamefont {Proux}\ \emph {et~al.}(2015)\citenamefont {Proux},
  \citenamefont {Maragkou}, \citenamefont {Baudin}, \citenamefont {Voisin},
  \citenamefont {Roussignol},\ and\ \citenamefont
  {Diederichs}}]{proux_measuring_2015}%
  \BibitemOpen
  \bibfield  {author} {\bibinfo {author} {\bibfnamefont {R.}~\bibnamefont
  {Proux}}, \bibinfo {author} {\bibfnamefont {M.}~\bibnamefont {Maragkou}},
  \bibinfo {author} {\bibfnamefont {E.}~\bibnamefont {Baudin}}, \bibinfo
  {author} {\bibfnamefont {C.}~\bibnamefont {Voisin}}, \bibinfo {author}
  {\bibfnamefont {P.}~\bibnamefont {Roussignol}},\ and\ \bibinfo {author}
  {\bibfnamefont {C.}~\bibnamefont {Diederichs}},\ }\href
  {https://doi.org/10.1103/PhysRevLett.114.067401} {\bibfield  {journal}
  {\bibinfo  {journal} {Phys. Rev. Lett.}\ }\textbf {\bibinfo {volume} {114}},\
  \bibinfo {pages} {067401} (\bibinfo {year} {2015})}\BibitemShut {NoStop}%
\bibitem [{\citenamefont {Scully}\ and\ \citenamefont
  {Zubairy}(1997)}]{scully_quantum_1997}%
  \BibitemOpen
  \bibfield  {author} {\bibinfo {author} {\bibfnamefont {M.~O.}\ \bibnamefont
  {Scully}}\ and\ \bibinfo {author} {\bibfnamefont {M.~S.}\ \bibnamefont
  {Zubairy}},\ }\href {https://doi.org/10.1017/CBO9780511813993} {\emph
  {\bibinfo {title} {Quantum optics}}}\ (\bibinfo  {publisher} {Cambridge
  University Press},\ \bibinfo {address} {Cambridge},\ \bibinfo {year}
  {1997})\BibitemShut {NoStop}%
\bibitem [{\citenamefont {Legero}\ \emph {et~al.}(2004)\citenamefont {Legero},
  \citenamefont {Wilk}, \citenamefont {Hennrich}, \citenamefont {Rempe},\ and\
  \citenamefont {Kuhn}}]{legero_quantum_2004}%
  \BibitemOpen
  \bibfield  {author} {\bibinfo {author} {\bibfnamefont {T.}~\bibnamefont
  {Legero}}, \bibinfo {author} {\bibfnamefont {T.}~\bibnamefont {Wilk}},
  \bibinfo {author} {\bibfnamefont {M.}~\bibnamefont {Hennrich}}, \bibinfo
  {author} {\bibfnamefont {G.}~\bibnamefont {Rempe}},\ and\ \bibinfo {author}
  {\bibfnamefont {A.}~\bibnamefont {Kuhn}},\ }\href
  {https://doi.org/10.1103/PhysRevLett.93.070503} {\bibfield  {journal}
  {\bibinfo  {journal} {Physical Review Letters}\ }\textbf {\bibinfo {volume}
  {93}},\ \bibinfo {pages} {070503} (\bibinfo {year} {2004})}\BibitemShut
  {NoStop}%
\bibitem [{\citenamefont {Baudin}\ \emph {et~al.}(2019)\citenamefont {Baudin},
  \citenamefont {Proux}, \citenamefont {Maragkou}, \citenamefont {Roussignol},\
  and\ \citenamefont {Diederichs}}]{baudin_correlation_2019}%
  \BibitemOpen
  \bibfield  {author} {\bibinfo {author} {\bibfnamefont {E.}~\bibnamefont
  {Baudin}}, \bibinfo {author} {\bibfnamefont {R.}~\bibnamefont {Proux}},
  \bibinfo {author} {\bibfnamefont {M.}~\bibnamefont {Maragkou}}, \bibinfo
  {author} {\bibfnamefont {P.}~\bibnamefont {Roussignol}},\ and\ \bibinfo
  {author} {\bibfnamefont {C.}~\bibnamefont {Diederichs}},\ }\href
  {https://doi.org/10.1103/PhysRevA.99.013842} {\bibfield  {journal} {\bibinfo
  {journal} {Physical Review A}\ }\textbf {\bibinfo {volume} {99}},\ \bibinfo
  {pages} {013842} (\bibinfo {year} {2019})}\BibitemShut {NoStop}%
\bibitem [{\citenamefont {Giesz}\ \emph {et~al.}(2015)\citenamefont {Giesz},
  \citenamefont {Portalupi}, \citenamefont {Grange}, \citenamefont {Ant\'on},
  \citenamefont {De~Santis}, \citenamefont {Demory}, \citenamefont {Somaschi},
  \citenamefont {Sagnes}, \citenamefont {Lema\^{\i}tre}, \citenamefont {Lanco},
  \citenamefont {Auff\`eves},\ and\ \citenamefont
  {Senellart}}]{giesz_cavity_2015}%
  \BibitemOpen
  \bibfield  {author} {\bibinfo {author} {\bibfnamefont {V.}~\bibnamefont
  {Giesz}}, \bibinfo {author} {\bibfnamefont {S.~L.}\ \bibnamefont
  {Portalupi}}, \bibinfo {author} {\bibfnamefont {T.}~\bibnamefont {Grange}},
  \bibinfo {author} {\bibfnamefont {C.}~\bibnamefont {Ant\'on}}, \bibinfo
  {author} {\bibfnamefont {L.}~\bibnamefont {De~Santis}}, \bibinfo {author}
  {\bibfnamefont {J.}~\bibnamefont {Demory}}, \bibinfo {author} {\bibfnamefont
  {N.}~\bibnamefont {Somaschi}}, \bibinfo {author} {\bibfnamefont
  {I.}~\bibnamefont {Sagnes}}, \bibinfo {author} {\bibfnamefont
  {A.}~\bibnamefont {Lema\^{\i}tre}}, \bibinfo {author} {\bibfnamefont
  {L.}~\bibnamefont {Lanco}}, \bibinfo {author} {\bibfnamefont
  {A.}~\bibnamefont {Auff\`eves}},\ and\ \bibinfo {author} {\bibfnamefont
  {P.}~\bibnamefont {Senellart}},\ }\href
  {https://doi.org/10.1103/PhysRevB.92.161302} {\bibfield  {journal} {\bibinfo
  {journal} {Phys. Rev. B}\ }\textbf {\bibinfo {volume} {92}},\ \bibinfo
  {pages} {161302} (\bibinfo {year} {2015})}\BibitemShut {NoStop}%
\bibitem [{\citenamefont {Iles-Smith}\ \emph {et~al.}(2017)\citenamefont
  {Iles-Smith}, \citenamefont {McCutcheon}, \citenamefont {M{\o}rk},\ and\
  \citenamefont {Nazir}}]{iles-smith_limits_2017}%
  \BibitemOpen
  \bibfield  {author} {\bibinfo {author} {\bibfnamefont {J.}~\bibnamefont
  {Iles-Smith}}, \bibinfo {author} {\bibfnamefont {D.~P.~S.}\ \bibnamefont
  {McCutcheon}}, \bibinfo {author} {\bibfnamefont {J.}~\bibnamefont
  {M{\o}rk}},\ and\ \bibinfo {author} {\bibfnamefont {A.}~\bibnamefont
  {Nazir}},\ }\href {https://doi.org/10.1103/PhysRevB.95.201305} {\bibfield
  {journal} {\bibinfo  {journal} {Physical Review B}\ }\textbf {\bibinfo
  {volume} {95}},\ \bibinfo {pages} {201305} (\bibinfo {year}
  {2017})}\BibitemShut {NoStop}%
\bibitem [{\citenamefont {Ficek}\ \emph
  {et~al.}(2005{\natexlab{a}})\citenamefont {Ficek}, \citenamefont {Asakura},
  \citenamefont {Swain}, \citenamefont {Brenner}, \citenamefont {H{\"a}nsch},
  \citenamefont {Kamiya}, \citenamefont {Krausz}, \citenamefont {Monemar},
  \citenamefont {Rhodes}, \citenamefont {Venghaus} \emph {et~al.}}]{Ficek2005}%
  \BibitemOpen
  \bibfield  {author} {\bibinfo {author} {\bibfnamefont {Z.}~\bibnamefont
  {Ficek}}, \bibinfo {author} {\bibfnamefont {T.}~\bibnamefont {Asakura}},
  \bibinfo {author} {\bibfnamefont {S.}~\bibnamefont {Swain}}, \bibinfo
  {author} {\bibfnamefont {K.}~\bibnamefont {Brenner}}, \bibinfo {author}
  {\bibfnamefont {T.}~\bibnamefont {H{\"a}nsch}}, \bibinfo {author}
  {\bibfnamefont {T.}~\bibnamefont {Kamiya}}, \bibinfo {author} {\bibfnamefont
  {F.}~\bibnamefont {Krausz}}, \bibinfo {author} {\bibfnamefont
  {B.}~\bibnamefont {Monemar}}, \bibinfo {author} {\bibfnamefont
  {W.}~\bibnamefont {Rhodes}}, \bibinfo {author} {\bibfnamefont
  {H.}~\bibnamefont {Venghaus}}, \emph {et~al.},\ }\href
  {https://books.google.co.uk/books?id=AYhF6-K2OKcC} {\emph {\bibinfo {title}
  {Quantum Interference and Coherence: Theory and Experiments}}},\ Springer
  Series in Optical Sciences\ (\bibinfo  {publisher} {Springer},\ \bibinfo
  {year} {2005})\BibitemShut {NoStop}%
\bibitem [{\citenamefont {Lebreton}\ \emph {et~al.}(2013)\citenamefont
  {Lebreton}, \citenamefont {Abram}, \citenamefont {Braive}, \citenamefont
  {Sagnes}, \citenamefont {Robert-Philip},\ and\ \citenamefont
  {Beveratos}}]{lebreton2013}%
  \BibitemOpen
  \bibfield  {author} {\bibinfo {author} {\bibfnamefont {A.}~\bibnamefont
  {Lebreton}}, \bibinfo {author} {\bibfnamefont {I.}~\bibnamefont {Abram}},
  \bibinfo {author} {\bibfnamefont {R.}~\bibnamefont {Braive}}, \bibinfo
  {author} {\bibfnamefont {I.}~\bibnamefont {Sagnes}}, \bibinfo {author}
  {\bibfnamefont {I.}~\bibnamefont {Robert-Philip}},\ and\ \bibinfo {author}
  {\bibfnamefont {A.}~\bibnamefont {Beveratos}},\ }\href
  {https://doi.org/10.1103/PhysRevA.88.013801} {\bibfield  {journal} {\bibinfo
  {journal} {Phys. Rev. A}\ }\textbf {\bibinfo {volume} {88}},\ \bibinfo
  {pages} {013801} (\bibinfo {year} {2013})}\BibitemShut {NoStop}%
\bibitem [{\citenamefont {Urbaszek}\ \emph {et~al.}(2013)\citenamefont
  {Urbaszek}, \citenamefont {Marie}, \citenamefont {Amand}, \citenamefont
  {Krebs}, \citenamefont {Voisin}, \citenamefont {Maletinsky}, \citenamefont
  {H\"{o}gele},\ and\ \citenamefont {Imamo{\u g}lu}}]{urbaszek_nuclear_2013}%
  \BibitemOpen
  \bibfield  {author} {\bibinfo {author} {\bibfnamefont {B.}~\bibnamefont
  {Urbaszek}}, \bibinfo {author} {\bibfnamefont {X.}~\bibnamefont {Marie}},
  \bibinfo {author} {\bibfnamefont {T.}~\bibnamefont {Amand}}, \bibinfo
  {author} {\bibfnamefont {O.}~\bibnamefont {Krebs}}, \bibinfo {author}
  {\bibfnamefont {P.}~\bibnamefont {Voisin}}, \bibinfo {author} {\bibfnamefont
  {P.}~\bibnamefont {Maletinsky}}, \bibinfo {author} {\bibfnamefont
  {A.}~\bibnamefont {H\"{o}gele}},\ and\ \bibinfo {author} {\bibfnamefont
  {A.}~\bibnamefont {Imamo{\u g}lu}},\ }\href
  {https://doi.org/10.1103/RevModPhys.85.79} {\bibfield  {journal} {\bibinfo
  {journal} {Reviews of Modern Physics}\ }\textbf {\bibinfo {volume} {85}},\
  \bibinfo {pages} {79} (\bibinfo {year} {2013})}\BibitemShut {NoStop}%
\bibitem [{\citenamefont {Malein}\ \emph {et~al.}(2016)\citenamefont {Malein},
  \citenamefont {Santana}, \citenamefont {Zajac}, \citenamefont {Dada},
  \citenamefont {Gauger}, \citenamefont {Petroff}, \citenamefont {Lim},
  \citenamefont {Song},\ and\ \citenamefont
  {Gerardot}}]{malein_screening_2016}%
  \BibitemOpen
  \bibfield  {author} {\bibinfo {author} {\bibfnamefont {R.}~\bibnamefont
  {Malein}}, \bibinfo {author} {\bibfnamefont {T.}~\bibnamefont {Santana}},
  \bibinfo {author} {\bibfnamefont {J.}~\bibnamefont {Zajac}}, \bibinfo
  {author} {\bibfnamefont {A.}~\bibnamefont {Dada}}, \bibinfo {author}
  {\bibfnamefont {E.}~\bibnamefont {Gauger}}, \bibinfo {author} {\bibfnamefont
  {P.}~\bibnamefont {Petroff}}, \bibinfo {author} {\bibfnamefont
  {J.}~\bibnamefont {Lim}}, \bibinfo {author} {\bibfnamefont {J.}~\bibnamefont
  {Song}},\ and\ \bibinfo {author} {\bibfnamefont {B.}~\bibnamefont
  {Gerardot}},\ }\href {https://doi.org/10.1103/PhysRevLett.116.257401}
  {\bibfield  {journal} {\bibinfo  {journal} {Physical Review Letters}\
  }\textbf {\bibinfo {volume} {116}},\ \bibinfo {pages} {257401} (\bibinfo
  {year} {2016})}\BibitemShut {NoStop}%
\bibitem [{\citenamefont {Fernandez}\ \emph {et~al.}(2009)\citenamefont
  {Fernandez}, \citenamefont {Volz}, \citenamefont {Desbuquois}, \citenamefont
  {Badolato},\ and\ \citenamefont {Imamo{\u g}lu}}]{fernandez_optically_2009}%
  \BibitemOpen
  \bibfield  {author} {\bibinfo {author} {\bibfnamefont {G.}~\bibnamefont
  {Fernandez}}, \bibinfo {author} {\bibfnamefont {T.}~\bibnamefont {Volz}},
  \bibinfo {author} {\bibfnamefont {R.}~\bibnamefont {Desbuquois}}, \bibinfo
  {author} {\bibfnamefont {A.}~\bibnamefont {Badolato}},\ and\ \bibinfo
  {author} {\bibfnamefont {A.}~\bibnamefont {Imamo{\u g}lu}},\ }\href
  {https://doi.org/10.1103/PhysRevLett.103.087406} {\bibfield  {journal}
  {\bibinfo  {journal} {Physical Review Letters}\ }\textbf {\bibinfo {volume}
  {103}},\ \bibinfo {pages} {087406} (\bibinfo {year} {2009})}\BibitemShut
  {NoStop}%
\bibitem [{\citenamefont {He}\ \emph {et~al.}(2013)\citenamefont {He},
  \citenamefont {He}, \citenamefont {Wei}, \citenamefont {Jiang}, \citenamefont
  {Chen}, \citenamefont {Xiong}, \citenamefont {Zhao}, \citenamefont
  {Schneider}, \citenamefont {Kamp}, \citenamefont {H\"{o}fling}, \citenamefont
  {Lu},\ and\ \citenamefont {Pan}}]{he_indistinguishable_2013}%
  \BibitemOpen
  \bibfield  {author} {\bibinfo {author} {\bibfnamefont {Y.}~\bibnamefont
  {He}}, \bibinfo {author} {\bibfnamefont {Y.-M.}\ \bibnamefont {He}}, \bibinfo
  {author} {\bibfnamefont {Y.-J.}\ \bibnamefont {Wei}}, \bibinfo {author}
  {\bibfnamefont {X.}~\bibnamefont {Jiang}}, \bibinfo {author} {\bibfnamefont
  {M.-C.}\ \bibnamefont {Chen}}, \bibinfo {author} {\bibfnamefont {F.-L.}\
  \bibnamefont {Xiong}}, \bibinfo {author} {\bibfnamefont {Y.}~\bibnamefont
  {Zhao}}, \bibinfo {author} {\bibfnamefont {C.}~\bibnamefont {Schneider}},
  \bibinfo {author} {\bibfnamefont {M.}~\bibnamefont {Kamp}}, \bibinfo {author}
  {\bibfnamefont {S.}~\bibnamefont {H\"{o}fling}}, \bibinfo {author}
  {\bibfnamefont {C.-Y.}\ \bibnamefont {Lu}},\ and\ \bibinfo {author}
  {\bibfnamefont {J.-W.}\ \bibnamefont {Pan}},\ }\href
  {https://doi.org/10.1103/PhysRevLett.111.237403} {\bibfield  {journal}
  {\bibinfo  {journal} {Physical Review Letters}\ }\textbf {\bibinfo {volume}
  {111}},\ \bibinfo {pages} {237403} (\bibinfo {year} {2013})}\BibitemShut
  {NoStop}%
\bibitem [{\citenamefont {Sun}\ \emph {et~al.}(2016)\citenamefont {Sun},
  \citenamefont {Delteil}, \citenamefont {Faelt},\ and\ \citenamefont {Imamo{\u
  g}lu}}]{sun_measurement_2016}%
  \BibitemOpen
  \bibfield  {author} {\bibinfo {author} {\bibfnamefont {Z.}~\bibnamefont
  {Sun}}, \bibinfo {author} {\bibfnamefont {A.}~\bibnamefont {Delteil}},
  \bibinfo {author} {\bibfnamefont {S.}~\bibnamefont {Faelt}},\ and\ \bibinfo
  {author} {\bibfnamefont {A.}~\bibnamefont {Imamo{\u g}lu}},\ }\href
  {https://doi.org/10.1103/PhysRevB.93.241302} {\bibfield  {journal} {\bibinfo
  {journal} {Physical Review B}\ }\textbf {\bibinfo {volume} {93}},\ \bibinfo
  {pages} {241302} (\bibinfo {year} {2016})}\BibitemShut {NoStop}%
\bibitem [{\citenamefont {Mahan}(2000)}]{Mahan}%
  \BibitemOpen
  \bibfield  {author} {\bibinfo {author} {\bibfnamefont {G.}~\bibnamefont
  {Mahan}},\ }\href {https://books.google.co.uk/books?id=xzSgZ4-yyMEC} {\emph
  {\bibinfo {title} {Many-Particle Physics}}},\ Physics of Solids and Liquids\
  (\bibinfo  {publisher} {Springer},\ \bibinfo {year} {2000})\BibitemShut
  {NoStop}%
\bibitem [{\citenamefont {Lang}\ and\ \citenamefont
  {Firsov}(1962)}]{lang_firsov_1962}%
  \BibitemOpen
  \bibfield  {author} {\bibinfo {author} {\bibfnamefont {I.}~\bibnamefont
  {Lang}}\ and\ \bibinfo {author} {\bibfnamefont {Y.~A.}\ \bibnamefont
  {Firsov}},\ }\href@noop {} {\bibfield  {journal} {\bibinfo  {journal}
  {Journal of Experimental and Theoretical Physics}\ }\textbf {\bibinfo
  {volume} {43}},\ \bibinfo {pages} {1301} (\bibinfo {year}
  {1962})}\BibitemShut {NoStop}%
\bibitem [{\citenamefont {Nazir}\ and\ \citenamefont
  {McCutcheon}(2016)}]{Nazir2016}%
  \BibitemOpen
  \bibfield  {author} {\bibinfo {author} {\bibfnamefont {A.}~\bibnamefont
  {Nazir}}\ and\ \bibinfo {author} {\bibfnamefont {D.~P.~S.}\ \bibnamefont
  {McCutcheon}},\ }\href {http://stacks.iop.org/0953-8984/28/i=10/a=103002}
  {\bibfield  {journal} {\bibinfo  {journal} {Journal of Physics: Condensed
  Matter}\ }\textbf {\bibinfo {volume} {28}},\ \bibinfo {pages} {103002}
  (\bibinfo {year} {2016})}\BibitemShut {NoStop}%
\bibitem [{\citenamefont {Scerri}\ \emph {et~al.}(2017)\citenamefont {Scerri},
  \citenamefont {Santana}, \citenamefont {Gerardot},\ and\ \citenamefont
  {Gauger}}]{Scerri2017}%
  \BibitemOpen
  \bibfield  {author} {\bibinfo {author} {\bibfnamefont {D.}~\bibnamefont
  {Scerri}}, \bibinfo {author} {\bibfnamefont {T.~S.}\ \bibnamefont {Santana}},
  \bibinfo {author} {\bibfnamefont {B.~D.}\ \bibnamefont {Gerardot}},\ and\
  \bibinfo {author} {\bibfnamefont {E.~M.}\ \bibnamefont {Gauger}},\ }\href
  {https://doi.org/10.1103/PhysRevB.95.165403} {\bibfield  {journal} {\bibinfo
  {journal} {Phys. Rev. B}\ }\textbf {\bibinfo {volume} {95}},\ \bibinfo
  {pages} {165403} (\bibinfo {year} {2017})}\BibitemShut {NoStop}%
\bibitem [{\citenamefont {Ramsay}\ \emph {et~al.}(2010)\citenamefont {Ramsay},
  \citenamefont {Gopal}, \citenamefont {Gauger}, \citenamefont {Nazir},
  \citenamefont {Lovett}, \citenamefont {Fox},\ and\ \citenamefont
  {Skolnick}}]{ramsay_damping_2010}%
  \BibitemOpen
  \bibfield  {author} {\bibinfo {author} {\bibfnamefont {A.~J.}\ \bibnamefont
  {Ramsay}}, \bibinfo {author} {\bibfnamefont {A.~V.}\ \bibnamefont {Gopal}},
  \bibinfo {author} {\bibfnamefont {E.~M.}\ \bibnamefont {Gauger}}, \bibinfo
  {author} {\bibfnamefont {A.}~\bibnamefont {Nazir}}, \bibinfo {author}
  {\bibfnamefont {B.~W.}\ \bibnamefont {Lovett}}, \bibinfo {author}
  {\bibfnamefont {A.~M.}\ \bibnamefont {Fox}},\ and\ \bibinfo {author}
  {\bibfnamefont {M.~S.}\ \bibnamefont {Skolnick}},\ }\href
  {https://doi.org/10.1103/PhysRevLett.104.017402} {\bibfield  {journal}
  {\bibinfo  {journal} {Physical Review Letters}\ }\textbf {\bibinfo {volume}
  {104}},\ \bibinfo {pages} {017402} (\bibinfo {year} {2010})}\BibitemShut
  {NoStop}%
\bibitem [{\citenamefont {Roy-Choudhury}\ and\ \citenamefont
  {Hughes}(2015{\natexlab{a}})}]{Hughes2015a}%
  \BibitemOpen
  \bibfield  {author} {\bibinfo {author} {\bibfnamefont {K.}~\bibnamefont
  {Roy-Choudhury}}\ and\ \bibinfo {author} {\bibfnamefont {S.}~\bibnamefont
  {Hughes}},\ }\href {https://doi.org/10.1364/OL.40.001838} {\bibfield
  {journal} {\bibinfo  {journal} {Opt. Lett.}\ }\textbf {\bibinfo {volume}
  {40}},\ \bibinfo {pages} {1838} (\bibinfo {year}
  {2015}{\natexlab{a}})}\BibitemShut {NoStop}%
\bibitem [{\citenamefont {Roy-Choudhury}\ and\ \citenamefont
  {Hughes}(2015{\natexlab{b}})}]{Hughes2015b}%
  \BibitemOpen
  \bibfield  {author} {\bibinfo {author} {\bibfnamefont {K.}~\bibnamefont
  {Roy-Choudhury}}\ and\ \bibinfo {author} {\bibfnamefont {S.}~\bibnamefont
  {Hughes}},\ }\href {https://doi.org/10.1103/PhysRevB.92.205406} {\bibfield
  {journal} {\bibinfo  {journal} {Phys. Rev. B}\ }\textbf {\bibinfo {volume}
  {92}},\ \bibinfo {pages} {205406} (\bibinfo {year}
  {2015}{\natexlab{b}})}\BibitemShut {NoStop}%
\bibitem [{\citenamefont {Breuer}\ and\ \citenamefont
  {Petruccione}(2007)}]{Breuer2007}%
  \BibitemOpen
  \bibfield  {author} {\bibinfo {author} {\bibfnamefont {H.}~\bibnamefont
  {Breuer}}\ and\ \bibinfo {author} {\bibfnamefont {F.}~\bibnamefont
  {Petruccione}},\ }\href {https://books.google.co.uk/books?id=DkcJPwAACAAJ}
  {\emph {\bibinfo {title} {The Theory of Open Quantum Systems}}}\ (\bibinfo
  {publisher} {OUP Oxford},\ \bibinfo {year} {2007})\BibitemShut {NoStop}%
\bibitem [{\citenamefont {Ficek}\ \emph
  {et~al.}(2005{\natexlab{b}})\citenamefont {Ficek}, \citenamefont {Asakura},
  \citenamefont {Swain}, \citenamefont {Brenner}, \citenamefont {H{\"a}nsch},
  \citenamefont {Kamiya}, \citenamefont {Krausz}, \citenamefont {Monemar},
  \citenamefont {Rhodes}, \citenamefont {Venghaus} \emph {et~al.}}]{Ficke2005}%
  \BibitemOpen
  \bibfield  {author} {\bibinfo {author} {\bibfnamefont {Z.}~\bibnamefont
  {Ficek}}, \bibinfo {author} {\bibfnamefont {T.}~\bibnamefont {Asakura}},
  \bibinfo {author} {\bibfnamefont {S.}~\bibnamefont {Swain}}, \bibinfo
  {author} {\bibfnamefont {K.}~\bibnamefont {Brenner}}, \bibinfo {author}
  {\bibfnamefont {T.}~\bibnamefont {H{\"a}nsch}}, \bibinfo {author}
  {\bibfnamefont {T.}~\bibnamefont {Kamiya}}, \bibinfo {author} {\bibfnamefont
  {F.}~\bibnamefont {Krausz}}, \bibinfo {author} {\bibfnamefont
  {B.}~\bibnamefont {Monemar}}, \bibinfo {author} {\bibfnamefont
  {W.}~\bibnamefont {Rhodes}}, \bibinfo {author} {\bibfnamefont
  {H.}~\bibnamefont {Venghaus}}, \emph {et~al.},\ }\href
  {https://books.google.co.uk/books?id=AYhF6-K2OKcC} {\emph {\bibinfo {title}
  {Quantum Interference and Coherence: Theory and Experiments}}},\ Springer
  Series in Optical Sciences\ (\bibinfo  {publisher} {Springer},\ \bibinfo
  {year} {2005})\BibitemShut {NoStop}%
\end{thebibliography}%

\end{document}

\title{Supplementary Material: Fundamental Limits to Coherent Photon Generation with Solid-State Atomlike Transitions}% Force line breaks with \\
\author{Z. X. Koong}
\email[Correspondence: ]{zk49@hw.ac.uk}
\affiliation{
 SUPA, Institute of Photonics and Quantum Sciences, Heriot-Watt University, Edinburgh EH14 4AS, Scotland, United Kingdom
}
\author{D. Scerri}
\affiliation{
 SUPA, Institute of Photonics and Quantum Sciences, Heriot-Watt University, Edinburgh EH14 4AS, Scotland, United Kingdom
}
\author{M. Rambach}
\affiliation{
 SUPA, Institute of Photonics and Quantum Sciences, Heriot-Watt University, Edinburgh EH14 4AS, Scotland, United Kingdom
}
\author{T. S. Santana}
\affiliation{%
Departamento de F\'{i}sica, Universidade Federal de Sergipe, Sergipe, 49100-000, Brazil
}%
\author{S. I. Park}
\affiliation{
 Center for Opto-Electronic Materials and Devices Research, Korea Institute of Science and Technology, Seoul 02792, Republic of Korea
}
\author{J. D. Song}
\affiliation{
 Center for Opto-Electronic Materials and Devices Research, Korea Institute of Science and Technology, Seoul 02792, Republic of Korea
}
\author{E. M. Gauger}
\affiliation{
 SUPA, Institute of Photonics and Quantum Sciences, Heriot-Watt University, Edinburgh EH14 4AS, Scotland, United Kingdom
}
\author{B. D. Gerardot}
\email[Correspondence: ]{b.d.gerardot@hw.ac.uk}
\affiliation{
 SUPA, Institute of Photonics and Quantum Sciences, Heriot-Watt University, Edinburgh EH14 4AS, Scotland, United Kingdom
}
\date{October 16, 2019}% It is always \today, today,

\iffalse
\begin{abstract}
In this supporting document we provide the following: details on the experimental set-up, details on the determination of two-photon interference visibility, model for the resonance fluorescence Raman spectrum of the $\mathrm{X^{1-}}$ transition under finite external magnetic field and non-zero nuclear field as well as the theoretical model for the implementation of the polaron master equation for the $\mathrm{X^{1-}}$ spin-$\Lambda$ system, taking into account of the finite exciton-phonon interaction.  
\end{abstract}
\fi
\maketitle

\section{Experimental Setup and Lifetime Measurements}
\begin{figure}
\includegraphics[width=0.5\textwidth]{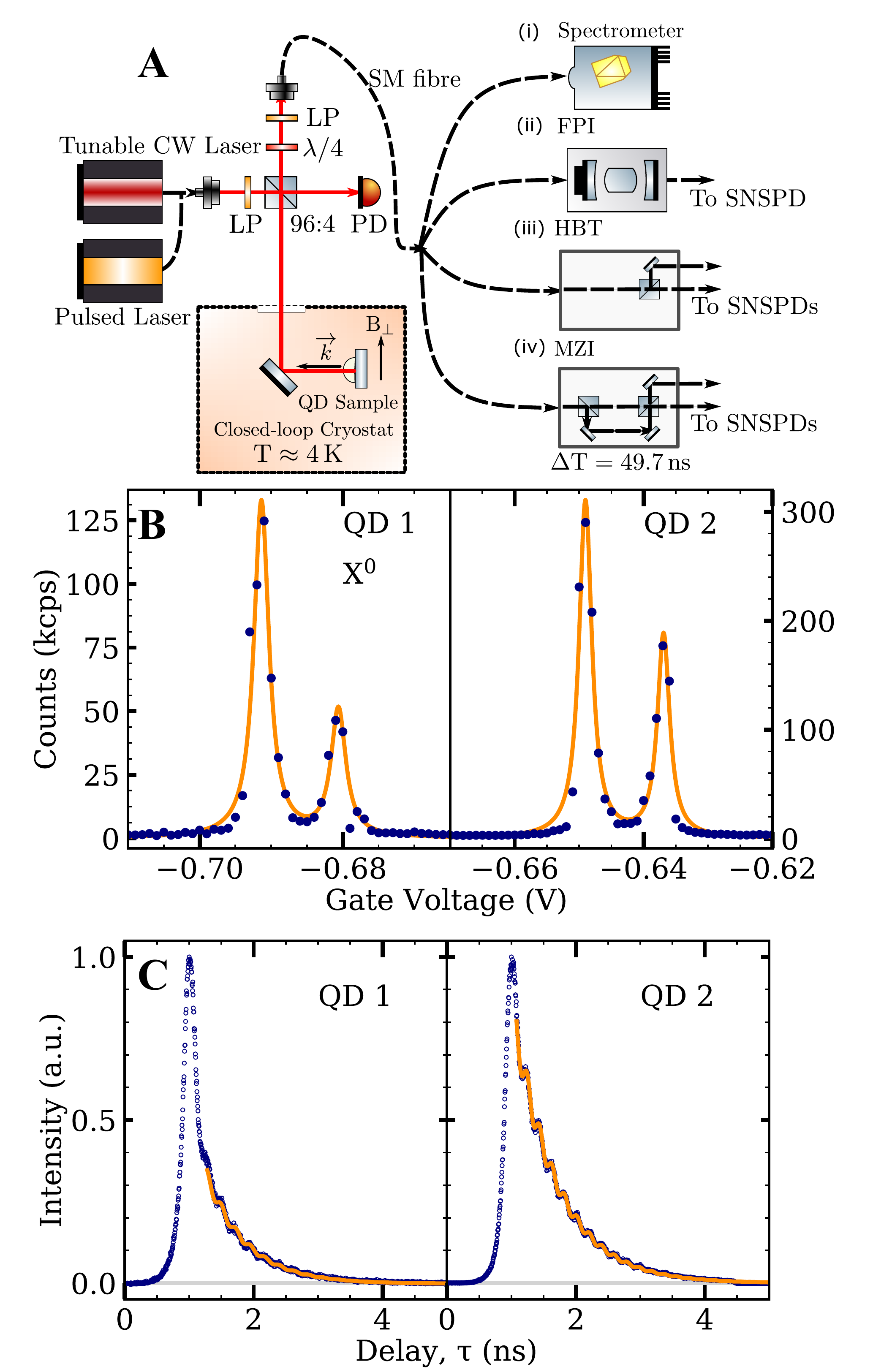}
\caption{\textbf{Measurement setup and lifetime of emitters.}
\textbf{(A)} Schematic of the experimental setup for the spectroscopy measurement on an InGaAs QD device.
%PD: photodiode, LP: linear polarizer, $\lambda/4$: quarter-wave plate, FC: fibre coupler, SM: single mode, APD : 
\textbf{(B)} Example of the RF voltage sweeps taken from QD1 and QD2 showing the neutral exciton $\mathrm{X^0}$ lines used in the main text. The excitation wavelength for QD1 and QD2 are $\lambda=963.45\,\mathrm{nm}$ and $\lambda=965.9\,\mathrm{nm}$ respectively.
\textbf{(C)} Lifetime measurement of the scattered photons from $\mathrm{ X^0}$ exhibit a single exponential decay and a quantum beating.
For QD1 (QD2), the lifetime is $T_1 = 625\,(2)\,\mathrm{ps}$ ($T_1 = 679\,(1)\,\mathrm{ps}$) and the fine structure splitting (given by the frequency of the beating) is $\Delta=18.3\,(1)\,\mu \mathrm{eV}$ ($\Delta=21.3\,(2)\,\mu \mathrm{eV}$).}

\label{fig:lifetime}
\end{figure}

Fig.~\ref{fig:lifetime}A shows the schematic of the experimental setup used to perform spectroscopy measurements on the emission from a semiconductor InGaAs Quantum dot (QD) mounted in the Voigt configuration.
A tunable continuous wave (CW) laser is used to excite the QD which is kept cold at $\mathrm{T}=4\,K$ in a closed-cycle helium flow cryostat. 
In resonance fluorescence (RF), a cross-polarization scheme using a pair of orthogonally oriented linear polarizers (LP) on the excitation and the collection arms of the confocal microscope is used to suppress the background scattering laser up to $10^7$.
A quarter-wave plate ($\lambda/4$) is used to correct any birefringence due to optics in the propagation path.
The excitation power is measured by the photodiode (PD).
The photons scattered from the QD are coupled into a single mode (SM) fibre which is directed to (i) a spectrometer, (ii) a $27.5\,\mathrm{MHz}$ resolution, $5.5\,\mathrm{GHz}$ free spectral range fiber Fabry–Pérot interferometer (FPI), (iii) a Hanbury-Brown and Twiss interferometer (HBT) or (iv) an unbalanced Mach-Zehnder interferometer (MZI) setup (with interferometric delay of $\Delta \mathrm{ T} = 49.7\,\mathrm{ns}$) to measure intensity correlations and two-photon interference.
An example of the RF spectrum as the QD is detuned across the laser resonance for the neutral exciton $\mathrm{ X^0}$ is shown in Fig.~\ref{fig:lifetime}B.
To perform a lifetime measurement on the emission from a neutral exciton, instead of a CW laser, the QD is excited with a mode-locked resonant pulse laser (with ps pulse width and $80.3\,\mathrm{ MHz}$ repetition rate).
The scattered photons are detected using superconducting nanowire single-photon detectors (SNSPD) (timing jitter $\sim 100\,\mathrm{ps}$).
The signal from the SNSPD is then sent to a time-correlated single photon counting module (PicoHarp 300) with the start channel synced to the pulsed laser.
The results from the lifetime measurement is shown in Fig.~\ref{fig:lifetime}C for both QDs studied in the paper.
The coincidence histogram shows a single exponential decay term (corresponding to the lifetime of the emitter, $T_1$) and a fast oscillation term (corresponding to the fine-structure splitting, $\Delta$ of the neutral exciton).
A simple exponential decay with a sinusoidal function gives $T_1=625\,(2)\,\mathrm{ps}$, $\Delta=18.3\,(1)\,\mu \mathrm{ eV}$ and $T_1=679\,(1)\,\mathrm{ps}$, $\Delta=21.3\,(2)\,\mu \mathrm{eV}$ for QD1 and QD2, respectively.

\begin{figure}
\includegraphics[width=0.35\textwidth]{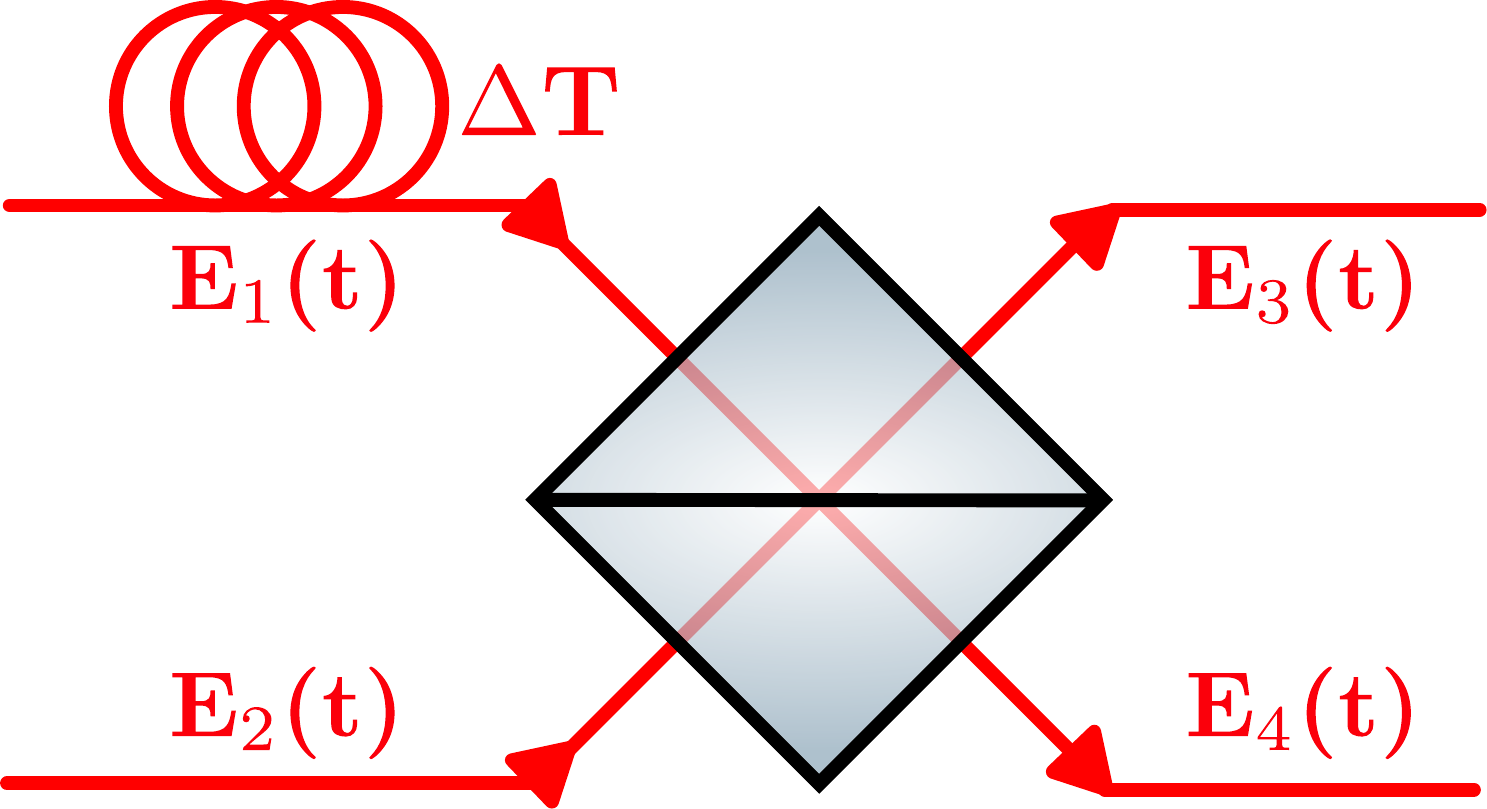}
\caption{Input ($E_{1,2}(t)$) and output ($E_{3,4}(t)$) field modes at the second beam splitter of the unbalanced Mach-Zehnder interferometer, with a temporal delay $\Delta T$ in the upper arm.}
\label{fig:BS}
\end{figure}

\section{Post-Select Two-Photon Interference}
\begin{figure*}
\includegraphics[width=0.8\textwidth]{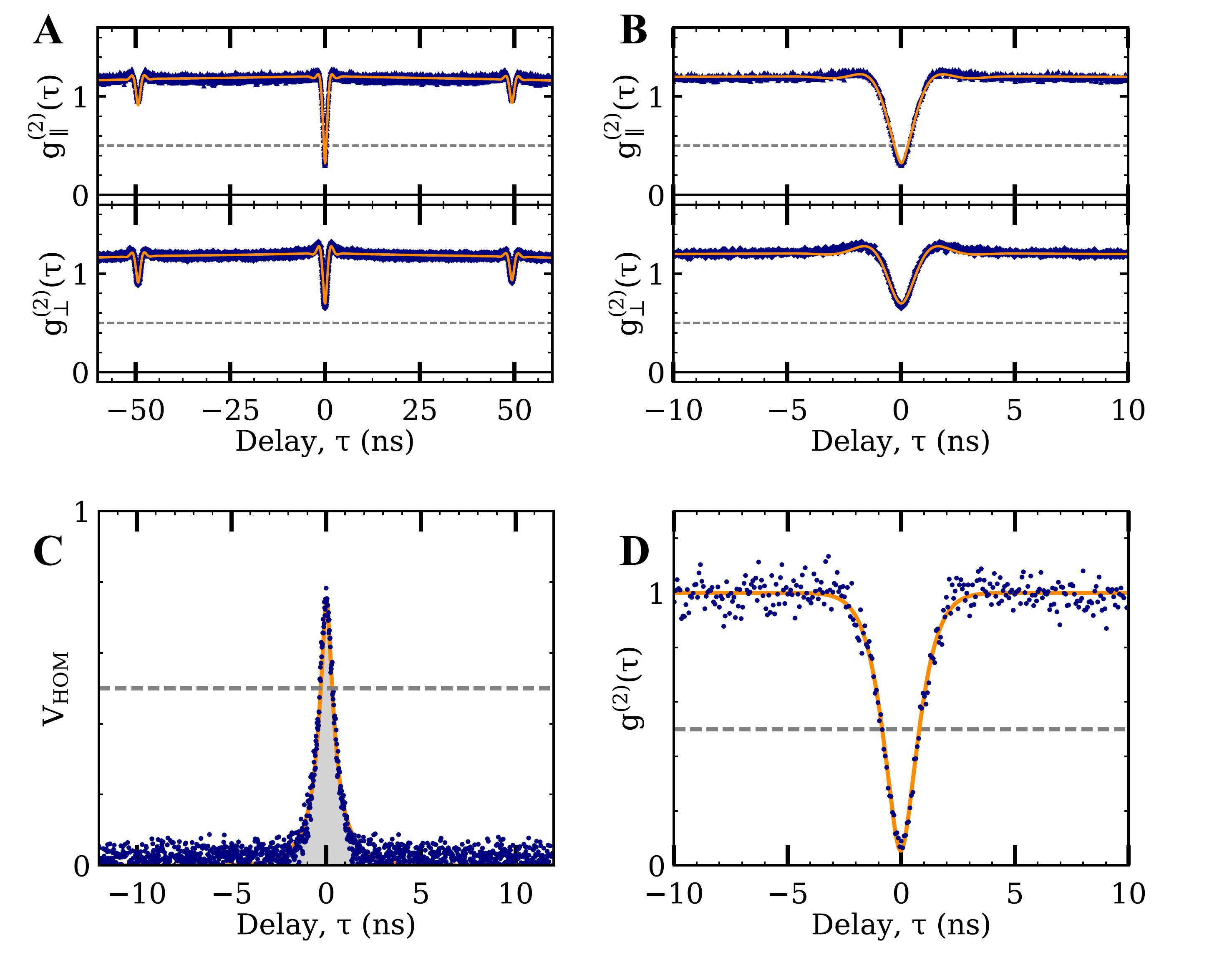}
\caption{\textbf{Autocorrelation measurement and Two-photon Interference}.
\textbf{(A, B)} Correlation measurement for the two-photon interference of indistinguishable $g^{(2)}_\parallel(\tau)$ and distinguishable photons $g^{(2)}_\perp(\tau)$  at saturation, $\Omega\approx\, \Omega_\mathrm{sat}$ with their respective fit functions (orange) within a correlation window of $\pm 100\,\mathrm{ns}$ (A) and $\pm 10\,\mathrm{ns}$ (B).
\textbf{(C)} Visibility of the resonantly scattered photons from QD1 extracted using Eq.~\ref{eqn:vhom} at $\Omega\approx\,\Omega_\mathrm{sat}$.
The shaded region in $V_\mathrm{HOM}(\tau)$ gives the coalescence time window (CTW) of $1.24\,(10)\,\mathrm{ns}$.
%\textbf{(D)} Autocorrelation measurement, $g^{(2)}(\tau)$  at $ \Omega\approx \, \Omega_\mathrm{sat}$ on the resonantly scattered photons exhibits a suppressed multi-photon emission probability of $g^{(2)}(0)=0.157\,(5)$.
\textbf{(D)} Autocorrelation measurement, $g^{(2)}(\tau)$  at $ \Omega\approx 0.1\, \Omega_\mathrm{sat}$ on the resonantly scattered photons exhibits a suppressed multi-photon emission probability of $g^{(2)}(0)=0.046\,(13)$.
}
\label{fig:tpi}
\end{figure*}
 
We perform Hong-Ou-Mandel type two-photon interference measurements~\cite{hong_measurement_1987} by sending the resonantly scattered photons to an unbalanced Mach-Zehnder interferometer (as shown in Fig.~\ref{fig:lifetime}A(iv)).
Here, the photons are split equally (transmission, $T\sim 53\,\%$) along two paths with one of the paths being delayed by $\Delta T = 49.7\,\mathrm{ns}$ before they interfere on a fibre beam splitter ($T\sim 51\,\%$), shown in Fig.~\ref{fig:BS}.
Here, we perform two sets of measurements: $g^{(2)}_\perp(\tau)$ and $g^{(2)}_\parallel(\tau)$.
% by applying a phase shift of $\pi/2$ in the polarization of one of the arms of the interferometer.
The visibility is obtained as follows:
\begin{equation}
V_\mathrm{HOM}(\tau)=(g^{(2)}_\perp(\tau)-g^{(2)}_\parallel(\tau))/g^{(2)}_\perp(\tau).
\label{eqn:vhom}
\end{equation}

Fig.~\ref{fig:tpi}(A-C) show the results of the two-photon interference measurement from a neutral exciton, $\mathrm{X^0}$ from QD1.
The bias applied to the QD device is selected such that only one of the fine structure of the neutral exciton is addressed.
The solid lines on $g^{(2)}_\parallel (\tau)$ and $g^{(2)}_\perp (\tau)$ are produced using the equations from Ref.~\cite{proux_measuring_2015}, convolved with the setup instrument response function (IRF), which is determined by a Gaussian peak with a full-width at half maximum (FWHM) of $160\,\mathrm{ps}$. 
We take into account the timescale for the additional bunching (caused by fluctuations in the electronic environment) in $g^{(2)}_{\perp,\parallel}(\tau)$ by modifying the autocorrelation function $g^{(2)}(\tau)$ shown in Fig.~\ref{fig:tpi}D. 
By introducing this additional bunching timescale $\tau_R$, the modified autocorrelation function is then given by
\begin{equation}
g^{(2)}(\tau) = g^{(2)}_{RF}(\tau)\times (1+ b\,e^{-|\tau|/\tau_R}),
\label{eqn:g2}
\end{equation}
where $b$ is the amplitude of the bunching and $g^{(2)}_{RF}(\tau)$ is the autocorrelation function for resonant excitation of a two-level system~\cite{scully_quantum_1997}.
Fitting Eq.~\eqref{eqn:g2} to the experimental data, we obtain $b = 0.245\,(1) $ and $\tau_R = 137\,(1)\,\mathrm{ns}$.

\iffalse
The visibility, $V_\mathrm{HOM}(\tau)$ is extracted from both $g^{(2)}_\parallel (\tau)$ and $g^{(2)}_\perp (\tau)$ using Eq.~\ref{eqn:vhom}. 
The unfiltered scattered photons from a neutral exciton at saturation, $\Omega\approx\,\Omega_\mathrm{sat}$ show a visibility, $V_\mathrm{HOM}(0)$ of $0.803\,(6)$ (deconvolved).
%This value is consistent with the maximum achievable visibility due to the finite exciton-phonon coupling, which gives an upper bound proportional to the square of the branching ratio, $\alpha_\mathrm{DW}$, i.e. ${ V_\mathrm{HOM}}\leq \alpha_\mathrm{DW}^2\approx 0.83$, given that $\alpha_\mathrm{DW}\approx 0.91$ for the particular QD.
The coalescence time window (CTW)~\cite{proux_measuring_2015,baudin_correlation_2019} obtained from integrating the visibility within temporal delay of $-10$ to $10\,\mathrm{ns}$, i.e. $\mathrm{CTW} = \int d\tau\, V_\mathrm{HOM}(\tau)$ gives $1.24\,(10)\,\mathrm{ns}$ as expected from the theoretical model (see Fig.~3(A) in the main text).
\fi

\section{Coalescence Time Window}
{
The coalescence time window (CTW) is defined as 
\begin{equation}\label{CTW_eqn}
\mathrm{CTW} = \int \mathrm{d}\tau\, V_\mathrm{HOM}(\tau),
\end{equation}
which gives the integrated area of the HOM visibility $V_\mathrm{HOM}(\tau)$. 
The CTW was first introduced in Ref.~\citen{proux_measuring_2015} as a figure of merit to quantify photon indistinguishability.
The motivation to use CTW rather than the two photon visibility at zero time delay is that, for continuous wave excitation, the latter is solely determined by the detector response time and is not indicative of indistinguishability of the photon wavepackets~\cite{legero_quantum_2004}.
%Any imperfection to $V_\mathrm{HOM}(\tau=0)$ is due to experimental imperfection $V_0$ (e.g. presence of background scattered laser, spatial overlap on the beam splitter, imperfection in distinction in polarization etc.).
The advantage of using the CTW to characterise the indistinguishability is that it is not affected by the timing jitter of the detectors: averaging over temporal delay, it gives the characteristic time beyond which no two photon interference can be observed. Indeed, one can see how Eq.~\eqref{CTW_eqn} does not depend on detector jitter by noting that the latter can be accurately described by a convolution of the signal $V_\mathrm{HOM}(\tau)$ with a normal distribution having width proportional to the detector resolution:

\begin{align}\label{CTWtheory}
\begin{split}
CTW &= \int^\infty_{-\infty} \mathrm{d}\tau~ V_\mathrm{HOM}(\tau)\star\phi_{\tau_d}(\tau) \\
&= \int^\infty_{-\infty} \mathrm{d}\tau~ V_\mathrm{HOM}(\tau)\int^\infty_{-\infty} \mathrm{d}\tau~\phi_{\tau_d}(\tau) \\
&= \int^\infty_{-\infty} \mathrm{d}\tau~ V_\mathrm{HOM}(\tau)~,
\end{split}
\end{align}
where $\tau_d$ is the detector resolution, and $\phi_{\tau_d}(\tau)$ is the normal distribution with FWHM $\tau_d$ centred at $\tau=0$. 
The CTW remains sensitive to experimental imperfections other than the detector jitter: for example, it will still be affected by the presence of background scattered light, spatial overlap on the beam splitter and imperfection in polarization~\cite{baudin_correlation_2019,giesz_cavity_2015}. 
%These imperfection can be accounted for, for instance using the approach found in Supplementary Information in Ref.~\citen{giesz_cavity_2015}.
%Whilst the CTW is still degraded by the experimental imperfections (e.g. presence of background scattered laser, spatial overlap on the beam splitter, imperfection in polarization etc.)~\cite{baudin_correlation_2019} by decreasing the HOM visibility $\mathrm{V_{HOM}}$, these imperfections can be accounted for, for instance using the approach found in Supplementary information in Ref.~\citen{giesz_cavity_2015}.

Having motivated the use of the CTW over the visibility at zero delay, we now briefly outline the derivation of our theoretical model, as well as an approximation leading to the simplified visibility model given in Ref.~\cite{iles-smith_limits_2017}. We denote the operators of the field inputs prior to the second beam splitter by $E_1 (t)$ and $E_2 (t)$, respectively (c.f. Fig.~\ref{fig:BS}). 
Since for an unbalanced  Mach-Zehnder setup the two input operators differ only by the temporal delay in one of the arms, we may write these as $E_1(t) = E(t+\Delta T)$ and $E_2(t) = E(t)$. Furthermore, we can relate the input field operators to the QD emission by using $E(t) = E_0 \sigma_-(t)$, where $E_0$ is the vacuum field, and $\sigma_-(t)$ is the Heisenberg picture dipole lowering operator~\cite{Ficek2005}. Thus, we may write the output field modes after the secondary beam splitter as

\begin{align}\label{output_ops}
\begin{split}
E_3(t) &= \frac{1}{\sqrt{2}} (E_1(t)+E_2(t)) = \frac{1}{\sqrt{2}} (\sigma_-(t+\Delta T)+\sigma_-(t))~, \\
E_4(t) &= \frac{1}{\sqrt{2}} (E_1(t)-E_2(t)) = \frac{1}{\sqrt{2}} (\sigma_-(t+\Delta T)-\sigma_-(t))~.
\end{split}
\end{align}

We are now in a position to fully expand the second order correlation function for the output field modes $E_3(\tau)$ and $E_4(\tau)$, given by

\begin{equation}
g^{(2)}(\tau) = \frac{\Braket{E^*_3(t) E^*_4(t+\tau) E_4(t+\tau) E_3(t)}}{\Braket{E^*_3(t+\tau) E_3(t+\tau) }\Braket{E^*_4(t+\tau) E_4(t+\tau) }}~.
\end{equation}

Expanding $g^{(2)}(\tau)$, we obtain 16 terms involving the dipole operators. Whilst we refrain from writing the full expression here, we note that each term has one of the three forms:

\noindent
\textbf{(1)} $\Braket{\sigma_+(t) \sigma_+(t+\tau+k\Delta T) \sigma_-(t+\tau+k\Delta T) \sigma_-(t)},~k\in\{ 0,\pm1 \}$: which is essentially the standard second order correlation function evaluated at different delays.

\noindent
\textbf{(2)} Correlation operators involving an odd number of dipole operator contributions from the two interferometer arms (such as $\Braket{\sigma_+(t+\Delta T) \sigma_+ (t+\tau) \sigma_- (t+\tau) \sigma_- (t)}$). Whilst these terms vanish for low laser coherence $T_L$ compared to $\Delta T$ \cite{lebreton2013, baudin_correlation_2019}, we are in the opposite regime, where $T_L \gg \Delta T$, and thus we keep these terms in our calculations.

\noindent
\textbf{(3)} $\Braket{\sigma_+(t+\Delta T) \sigma_+ (t+\tau) \sigma_- (t+\tau+\Delta T) \sigma_- (t)}$ and its complex conjugate. If we treat the output from both arms as independent, we can write this term as 

\begin{align}
\begin{split}
&\Braket{\sigma_+(t+\Delta T) \sigma_+ (t+\tau) \sigma_- (t+\tau+\Delta T) \sigma_- (t)} \\
&\approx \Braket{\sigma_+(t+\Delta T) \sigma_- (t+\tau+\Delta T)} \Braket{\sigma_+ (t+\tau) \sigma_- (t)} \\
&= |g^{(1)}(\tau)|^2~.
\end{split}
\end{align}
We find that this approximation is valid for a delay $\Delta T = 49.7$~ns.
In the absence of the vibrational environment, it has been shown that the CTW increases beyond $2\,T_1$ as $\Omega\rightarrow 0 $ due to the increasing elastic fraction of scattered photons~\cite{proux_measuring_2015,baudin_correlation_2019}.

}

\section{Spin-$\Lambda$ System Coupled to Nuclear Spin Bath}\label{sec:raman_coupling}
\begin{figure*}
\includegraphics[width=0.8\textwidth]{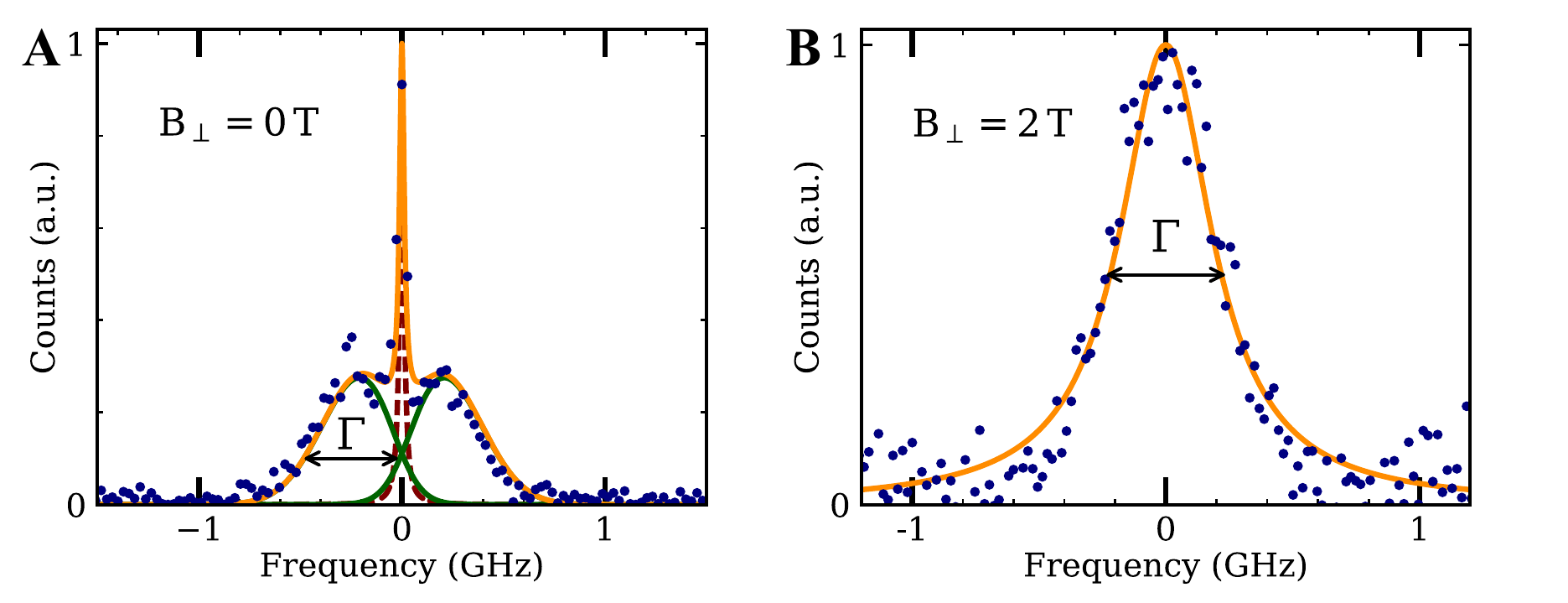}
\caption{\textbf{Resonance fluorescence spectra from the same negatively-charged exciton, $\mathrm{X^{1-}}$ transition at below saturation, $\Omega\approx 0.1\,\Omega_{sat}$.}
\textbf{(A)}~At $B_\perp = 0\,\mathrm{T}$, the fluorescence spectra of the photons scattered from the negatively-charged exciton ($\mathrm{X^{1-}}$) shows an elastic peak (red dashed line) and two displaced inelastic Raman peaks (green solid line). 
The origin of the splitting is due to the coupling of the electron spin to the nuclear spin bath, whereby it induces a Zeeman splitting proportional to the mean nuclear spin field, $\left< B_N\right>$. 
The linewidth of the Raman peaks gives $\Gamma=421\,(11)\,\mathrm{MHz}$ and is proportional to the nuclear spin fluctuation, $\delta B$.
\textbf{(B)} At $B_\perp = 2\,\mathrm{T}$, the emission spectra from the Raman transition shows a single Gaussian line-shape.
The linewidth $\Gamma$, extracted from the fit (orange) is ${ \Gamma} = 439\,(20)\,\mathrm{MHz}$, which shows agreement with the linewidth obtained in the zero field.
}
\label{fig:nuc_spin}
\end{figure*}

Fig.~\ref{fig:nuc_spin} shows the emission spectra from a resonantly driven negatively charged exciton, $\mathrm{X^{1-}} $ at zero field (A) and Voigt field (4 T) (B) in the weak excitation regime, $ \Omega\approx 0.1\, \Omega_\mathrm{sat}$.
At zero field, the emission spectra of $\mathrm{X^{1-}} $ consists of an elastic peak (given by the red dashed line) and two Raman peaks (given by the green dashed-dotted line).
The origin of the presence of the Raman peaks is due to the non-zero coupling of the ground state electron spins to the nuclear spin bath~\cite{urbaszek_nuclear_2013,malein_screening_2016}.
This gives rise to non-zero nuclear-spin induced magnetic field (Overhauser field), $\left< B_N\right>\approx 346\,(11)\,\mathrm{MHz}\approx 25\,\mathrm{mT}$, which splits the ground state of $\mathrm{X^{1-}} $.
The broken symmetry forms a four-level double-$\Lambda$ system (see Fig.~1(D) in main text).
This allows the previously forbidden diagonal transition (in the spin-z ($\sigma_\mathrm{z}$) basis) which gives the Raman side peaks~\cite{malein_screening_2016}.
The ground state spin flip rate is proportional to the Overhauser field fluctuations, $\delta B_N = 188\,(5)\,\mathrm{MHz}$.
In the absence of the finite exciton-phonon coupling, since the coherence of the Raman photons is determined by the ground state dephasing rate~\cite{fernandez_optically_2009,he_indistinguishable_2013,sun_measurement_2016}, the linewidth of the Raman side peaks is given by the spin flip rate, which is $2\sqrt{2 \log 2}\, \delta B_{N}=421\,(11)\,\mathrm{MHz}$.
This result matches the linewidth of the scattered Raman photons at $B_\mathrm{ext} = B_\perp = 2\,\mathrm{T}$,  $ \Gamma = 439\,(20)\,\mathrm{MHz}$, within one standard deviation.

Alternatively, one could solve the rate equations of a three level spin-$\Lambda$ system analytically for a strong external magnetic field, $B_\mathrm{ext} \gg \delta B_N$ such that the Overhauser field fluctuations are partially suppressed~\cite{malein_screening_2016}.
In this regime, the component of $\delta B_N$ perpendicular to the applied field, $B_\mathrm{ext}$ can be ignored to a good approximation. 
%In the presence of $B_\mathrm{ext} \gg \delta B_N$ and 
Assuming both ground states are equally populated initially, the resulting emission spectrum, $S(\omega)$, contains two delta functions, one being the elastic peak with the same frequency as the laser, $\omega$, and the other being a shifted inelastic peak with frequency given by $\omega-B$,
\begin{equation}
S(\omega)=\frac{\Omega^{2}}{2 \gamma}[\delta(\omega)+\delta(\omega-B)],
\end{equation}
where $B = B_\mathrm{ext} + B_N$, $\Omega$ is the driving Rabi frequency and $\gamma$ is the spontaneous emission rate.
Averaging the emission spectrum $S(\omega)$ over the Overhauser field fluctuations, the resulting spectrum, $\langle S(\omega)\rangle$ is
\begin{equation}
\label{eqn:spectral}
\langle S(\omega)\rangle=\frac{\Omega^{2}}{2 \gamma}\left[\delta(\omega)+\frac{e^{\left(\omega-B_\mathrm{ext}\right)^{2} / 2 \delta B_{N}^{2}}}{\delta B_{N} \sqrt{2 \pi}}\right],
\end{equation}
where the first term represents the elastically scattered Rayleigh photons, given by the delta function and the second term describes the inelastically scattered Raman photons which is given by a Gaussian with a full-width at half maximum (FWHM) of $2\sqrt{2 \log 2}\, \delta B_{N} = 439\,(20)\,\mathrm{MHz}$.
In the experiment, since a cross polarization technique is used to collect only the orthogonally polarized Raman photons, we could neglect the first term in Eq.~\ref{eqn:spectral}.
Hence we are left with the Gaussian term which describes the spectrum of the scattered Raman photons.
Using the linewidth of the Gaussian fit on the spectrum in Fig.~\ref{fig:nuc_spin}B, we get $\delta B_N = 186 \,(8)\,\mathrm{MHz}$ which shows agreement with the results obtained from fitting the zero field spectra with the model described in Ref.~\cite{malein_screening_2016}. 

Disregarding phonon interactions, this agrees with the previous predictions that the coherence of the Raman zero phonon line is given by the ground (spin) state dephasing rate, proportional to the Overhauser field fluctuations (Refs.~\cite{fernandez_optically_2009,malein_screening_2016,sun_measurement_2016}).
However, the presence of the phonon sideband that accompanied by the Raman zero phonon line (as shown in Fig.~4A) modifies this picture and renders the scattered Raman photons partially incoherent.
%Our results confirm the findings shown in previous literature , that the coherence of the Raman zero phonon line is given by the ground (spin) state dephasing rate, proportional to the Overhauser field fluctuations.
%This indicates that the Raman photons can never be perfectly coherent, as opposed to the claim in Ref.~\cite{sun_measurement_2016}.

\section{X$^{1-}$ spin-$\Lambda$ system dynamics}

In this section we present the model used for the theoretical spectra using the polaron frame master equation formalism. We focus on the $\mathrm{X^{1-}}$ $\Lambda$ system formed by driving the spin-conserving $\Ket{\ua} \rightarrow \Ket{T_\Ua}$ transition and collect emission from the spin-flipping $\Ket{T_\Ua} \rightarrow \Ket{\da}$ transition. The full Hamiltonian for the four level system discussed in the main text can be written as:

\begin{align}
\begin{split}
H_{4LS} =& H_0 + H_\Omega\\
H_0 =& \frac{\delta_e}{2} \left(\Ket{\da} \Bra{\da} - \Ket{\ua} \Bra{\ua} \right) +  \frac{\delta_h}{2} \left(\Ket{T_\Ua} \Bra{T_\Ua} - \Ket{T_\Da} \Bra{T_\Da} \right)  \\
&-\frac{\omega_0}{2} \left(\Ket{\ua} \Bra{\ua} + \Ket{\ua} \Bra{\ua} \right) \\
&+\frac{\omega_0}{2} \left(\Ket{T_\Ua} \Bra{T_\Ua} + \Ket{T_\Da} \Bra{T_\Da} \right) \\
H_\Omega =& ~\Omega \cos(\omega_L t) \left( \Ket{T_\Ua} \Bra{\ua}  +   \Ket{\ua} \Bra{T_\Ua}  \right) ~,
\end{split}
\end{align}
where $\Ket{\ua, \da}$ and $\Ket{T_{\Ua,\Da}} = \Ket{\ua \da \Ua, \ua \da \Da}$ are the negatively charged ground and trion states, respectively. $\omega_0$ and $\omega_L$ are the exciton transition and laser frequencies, respectively. 

Starting with the system initialized in the $\Ket{\ua}$ ground state, we drive using vertically polarized light (driving the $\Ket{\ua} \rightarrow \Ket{T_\Ua}$ transition), and collect cross polarized light ($\Ket{T_\Ua} \rightarrow \Ket{\da}$). After moving to the rotating frame with respect to the driven transition, and subtracting a term proportional the identity, we focus on the spin-$\Lambda$ system formed by the states $\{ \Ket{\ua},\Ket{\da},\Ket{T_\Ua}\}$ in order to obtain the Hamiltonian

\begin{align}
\begin{split}
H =& \frac{\delta}{2} \Ket{\da} \Bra{\da} + \frac{\Omega}{2} \left( \Ket{T_\Ua} \Bra{\ua}  + \Ket{\ua} \Bra{T_\Ua} \right)~,\\
\end{split}
\end{align}
where $\delta = 2 \delta_e$.

\subsection{Exciton--photon interaction}

The photonic environment can be modelled by the Hamiltonian
\begin{equation}
H^{pt}_E = \sum_{\mathbf{q}, \, \lambda} \nu_\mathbf{q} a^\dagger_{\mathbf{q}\lambda} a_{\mathbf{q}\lambda}~, 
\end{equation}
where $a^\dagger_{\mathbf{q}\lambda}$ ($a_{\mathbf{q}\lambda}$) is the creation (annihilation) operator for a photon of momentum $\mathbf{q}$ and polarization $\lambda$. In the dipole approximation, the photon interaction Hamiltonian is of the form 
\begin{align}\label{eq:hpt0}
\begin{split}
H^{pt}_I =&  -\mathbf{d}_{\Ua \da} \cdot \mathbf{E}(\mathbf{r}_d) (\Ket{\da} \Bra{T_\Ua} + \Ket{T_\Ua} \Bra{\da}) \\
 & -\mathbf{d}_{\Ua \ua} \cdot \mathbf{E}(\mathbf{r}_d) (\Ket{\ua} \Bra{T_\Ua} + \Ket{T_\Ua} \Bra{\ua})~,
\end{split}
\end{align}
with $\mathbf{E}(\mathbf{r}_d)$ being the Schr{\"o}dinger picture electric field in free space at the location $\mathbf{r}_d$ of the dipole, and $\mathbf{d}_{\Ua \da} = \Braket{\da \vert \mathbf{D} \vert T_\Ua}$ and $\mathbf{d}_{\Ua \ua} = \Braket{\ua \vert \mathbf{D} \vert T_\Ua}$ are the transition matrix element of the dipole operator for the $\Ket{T_\Ua} \rightarrow \Ket{\da}$ and $\Ket{T_\Ua} \rightarrow \Ket{\ua}$ transitions respectively.
%
%\begin{equation}\label{electricfield}
%\mathbf{E}(\mathbf{r}) = i \sum_{\mathbf{q}, \lambda} \left[ \mathbf{u}_{\mathbf{q} \lambda}(\mathbf{r}) a_{\mathbf{q} \lambda} - \mathrm{H.c.} \right] ~.
%\end{equation}
%%
%The spatial mode functions $\mathbf{u}_{\mathbf{q} \lambda}(\mathbf{r})$ for an ideal half-sided cavity (of perfect reflectivity) are given by
%
%\begin{equation}\label{cavity_spatial_fns}
%\mathbf{u}_{\mathbf{q} \lambda}(\mathbf{r}) = \sqrt{\frac{\omega_{\mathbf{q} \lambda}}{2 \epsilon V}}\left( \mathbf{e}_{\mathbf{q}_- \lambda} \mathrm{e}^{i \mathbf{q}_- r} - \mathbf{e}_{\mathbf{q}_+ \lambda} \mathrm{e}^{i \mathbf{q}_+ r} \right)~.
%\end{equation}

\subsection{Exciton--phonon interaction}

Unlike for atomic systems, the vibrational environment plays a key role in the dynamics of a confined electron in a QD. As an electron is excited from the valence to the conduction band, the charge configuration of the semiconductor is modified accordingly. This results in a shift of the lattice ions' equilibrium positions, giving rise to an exciton--phonon coupling depending on the excitons's state. The phonon bath can be described via the Hamiltonian

\begin{equation}
H^{pn}_E = \sum_\mathbf{k} \omega_\mathbf{k} b^\dagger_\mathbf{k} b_\mathbf{k}~,
\end{equation}
where $b^\dagger_\mathbf{k}$ ($b_\mathbf{k}$) is the creation (annihilation) operator for a phonon of momentum $\mathbf{k}$. The corresponding  Hamiltonian governing the exciton--phonon interaction dynamics is then given by

\begin{equation}
H^{pn}_I = \Ket{T_\Ua} \Bra{T_\Ua}\sum_{\mathbf{k}} g_\mathbf{k} ( b^\dagger_\mathbf{k} + b_\mathbf{k} ) ~,
\end{equation}
where $g_\mathbf{k}$ is the coupling strength of the excited electronic configuration with phonon mode $\mathbf{k}$, given by \cite{Mahan}

\begin{equation}
g_\mathbf{k} = \left( \frac{\hbar}{2\rho \nu \omega_\mathbf{k}} \right)^{1/2} \left[ \tilde{M}_X (\mathbf{k}) \mathrm{e}^{-d^2_X |\mathbf{k}|^2 / 4}  - \tilde{M}_0 (\mathbf{k})  \mathrm{e}^{-d^2_0 |\mathbf{k}|^2 / 4} \right]~,
\end{equation}
where $\rho$ is the mass density of the solid, $\nu$ is the lattice volume, $\tilde{M}_X(\mathbf{k})$ and $\tilde{M}_0(\mathbf{k})$ are the long-wavelength phonon coupling matrix elements for the excited and ground state, respectively, and $d_X$ and $d_0$ characterize the size of the electron and hole wavefunctions, respectively.

\subsection{Master equation dynamics: strong vibrational coupling}

To summarize, our system consists of the following Hamiltonian:

\begin{align}
\begin{split}
H_{tot} =& H + H^{pt}_E + H^{pn}_E + H^{pt}_I + H^{pn}_I~; \\
\vspace{5mm}
H =& \delta \Ket{\da} \Bra{\da} + \frac{\Omega}{2} \left( \Ket{T_\Ua} \Bra{\ua}  + \Ket{\ua} \Bra{T_\Ua} \right) ~,\\
H^{pt}_E =& \sum_{\mathbf{q}, \, \lambda} \nu_\mathbf{q} a^\dagger_{\mathbf{q}\lambda} a_{\mathbf{q}\lambda}~,\\
H^{pn}_E =& \sum_\mathbf{k} \omega_\mathbf{k} b^\dagger_\mathbf{k} b_\mathbf{k}~,\\
H^{pt}_I =&  -\mathbf{d}_{\Ua \da} \cdot \mathbf{E}(\mathbf{r}_d) (\Ket{\da} \Bra{T_\Ua} + \Ket{T_\Ua} \Bra{\da}) \\
 & -\mathbf{d}_{\Ua \ua} \cdot \mathbf{E}(\mathbf{r}_d) (\Ket{\ua} \Bra{T_\Ua} + \Ket{T_\Ua} \Bra{\ua}) ~,\\
H^{pn}_I =& \Ket{T_\Ua} \Bra{T_\Ua}\sum_{\mathbf{k}} g_\mathbf{k} ( b^\dagger_\mathbf{k} + b_\mathbf{k} ) ~.
\end{split}
\end{align}

Having described full exciton, vibrational and photonic environment systems, we are now in a position to briefly go through the details of the master equation describing the excitonic dynamics. For a solid state system, we expect a significantly strong coupling to the vibrational environment. Hence, the polaron, or Lang-Firsov \cite{lang_firsov_1962}, frame adequately captures the dynamics of our system \cite{Nazir2016,Scerri2017}. The two transitions $\Ket{T_\Ua} \rightarrow \Ket{\da}$ and $\Ket{T_\Ua} \rightarrow \Ket{\ua}$ couple to the same vibrational environment, and hence the polaron transformation can be written as $U_p = \mathrm{e}^{S_{\Ua}}$, where

\begin{equation}
S_{\Ua} = \Ket{T_\Ua}\Bra{T_\Ua} \sum_\mathbf{k} \frac{g_\mathbf{k}}{\omega_\mathbf{k}} (b^\dagger_\mathbf{k}-b_\mathbf{k})~.
\end{equation}
It can be easily shown that $U_p$ can be simplified to
\begin{align}
\begin{split}
U_P &= \Ket{\ua}\Bra{\ua} + \Ket{\da}\Bra{\da} + \Ket{T_\Ua}\Bra{T_\Ua} B_+~; \\
B_+ &= \prod_k \mathrm{e}^{\frac{g_\mathbf{k} }{ \omega_\mathbf{k}} (b^\dagger_\mathbf{k} - b_\mathbf{k})}~,
\end{split}
\end{align}
and that, in the absence of driving, diagonalizes the phonon interaction $H^{pn}_{I}$. Using this transformation, we obtain the Hamiltonian in the polaron frame, indexed by the subscript $P$:

\begin{align}
\begin{split}
H_{tot \hspace{1mm} P} =& H_P + H^{pt}_E + H^{pn}_{EP} + H^{pt}_{IP} + H^{pn}_{IP}~; \\
\vspace{5mm}
H_P =& \delta_P \Ket{\da} \Bra{\da} + \frac{\Omega_P}{2} \left( \Ket{T_\Ua} \Bra{\ua}  + \Ket{\ua} \Bra{T_\Ua} \right)~,\\
H^{pt}_{EP} =& \sum_{\mathbf{q}, \, \lambda} \nu_\mathbf{q} a^\dagger_{\mathbf{q}\lambda} a_{\mathbf{q}\lambda}~,\\
H^{pn}_{EP} =& \sum_\mathbf{k} \omega_\mathbf{k} b^\dagger_\mathbf{k} b_\mathbf{k}~,\\
H^{pt}_{IP} =&  -\mathbf{d}_{\Ua \da} \cdot \mathbf{E}(\mathbf{r}_d) (B_-\Ket{\da} \Bra{T_\Ua} + B_+ \Ket{T_\Ua} \Bra{\da}) \\
 & -\mathbf{d}_{\Ua \ua} \cdot \mathbf{E}(\mathbf{r}_d) ( B_- \Ket{\ua} \Bra{T_\Ua} + B_+ \Ket{T_\Ua} \Bra{\ua}) ~,\\
H^{pn}_{IP} =& \frac{\Omega}{2}\Big[\Ket{\ua}\Bra{T_\Ua} (B_- - \Braket{B}) + \Ket{T_\Ua}\Bra{\ua} (B_+ - \Braket{B}) \Big] ~.
\end{split}
\end{align}
where $B_- = B^\dagger_+$, $\Braket{B} = \Braket{B_+} = \Braket{B_-}$, $\Omega_P = \Braket{B} \Omega$, and $\delta_{P} = \delta - \sum_\mathbf{k} g^2_\mathbf{k} / \omega_\mathbf{k} \rightarrow \delta - \int^\infty_0 \mathrm{d}\omega J_{pn}(\omega)/\omega$, with $J_{pn}(\omega)$ being the phonon spectral density, i.e. $J_{pn}(\omega) = \alpha \omega^3 \mathrm{e}^{-\frac{\omega^2}{\omega^2_c}}$, where $\alpha$ is a measure of the coupling strength and $\omega_c$ is the phonon frequency cut-off, which depends on the size and confinement of the quantum dot.  The parameters were chosen to be $\alpha = 0.03\, \mathrm{ps}^2$ and $\omega_c = 2.2\, \mathrm{ps}^{-1}$, which agree with the standard values for self-assembled GaAs quantum dots \cite{ramsay_damping_2010}. The new vibrational interaction Hamiltonian $H^{pn}_{IP}$ appears in the total Hamiltonian due to the driving term. The details of the derivation can be found in Refs.~\citen{Nazir2016, Scerri2017, Hughes2015a, Hughes2015b}, the main result being that the new interaction can now be treated perturbatively in the master equation (ME) derivation.

In order to derive the corresponding emission rates, it would be beneficial to write the two interaction Hamiltonians in a more compact form. Thus, with the definitions $A_{1,\ua}^{pt} = \Ket{\ua} \Bra{T_\Ua}$ $\left(A_{1,\da}^{pt} = \Ket{\da} \Bra{T_\Ua}\right)$, $A_{2,\ua}^{pt} = A_{1,\ua}^{pt \dagger}$ $\left(A_{2,\da}^{pt} = A_{1,\da}^{pt \dagger} \right)$, $B^{pt}_{1/2} \equiv B_\mp$, $C_1 = i \sum_{\mathbf{q}, \lambda} \mathbf{d}\cdot\mathbf{u}^*_{\mathbf{q}\lambda}(\mathbf{r}_d) a^\dagger_{\mathbf{q}\lambda}$, and $C_2 = C^\dagger_1$, and expanding $\mathbf{E}(\mathbf{r}_d)$ into the corresponding photonic creation and annihilation operators \cite{Breuer2007}, we can write the Hamiltonian $H^{pt}_{IP}$ \cite{Scerri2017} as
\begin{equation}
H^{pt}_{IP} = \sum_{j\in\{ \ua,\da \}} \sum^2_{i=1} A^{pt}_{i,j} \otimes B^{pt}_i \otimes C_i~.
\end{equation}
Similarly we can write a more compact phonon interaction Hamiltonian 
\begin{equation}
H^{pn}_{IP} = \sum_{i=1}^2 A^{pn}_{i,\ua} \otimes B^{pn}_i ~,
\end{equation}
where $B^{pn}_{1/2} = \mathcal{B}_\mp = B_\mp - \Braket{B}$, $A^{pn}_{1,\ua} =  \Omega /2 \, \Ket{\ua} \Bra{T_\Ua}$ and $A^{pn}_{2, \ua} = A_{1,\ua}^{pn \dagger}$. Following the steps of Refs.~\citen{Nazir2016,Scerri2017}, we can now obtain the following ME in the polaron frame:
\begin{align}\label{addbathsME}
\begin{split}
\diff{}{t} &\rho_{SP}(t) = \\
 &-\int_0^\infty \mathrm{d}\tau \mathrm{Tr}^{pn}_E [ H^{pn}_{IP}(t), [ H^{pn}_{IP}(t-\tau), \rho_{SP}(t)\otimes\rho^{pn}_E(0) ] ]~  \\
 &-\int_0^\infty \mathrm{d}\tau \mathrm{Tr}_E [ H^{pt}_{IP}(t), [ H^{pt}_{IP}(t-\tau), \rho_{SP}(t)\otimes\rho_E(0) ] ]~,
\end{split}
\end{align}
where $\rho_{SP}(t)$ is the density matrix of the $\Lambda$-system in the polaron frame, and $\rho_E$ is the joint optical-vibrational density matrix. The above was derived under the assumption that the (initial) environmental state is thermal, hence $\rho_E(0) $ factorizes as $\rho_E(0) = \rho^{pn}_E(0) \otimes \rho^{pt}_E(0)$.

In the ME formalism, the rate $\gamma(\omega)$ of a dissipative process is given by $\gamma(\omega) = 2 \mathrm{Re}\left[ \int_0^\infty \mathrm{d}s K(s) \right]$, where $K(s)$ is the relevant correlation function [{\it c.f.}~Eq.~(3.137) in Ref.~\cite{Breuer2007}]. For our phonon dissipator (first term in Eq.~\ref{addbathsME}), these functions are given by
\begin{align}\label{phonon_cor_fns:reg}
\begin{split}
C^{pn}_{ii}(\tau) &= \mathrm{Tr}^{pn}_{E} \left[ \mathcal{B}^\dagger_\pm(\tau) \mathcal{B}_\pm(0) \rho^{pn}_E(0)\right] \\ 
& = \langle B \rangle^2 (\mathrm{e}^{\phi(\tau)} -1)~, 
\end{split}
\end{align}
\begin{align}\label{phonon_cor_fns:cross}
\begin{split}
C^{pn}_{ij}(\tau) &= \mathrm{Tr}^{pn}_{E} \left[ \mathcal{B}^\dagger_\pm(\tau) \mathcal{B}_\mp(0) \rho^{pn}_E(0)\right] \\
&= \langle B \rangle^2 (\mathrm{e}^{-\phi(\tau)} -1)~, 
\end{split}
\end{align}
where $i, j \in \{ 1,2 \}$, $i \neq j$. {The temperature-dependent phonon propagator $\phi(\tau)$ is given by

\begin{equation}\label{phon_propagator}
\phi(\tau) = \int^\infty _0  \frac{J_{pn} (\omega)}{\omega^2} \left( \cos(\omega \tau) \coth(\beta \omega / 2) - i \sin(\omega \tau) \right)~,
\end{equation}
where $\beta = 1/k_B T$, with $k_B$ being the Boltzmann constant and $T$ the temperature. This allows us to write the operator expectation $\langle B \rangle$ as 

\begin{align}\label{phon_propagator}
\begin{split}
\langle B \rangle &= \exp\left[-\frac{1}{2} \int^\infty _0  \frac{J_{pn} (\omega)}{\omega^2} \coth(\beta \omega / 2) \right] \\
&= \exp\left[ -\frac{1}{2} \phi(0) \right]~.
\end{split}
\end{align}
}
After some algebra, we obtain a phonon dissipator of the form
\begin{align}\label{phonon_diss}
\begin{split}
&\gamma^{pn}(\omega') (\mathcal{L}[\sigma^\ua_-]+\mathcal{L}[\sigma^\da_-])  + \gamma^{pn}(-\omega') (\mathcal{L}[\sigma^\ua_+] + \mathcal{L}[\sigma^\da_+]) \\[10pt]
&- \gamma^{pn}_{cd}(\omega')  (\mathcal{L}_{cd}[\sigma^\ua_-]+\mathcal{L}_{cd}[\sigma^\da_-]) \\[10pt]
&- \gamma^{pn}_{cd}(-\omega') (\mathcal{L}_{cd}[\sigma^\ua_+] +\mathcal{L}_{cd}[\sigma^\da_+])  ~,
\end{split}
\end{align}
where $\sigma^{\ua,\da}_- = \Ket{\ua,\da}\Bra{T_\Ua}$, $\sigma^{\ua,\da}_+ = \Ket{T_\Ua}\Bra{\ua,\da}$, $\mathcal{L}[C] = C \rho_{SP} C^\dagger - \frac{1}{2}\{C^\dagger C, \rho_{SP} \}$ and $\mathcal{L}_{cd}[C] = C \rho_{SP} C - \frac{1}{2}\{C^2, \rho_{SP} \}$. The phonon absorption ($\gamma^{pn}(\omega')$), emission ($\gamma^{pn}(-\omega')$) and cross-dephasing rates ($\gamma^{pn}_{cd}(-\omega')$) \cite{Scerri2017, Hughes2015a, Hughes2015b} are given by
\begin{align*} 
\gamma^{pn}(\pm \omega') &= \frac{\left| \Omega^{pn} \right|^2}{4} \int_{-\infty}^\infty \mathrm{d}\tau \; \mathrm{e}^{\pm i \omega' \tau} \left( \mathrm{e}^{\phi(\tau)} - 1 \right)~, \\
\gamma^{pn}_{cd}(\omega') &= \frac{\left( \Omega^{pn} \right)^2}{4} \int_{-\infty}^\infty \mathrm{d}\tau \; \cos(\omega' \tau) \left( 1- \mathrm{e}^{-\phi(\tau)} \right)~, \\
\gamma^{pn}_{cd}(-\omega') &= \frac{\left( \Omega^{pn} \right)^2}{4} \int_{-\infty}^\infty \mathrm{d}\tau \; \cos(\omega' \tau) \left( 1- \mathrm{e}^{-\phi(\tau)} \right)~,
\end{align*}

Having derived the corresponding phonon absorption/emission rates, we now briefly turn our attention to the second term in Eq.~\eqref{addbathsME}. This term governs the system's interaction with the electromagnetic environment in the polaron frame, and thus gives the corresponding correlation functions 
\begin{align}\label{photon_cor_fns}
&C^{pt}_{ij,\ua}(\tau) = C^{pt}_{ij, \da}(\tau) \coloneqq C^{pt}_{ij}(\tau) \\
&= \mathrm{Tr}_{E} \left[ \left(B^{pt \dagger}_i(\tau) \otimes C^\dagger_i(\tau) \right) \left( B^{pt}_j(0) \otimes C_j(0) \right) \rho_E(0)\right]~ \nonumber \\
&= \mathrm{Tr}^{pn}_{E} \left[ B^{pt \dagger}_i(\tau) B^{pt}_j(0)  \rho^{pn}_E(0)\right]  \mathrm{Tr}^{pt}_{E} \left[ C^\dagger_i(\tau) C_j(0) \rho^{pt}_E(0)\right]~, \nonumber
\end{align}
where $i, j \in \{ 1,2 \}$. As the two transitions $\Ket{\ua} \leftrightarrow \Ket{T_\Ua}$ and $\Ket{\da} \leftrightarrow \Ket{T_\Ua}$ are coupled to both the vibrational and electromagnetic baths, the above correlation function is the same for both transitions of the $\Lambda$ system. We also note that the cross correlation terms involving both transitions will vanish due to the orthogonal dipole moments of the two transitions \cite{Ficke2005}. After substituting for the bath operators, and making use of the creation/annihilation commutation relations, we get that the only the $C^{pt}_{11}$ correlation function is non zero, giving the standard photon emission rates $\gamma_{11,j} = \omega^3_j d^2_j / (3\pi \epsilon_0 \hbar c^3)$, $j \in \{ \ua, \da \}$ \cite{Ficke2005}.

\section{X$^0$ two-level system dynamics}

In this section, we briefly sketch the derivation for the X$^0$ exciton master equation, analogous to the charged exciton master equation from the previous section. Denoting the ground and excited states for this two-level system as $\Ket{0}$ and $\Ket{X}$, respectively, the Hamiltonian in the frame rotating at the exciton transition frequency $\omega_0$ is given by

\begin{align}
\begin{split}
H_{tot} =& H + H^{pt}_E + H^{pn}_E + H^{pt}_I + H^{pn}_I~; \\
\vspace{5mm}
H =&  \frac{\Omega}{2} \left( \Ket{X} \Bra{0}  + \Ket{0} \Bra{X} \right) ~,\\
H^{pt}_E =& \sum_{\mathbf{q}, \, \lambda} \nu_\mathbf{q} a^\dagger_{\mathbf{q}\lambda} a_{\mathbf{q}\lambda}~,\\
H^{pn}_E =& \sum_\mathbf{k} \omega_\mathbf{k} b^\dagger_\mathbf{k} b_\mathbf{k}~,\\
H^{pt}_I =&  -\mathbf{d}_{X0} \cdot \mathbf{E}(\mathbf{r}_d) (\Ket{0} \Bra{X} + \Ket{X} \Bra{0}) \\
 & -\mathbf{d}_{X0} \cdot \mathbf{E}(\mathbf{r}_d) (\Ket{0} \Bra{X} + \Ket{X} \Bra{0}) ~,\\
H^{pn}_I =& \Ket{X} \Bra{X}\sum_{\mathbf{k}} g_\mathbf{k} ( b^\dagger_\mathbf{k} + b_\mathbf{k} ) ~,
\end{split}
\end{align}
where $\mathbf{d}_{X0}$ is the transition matrix element for the exciton transition. Thus, using the polaron transformation $S_{X} = \Ket{X}\Bra{X} \sum_\mathbf{k} \frac{g_\mathbf{k}}{\omega_\mathbf{k}} (b^\dagger_\mathbf{k}-b_\mathbf{k})$ and following the derivation in the previous section and Ref.~\citep{Nazir2016}, we arrive at a similar master equation for the neutral exciton, with the phonon dissipator given by

\begin{align}\label{X0_dissipator}
\begin{split}
&\gamma^{pn}(\omega') \mathcal{L}[\sigma_-]  + \gamma^{pn}(-\omega') \mathcal{L}[\sigma_+]  \\[10pt]
&- \gamma^{pn}_{cd}(\omega') \mathcal{L}_{cd}[\sigma_-] - \gamma^{pn}_{cd}(-\omega') \mathcal{L}_{cd}[\sigma_+]  ~,
\end{split}
\end{align}
where $\sigma_- = \Ket{0}\Bra{X}$, whilst the electromagnetic environment dissipation can be described by a Lindblad superoperator $\mathcal{L}[\sigma_-]$ and a dissipation rate $\gamma_{0} = \omega^3_0 d^2_j / (3\pi \epsilon_0 \hbar c^3)$, that is, $\gamma_0 \mathcal{L}[\sigma_-]$.

{
\section{Polaron frame correlation operators}

Without going into the explicit form of every term of the HOM $g^{(2)}(\tau)$ in the polaron frame, we note that a general correlation operator of the form $\Braket{\sigma_+(t_1)\sigma_+(t_2)\sigma_-(t_3)\sigma_-(t_4)}$ in the polaron frame is given by

\begin{align}
\begin{split}
&\Braket{\sigma_+(t_1)B_+(t_1)\sigma_+(t_2)B_+(t_2)\sigma_-(t_3)B_-(t_3)\sigma_-(t_4)B_-(t_4)} \\
&\hspace{10mm}\approx\Braket{B_+(t_1)B_+(t_2)B_-(t_3)B_-(t_4)}  \\
&\hspace{15mm}\times\Braket{\sigma_+(t_1)\sigma_+(t_2)\sigma_-(t_3)\sigma_-(t_4)}~,
\end{split}
\end{align}
where the approximation in the second holds due to the difference in timescales associated with the exciton lifetime (nanoseconds) and the phonon bath relaxation time (picoseconds) \cite{iles-smith_limits_2017}. The phonon correlation operator can be easily expanded as

\begin{align}\label{phonon_corr}
\begin{split}
&\Braket{B_+(t_1)B_+(t_2)B_-(t_3)B_-(t_4)} \\
&= \Braket{\mathcal{B}^{t_1, t_2}_{t_3, t_4}} \\
&\quad \times \exp \left[ i \Im \{ \phi (t_2-t_1) \} \right] \exp \left[ i \Im \{ \phi (t_4-t_3) \} \right] \\
&\quad \times \exp \left[ i \Im \{ \phi (t_1+t_2 - t_3 - t_4) \}) \}\right]~,
\end{split}
\end{align}
where $\Im$ denotes the imaginary part, and $\Braket{\mathcal{B}^{t_1, t_2}_{t_3, t_4}}$ is given by

\begin{align}
\begin{split}
&\Braket{\mathcal{B}^{t_1, t_2}_{t_3, t_4}} \\
&=\exp\left[-\frac{1}{2} \int^\infty _0  \frac{J_{pn} (\omega)}{\omega^2} |\mathcal{K}^{t_1,t_2}_{t_3,t_4}|^2 \coth(\beta \omega / 2) \right]~,
\end{split}
\end{align}
where $\mathcal{K}^{t_1,t_2}_{t_3,t_4} = \mathrm{e}^{i \omega t_1}+\mathrm{e}^{i \omega t_2}-\mathrm{e}^{i \omega t_3}-\mathrm{e}^{i \omega t_4}$. These expressions can be simplified further by noting that the phonon propagator vanishes on a picosecond timescale, resulting in an easier numerical evaluation of Eq.~\eqref{phonon_corr}.
}

\vspace{1cm}
\section{Resonance fluorescence spectrum}

In this section, we outline the derivation of the RF spectral function in the presence of strong vibrational coupling. In the polaron frame, this quantity is simply given by the Fourier transform of the (steady-state) first order correlation function $\mathrm{lim}_{t \rightarrow \infty}\langle \mathbf{E}^{(-)}(\mathbf{R}, t) \mathbf{E}^{(+)}(\mathbf{R}, t + \tau) \rangle$, where $\mathbf{E}^{(-)}(\mathbf{R}, t)$ and $\mathbf{E}^{(+)}(\mathbf{R}, t)$ are, respectively, the negative and positive components of the electric field operator evaluated at the position $\mathbf{R}$ of the detector \cite{Ficke2005}. 
\begin{align}\label{spectrum}
\begin{split}
S(\omega) \propto \int_{-\infty}^\infty \mathrm{d}\tau &\sum_{j \in {\ua, \da}} \mathrm{e}^{-i (\omega - \omega') \tau} \times \\
&\langle \sigma^j_+(\tau) B_+(\tau) \sigma^j_-(0) B_-(0) \rangle_{s}~, \\
\end{split}
\end{align}
where we have exploited the temporal homogeneity of the stationary correlation function, and where the subscript `s' denotes the trace taken with respect the steady-state density matrix. Due to the different timescales \cite{Scerri2017} for the phonon and photon processes, Eq.~\eqref{spectrum} simplifies to
\begin{align}\label{spectrum_simplified}
\begin{split}
S(\omega) \propto \langle B \rangle^2\int_{-\infty}^\infty \mathrm{d}\tau &\sum_{j \in {\ua, \da}} \mathrm{e}^{-i (\omega - \omega') \tau} \mathrm{e}^{\phi(\tau)} \times  \\ 
&\langle \sigma^j_+(\tau) \sigma^j_-(0) \rangle_{s}~. \\
\end{split}~.
\end{align}
Similarly, for the neutral exciton, the spectral function is given by

\begin{align}\label{spectrum_simplified_X0}
\begin{split}
S(\omega) \propto \langle B \rangle^2\int_{-\infty}^\infty \mathrm{d}\tau & \mathrm{e}^{-i (\omega - \omega') \tau} \mathrm{e}^{\phi(\tau)} \times  \\ 
&\langle \sigma_+(\tau) \sigma_-(0) \rangle_{s}~. \\
\end{split}~
\end{align}

\bibliographystyle{apsrev4-2}
\bibliography{reference2}% Produces the bibliography via BibTeX.